\documentclass [] {aa} 
\usepackage{txfonts,graphicx}

\begin{document}

\title{C, S, Zn and Cu abundances in planet-harbouring stars\thanks{Based on
    observations collected at the La Silla Observatory, ESO (Chile),
    with the {\footnotesize CORALIE} spectrograph at the 1.2-m
    Euler Swiss telescope and with the {\footnotesize FEROS} spectrograph    
    at the 1.52-m and 2.2-m ESO telescopes, at the Paranal Observatory, ESO (Chile), using the UVES spectrograph
    at the {\footnotesize VLT/UT2} Kueyen telescope, and with the UES and SARG spectrographs at the 4-m William 
    Hershel Telescope (WHT) and at the 3.5-m TNG telescope, respectively, both 
    at La Palma (Canary Islands).}}

\author{A.~Ecuvillon,\inst{1} G.~Israelian,\inst{1} N. C.~ Santos,\inst{2,3} 
 M.~Mayor,\inst{3} V.~Villar\inst{4} \and G.~Bihain\inst{1}}

\offprints{ \email{aecuvill@ll.iac.es}}

\institute{Instituto de Astrof\'{\i}sica de Canarias, E-38200 La Laguna, Tenerife, Spain \and Centro de 
Astronomia e Astrofisica de Universidade de Lisboa, Observatorio Astronomico de Lisboa, Tapada de Ajuda, 1349-018 Lisboa, 
Portugal \and Observatoire de Gen\`eve, 51 ch.  des  Maillettes, CH--1290 Sauverny, Switzerland \and Departamento de
Astrof\'{\i}sica, Fac. de F\'{\i}sicas, Universidad Complutense de Madrid, Ciudad Universitaria, E-28040 Madrid, Spain}

\date{Received 22 April 2004 / Accepted 17 June 2004} 

\titlerunning{C, S, Zn and Cu abundances in Planet-harbouring stars} 
\authorrunning{A. Ecuvillon et al.}

\abstract{We present a detailed and uniform study of C, S, Zn and Cu abundances in a large set of planet host stars, as 
well as in a homogeneous comparison sample of solar-type dwarfs with no known planetary-mass companions. Carbon 
abundances were 
derived by {EW} measurement of two \ion{C}{i} optical lines, while spectral syntheses were 
performed for S, Zn and Cu. We investigated possible differences in the behaviours of the volatiles C, S and Zn and in the
refractory Cu in targets with and without known planets in order to check possible anomalies due
to the presence of planets. We found that the abundance distributions in stars with exoplanets are the high [Fe/H]
extensions of the trends traced by the comparison sample. All volatile elements we studied show [$X$/Fe] trends
decreasing with [Fe/H] in the metallicity range $-0.8<$ [Fe/H] $<$ 0.5, with significantly negative slopes of 
$-0.39\pm0.04$ 
and $-0.35\pm0.04$ for C and S, respectively. A comparison of our abundances with those available in the literature 
shows good agreement in most cases.  
	  }
\maketitle

\section{Introduction}
Since the first success in the search for exoplanets  with the discovery of the planet orbiting 51 Peg (Mayor \& Queloz 
\cite{May95}),
studies on the formation and evolution of planetary systems have  increased. Several 
spectroscopic analyses of stars with planets have recently been carried out. Most of them used iron as the reference element 
(Gonzalez \cite{Gonz97}; Murray \& Chaboyer \cite{Mur02}; Laws et al.\ \cite{Law03}; Santos et al.\ \cite{San01}, 
\cite{San03b}, \cite{San04a}; for a review see Santos et al.\ \cite{San03a}), while only a few studies determined abundance 
trends for other metals (Gonzalez \& Laws \cite{Gonz00}; Gonzalez et al.\ \cite{Gonz01}; Santos et al.\ \cite{San00}; 
Bodaghee et al.\ \cite{Bod03}; Takeda et al.\
\cite{Tak01}; Sadakane et al.\ \cite{Sad02}; for a review see Israelian \cite{Isr03b}). 

One of the most remarkable results is that planet-harbouring stars are on average 
more metal-rich than solar-type disc stars (Gonzalez \cite{Gonz97}; Gonzalez et al.\ \cite{Gonz01}; Laws et al.\ 
\cite{Law03}; Santos et al.\ \cite{San01}, \cite{San03b}, \cite{San04a}).  Two main explanations have been suggested
for linking this metallicity excess with the 
presence of
planets. The first of these, the ``self-enrichement'' hypothesis, attributes 
the origin 
of the observed overabundance of metals to the accretion of large amounts of metal-rich H- and He-depleted rocky 
planetesimal materials on to the star (Gonzalez \cite{Gonz97}). The opposite view, the ``primordial'' hypothesis, considers 
the metallicity enhancement to be caused by the high metal content of the protoplanetary cloud from which the planetary 
system formed (Santos et al.\ \cite{San00}, \cite{San01}). Discriminating between these two possibilities will help in 
understanding how planetary systems form. 

Light elements may give fundamental information about the mixing, diffusion and
angular momentum history of exoplanets hosts, as well as stellar activity caused by interaction with exoplanets (Santos et 
al.\ \cite{San02}, \cite{San04b}; Israelian et al.\ \cite{Isr04}). Studies of Be, Li and the isotopic ratio
\element[][6]{Li}/\element[][7]{Li} could give evidences to distinguish between  different planet formation
theories (Sandquist et al.\ \cite{Sand02}).

Murray \& Chaboyer (\cite{Mur02}) concluded that an accretion of 5.0 $M_\oplus$ of iron while on the main sequence would 
explain the overabundance of metals found in planet host stars. On the other hand, Pinsonneault et al. (\cite{Pin01})
rejected the ``self-enrichement'' scenario since the metallicity enhancement does not show the expected 
$T_\mathrm{eff}$ dependence; recent results by Vauclair (\cite{Vau04}) may have refuted, however, this argument. Furthermore, several studies have shown that  planetary frequency is rising as a function of
 [Fe/H] (Santos et al.\ \cite{San01}, \cite{San04a}; Reid \cite{Reid02}), result which seems to support the primordial 
metallicity enhancement scenario. However, evidences of pollution have been found for a few cases (Israelian et al.\ 
\cite{Isr01}, \cite{Isr03a}; Laws \& Gonzalez \cite{Law01}). 

\begin{figure*}
\centering 
\includegraphics[height=6cm]{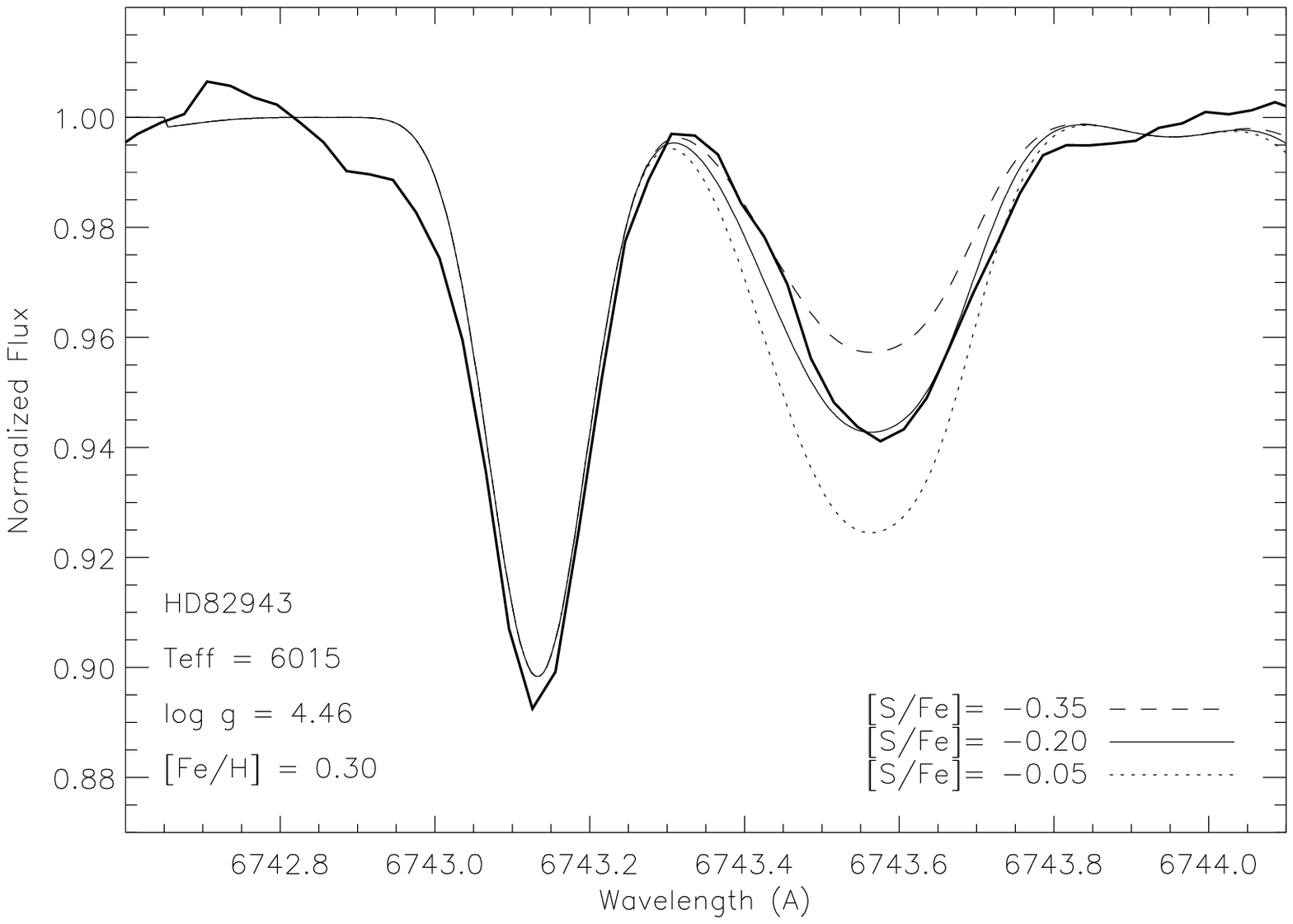}
\includegraphics[height=6cm]{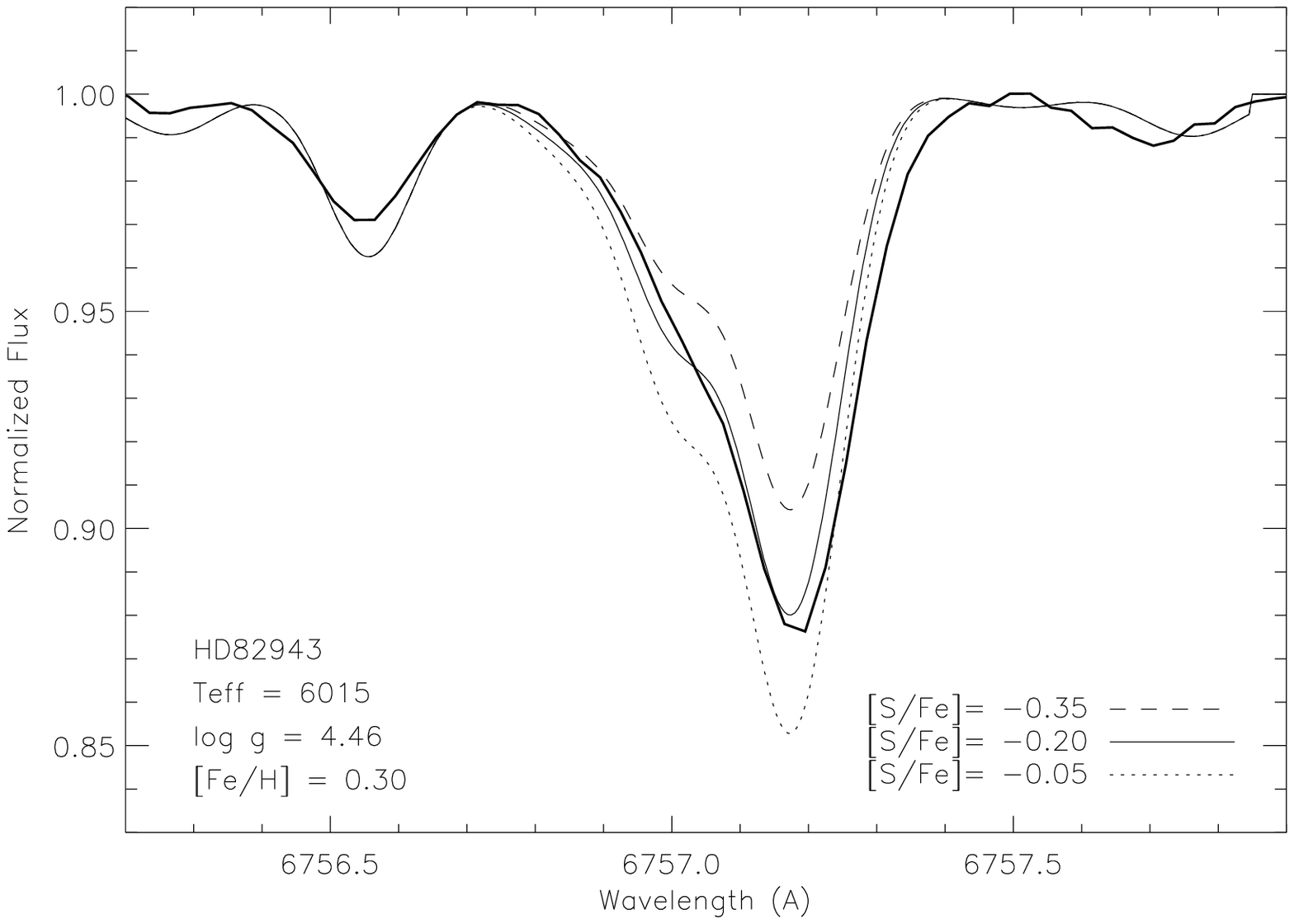}
\centering
\includegraphics[height=6cm]{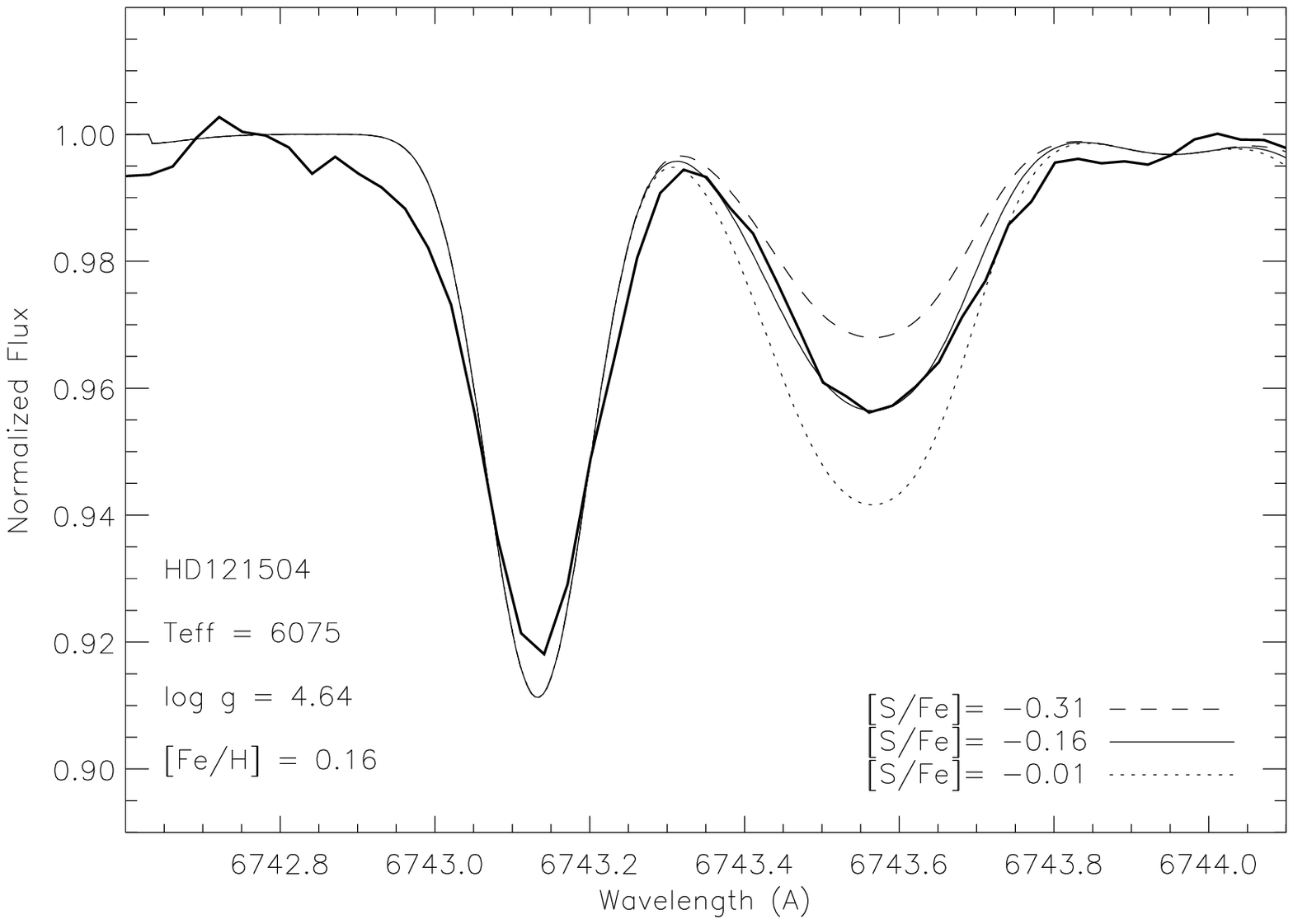}
\includegraphics[height=6cm]{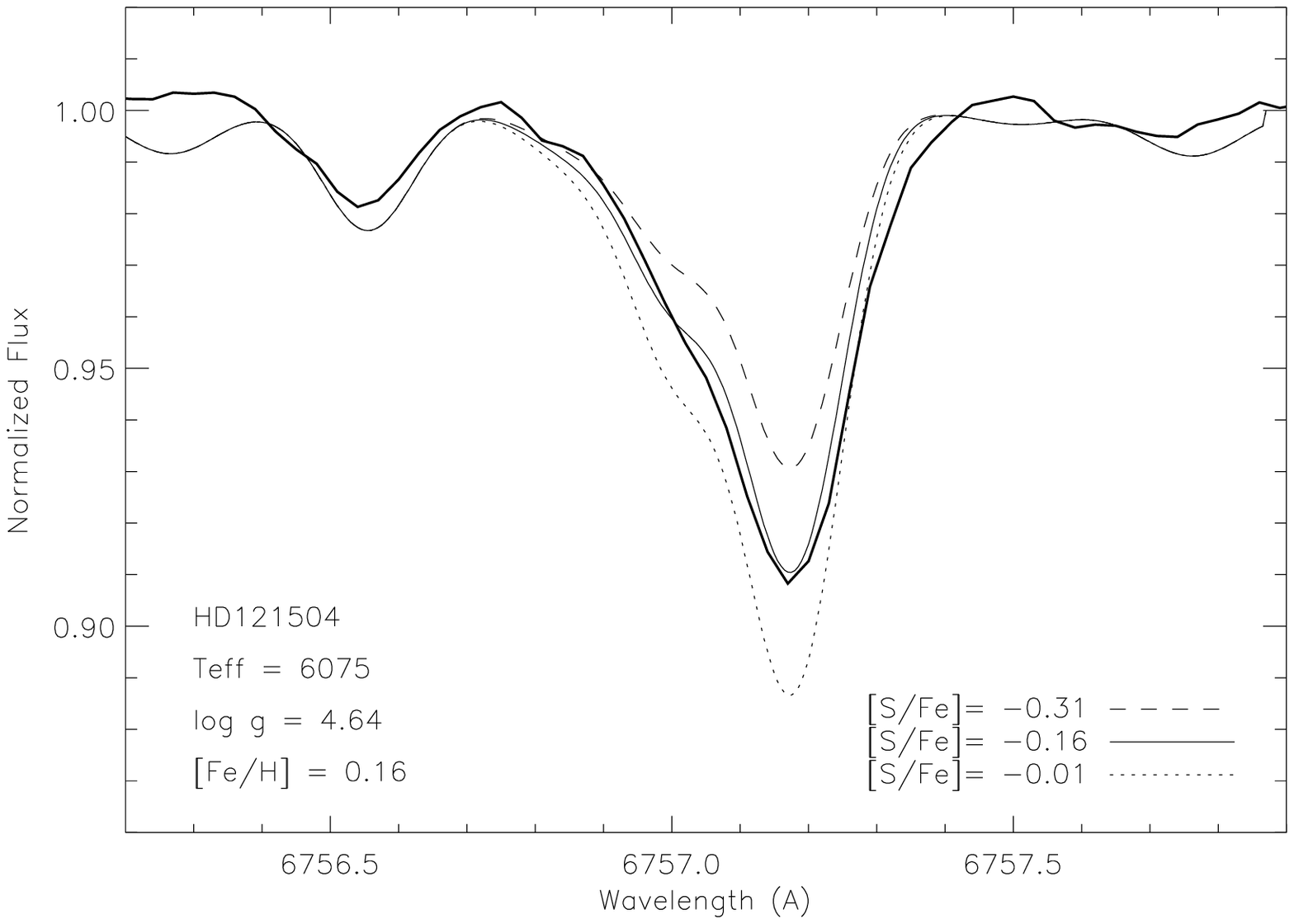}
\caption[]{The observed spectrum (thick solid line) and three synthetic spectra (dotted, dashed and solid lines) 
for different values of [S/Fe], in the spectral regions 6742.6--6744.1 \AA\ and 6756.2--6757.9 \AA, for two targets.}   
\label{fig1}
\end{figure*}

\begin{figure*}
\centering
\includegraphics[height=6cm]{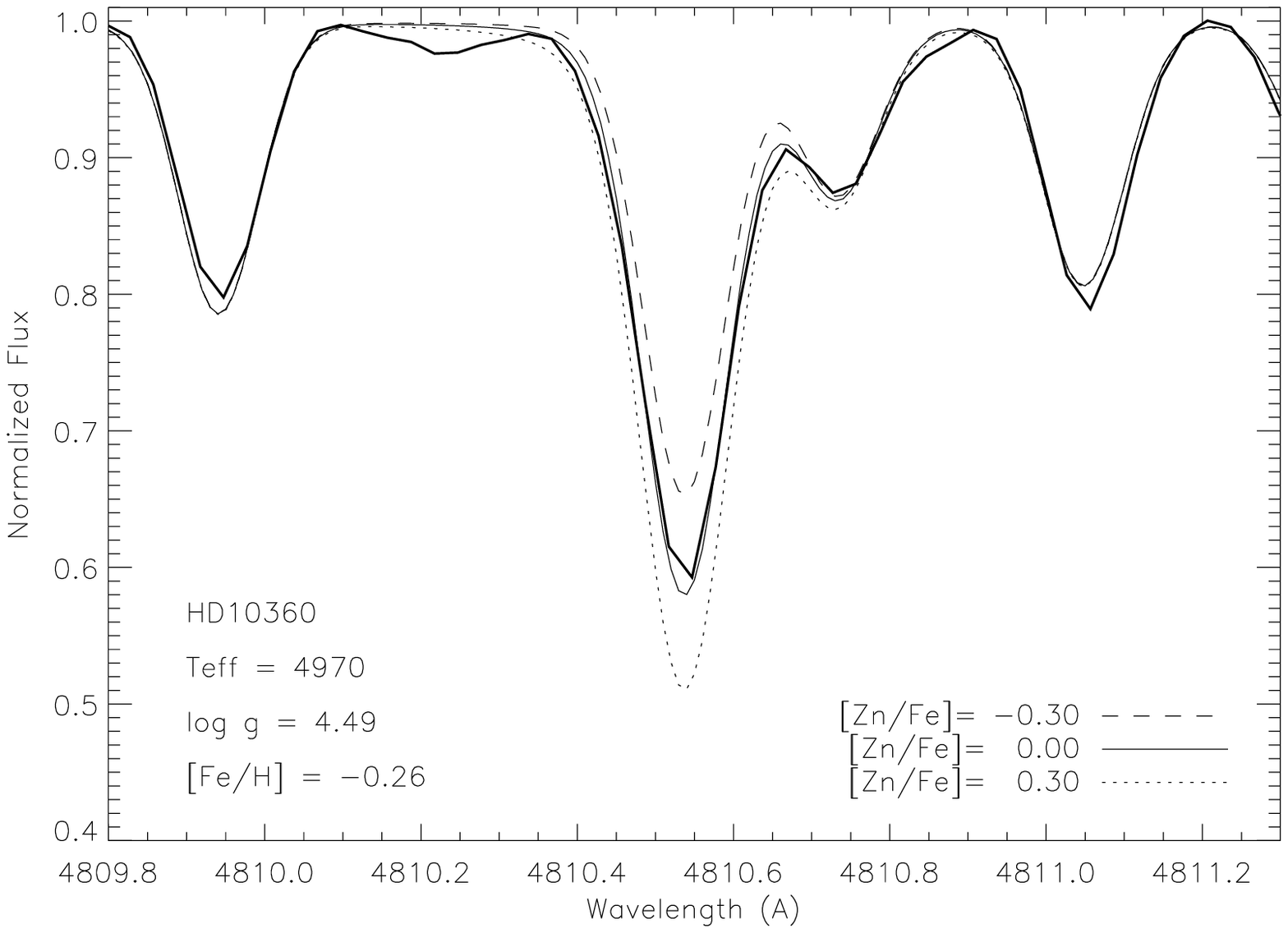}
\includegraphics[height=6cm]{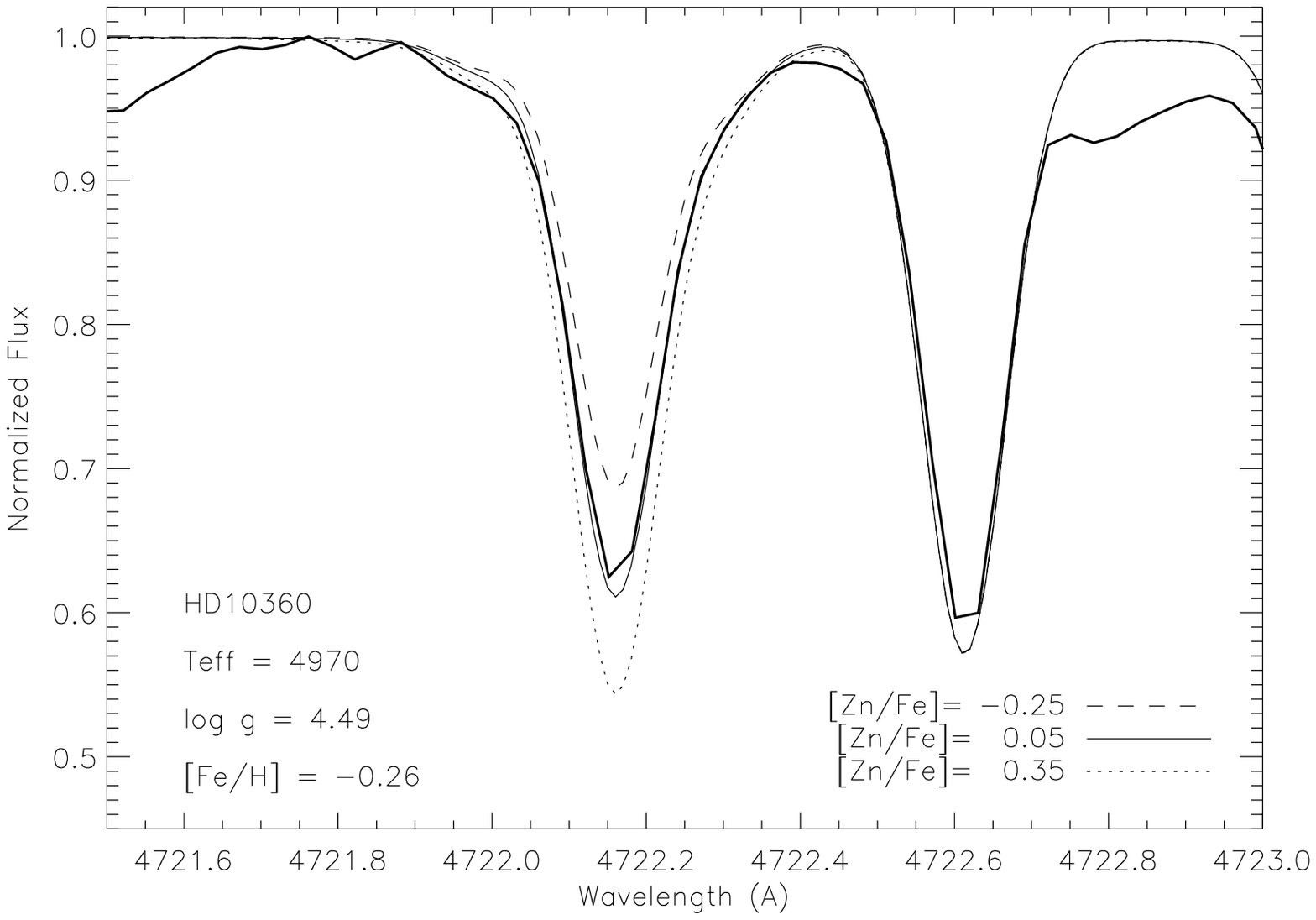}
\centering
\includegraphics[height=6cm]{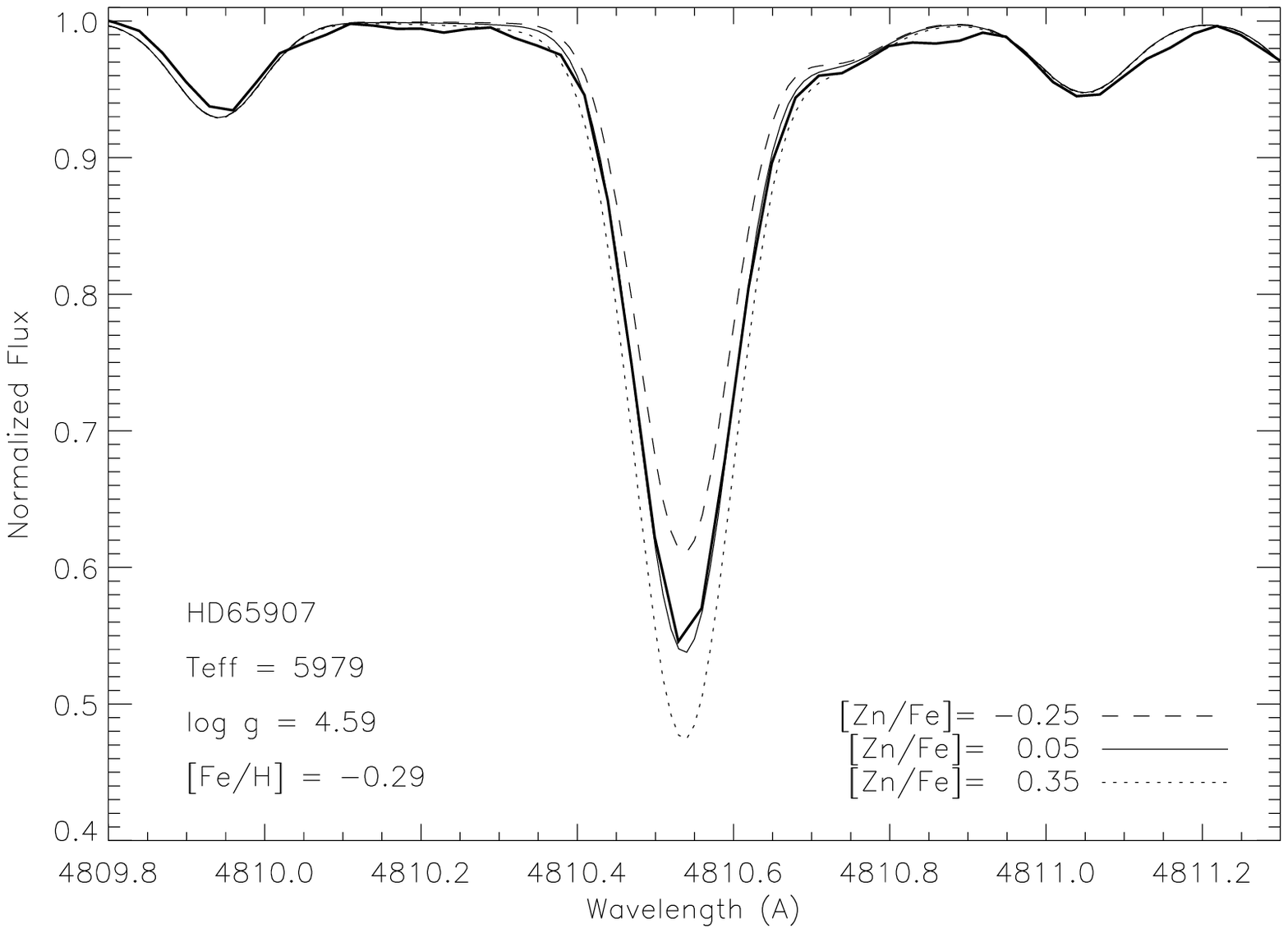}
\includegraphics[height=6cm]{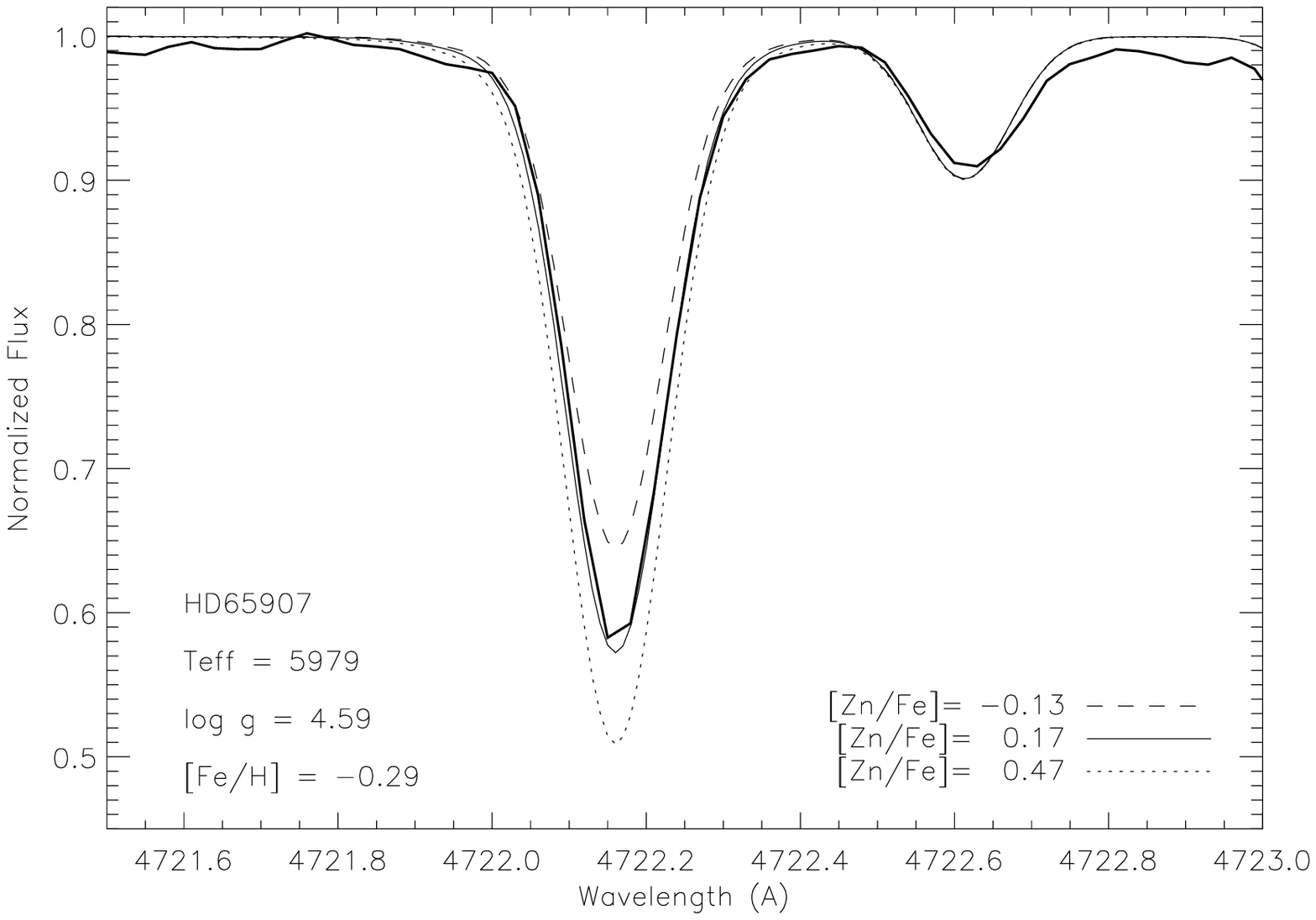}
\caption[]{The observed spectrum (thick solid line) and three synthetic spectra (dotted, dashed and solid lines) 
for different values of [Zn/Fe], in the spectral regions 4809.8--4811.3 \AA\ and 4721.5--4723.0 \AA, for two targets.}   
\label{fig2}
\end{figure*}

\begin{figure*}
\centering
\includegraphics[height=6cm]{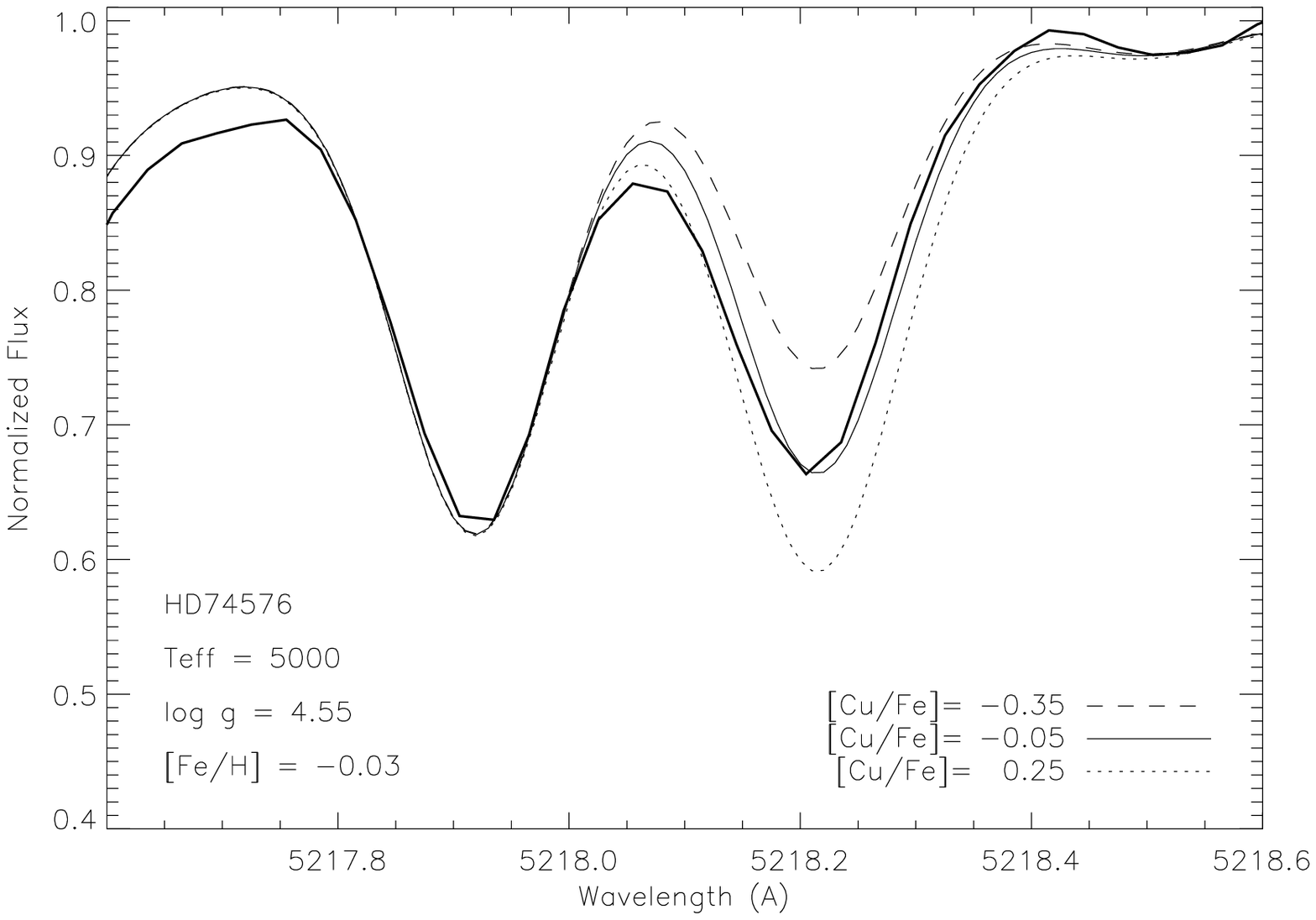}
\includegraphics[height=6cm]{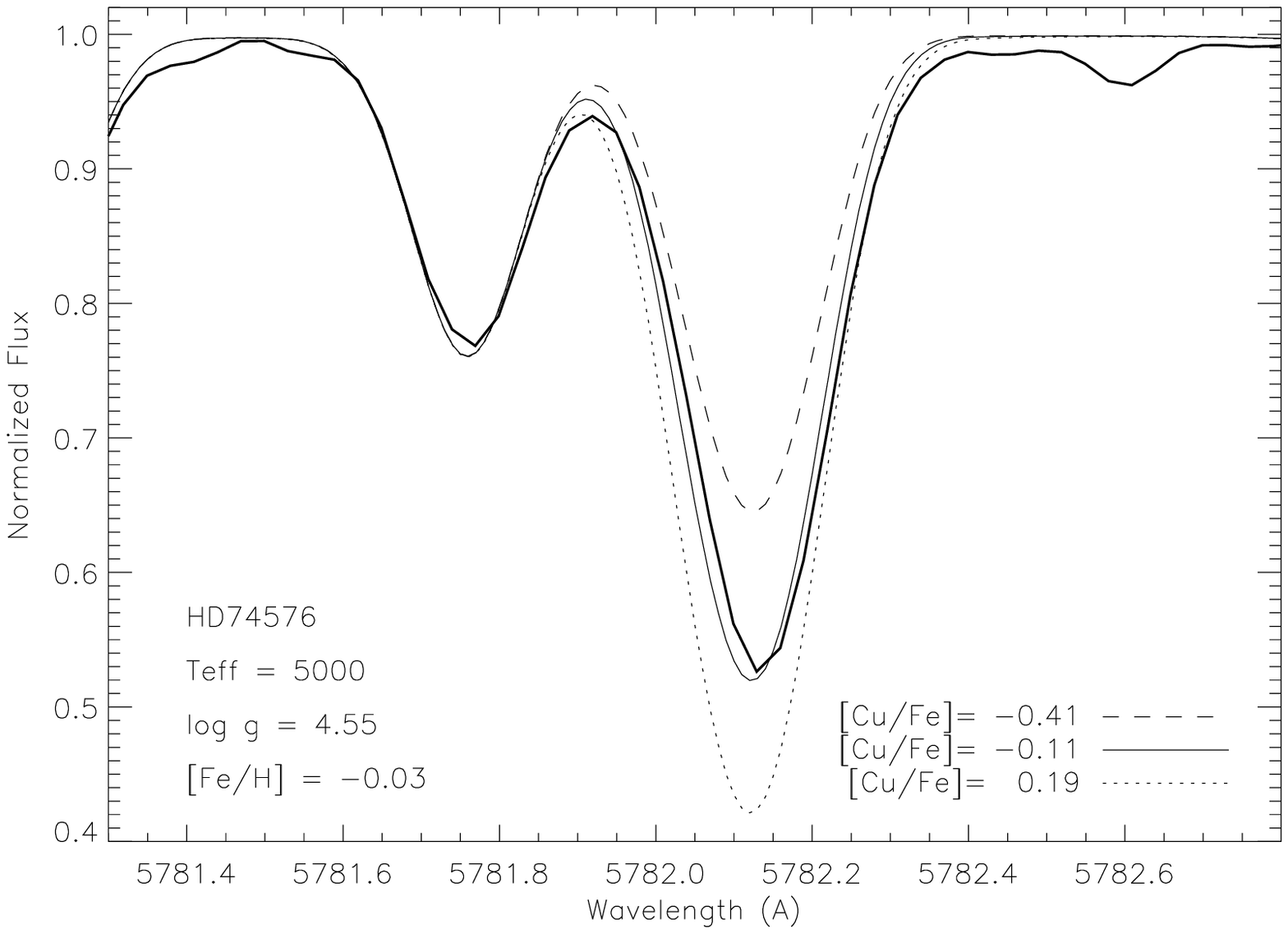}
\centering
\includegraphics[height=6cm]{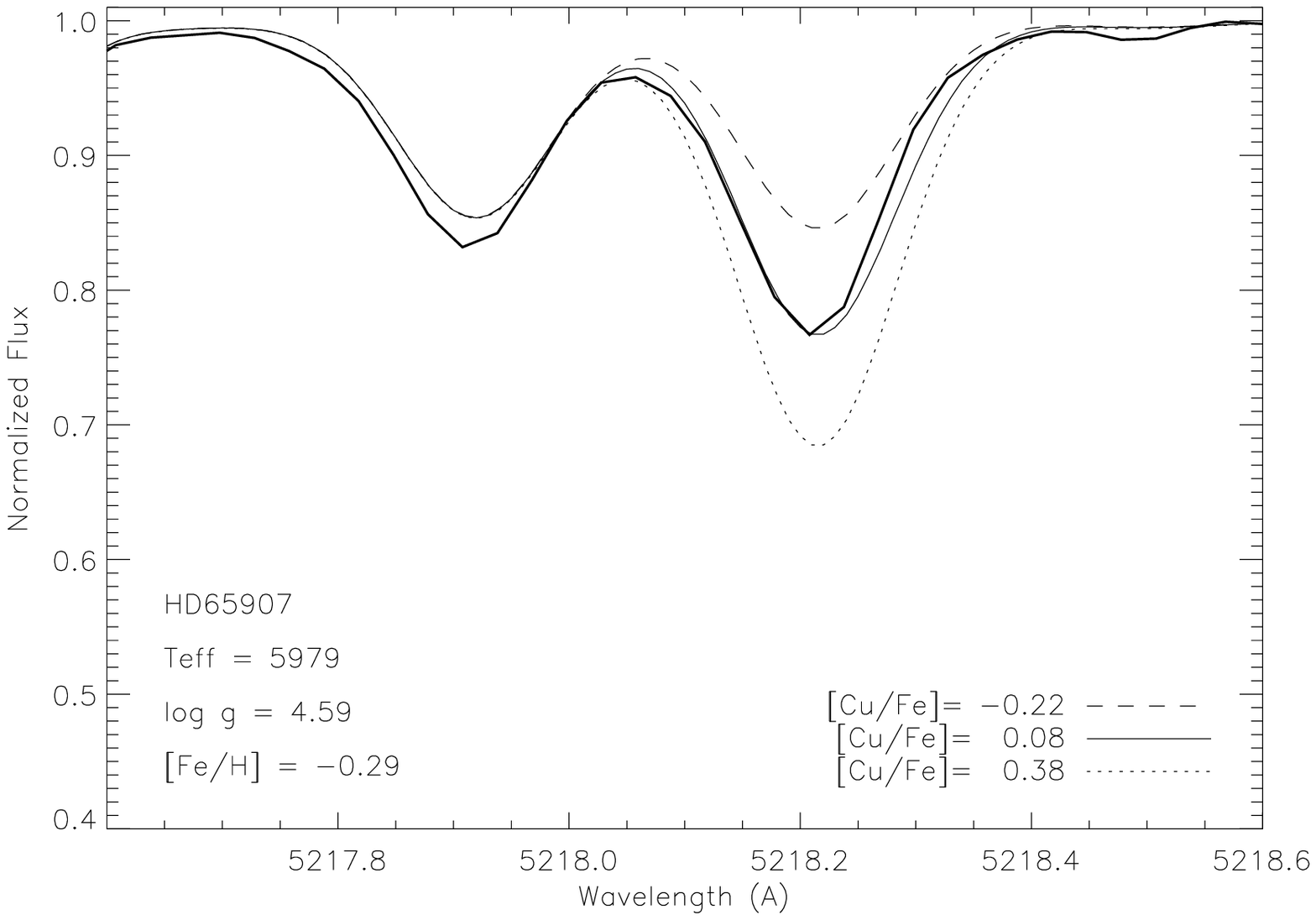}
\includegraphics[height=6cm]{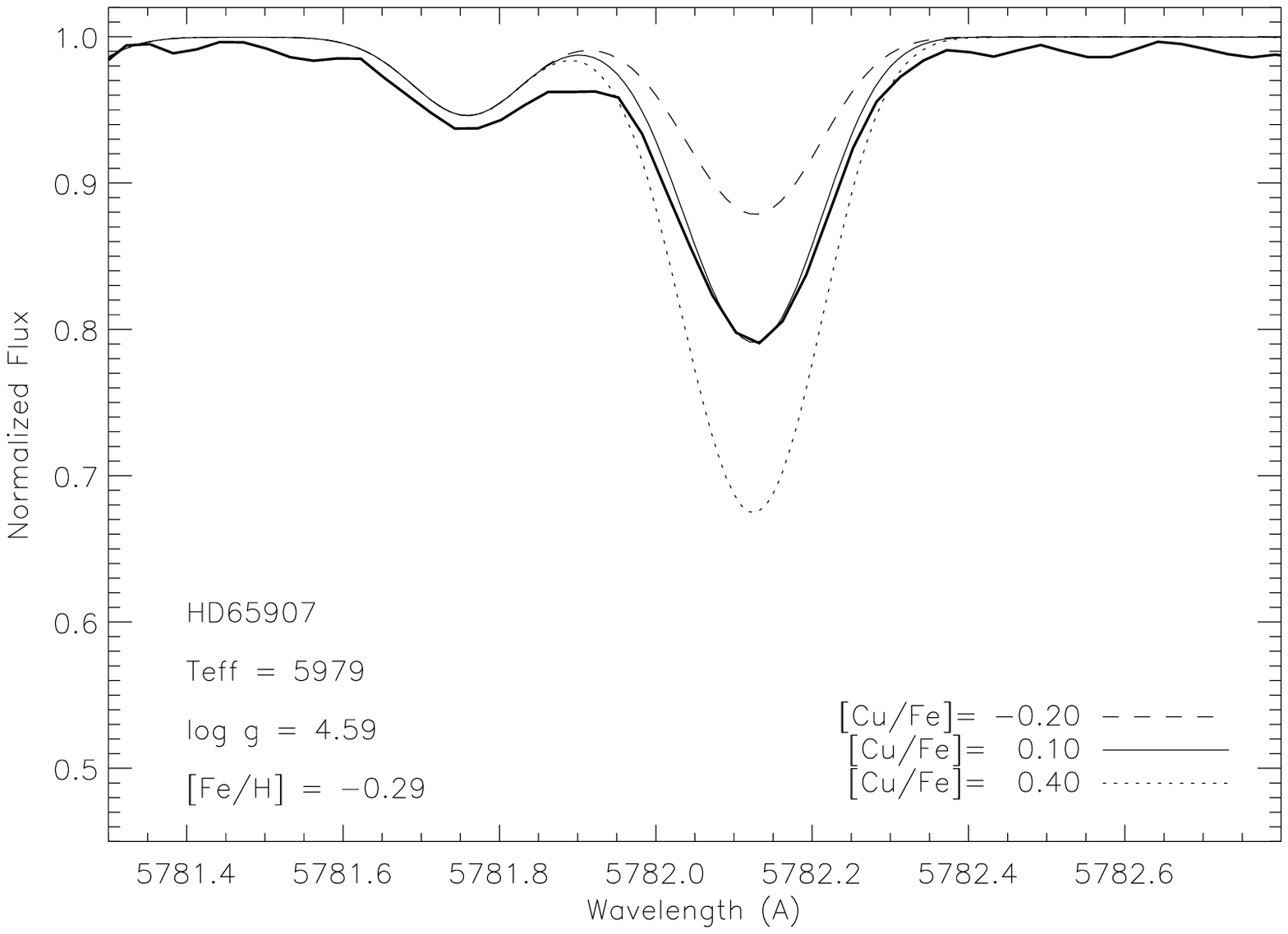}
\caption[]{The observed spectrum (thick solid line) and three synthetic spectra (dotted, dashed and solid lines)
for different values of [Cu/Fe], in the spectral regions 5217.6--5218.6 \AA\ and 5781.3--5782.8 \AA, for two targets.}
\label{fig3}
\end{figure*}
 
Abundance trends of different elements may provide important clues to the question. Volatile elements (with low 
condensation temperature $T_{\rm C}$) are expected to be deficient in accreted materials because of the high temperatures near 
the star. Therefore, the ``self-enrichement'' scenario should lead to a relative overabundance of refractories, such as the
$\alpha$-elements Si, Mg, Ca, Ti and the iron-group elements compared to volatiles, such as CNO, S and Zn. 
Smith et al. (\cite{Smi01}) found a small subset of planet host stars which bore this accretion signature since these 
stars exhibited an accentuated trend of increasing [$X$/H] with increasing $T_{\rm C}$ for a given element $X$. 
Nevertheless, subsequent studies 
have obtained a similar behaviour in volatiles as in refractory elements (Takeda et al.\ \cite{Tak01}; Sadakane et al.\ 
\cite{Sad02}). 

Other evidence in favour of a ``primordial'' 
high metallicity has been provided by several studies, which conclude that the
 abundances of volatiles in planet host stars do not 
reveal significant differences from those in field dwarfs of the same metallicity (Santos et al.\ \cite{San00}; Gonzalez et 
al.\ \cite{Gonz01}; Takeda et al.\ \cite{Tak01}; Sadakane et al.\ \cite{Sad02}). However, all these authors used
inhomogeneous comparison samples of field dwarfs from the literature, which could be a source of systematic errors.
A sample of stars with no known planets has been prepared by Santos et al.\ (\cite{San01}) and analysed in the same way
as the planet host set in several studies of the  stellar and kinematic parameters of stars with planets (Santos et al.\ 
\cite{San03b}) and about their metal-rich nature (Santos et al.\ \cite{San01}, \cite{San04a}). Other uniform and unbiased 
comparisons have been carried out for abundances of elements other than iron -- some $\alpha$- and Fe-group elements 
(Bodaghee et al.\ \cite{Bod03}) and  the volatile nitrogen (Ecuvillon et al.\ \cite{Ecu04}) -- and conclude that the 
abundance trends observed in planet host stars are almost identical to those in the comparison sample. 

Sulphur is a volatile $\alpha$ element which behaves as true primary; it is well established that it is produced 
mainly by Type II supernovae (SNe II) of massive stars (e.g. Timmes, Woosley \& Weaver \cite{Tim95}; Carretta, Gratton
\& Sneden \cite{Car00}). For Cu and Zn the situation is rather confused; massive (M$>$8M$_\odot$), low- (M$<$4M$_\odot$) and
intermediate-mass stars (4M$_\odot$$<$M$<$8M$_\odot$) 
are likely contributors, although their relative weights are still uncertain (e.g. Luck \& Bond \cite{Luc85};
Sneden et al.\ \cite{Sne91}; Mishenina et al.\ \cite{Mis02}; Bihain et al.\ \cite{Bih04}). A number of
different production sites have been suggested for carbon: supernovae, novae, Wolf-Rayet stars, intermediate-
and low-mass stars. Recently, Kobulnicky \& Skillman (\cite{Kob98}) concluded that low- and intermediate-
mass stars are the significant contributors of carbon. On the other hand, Prantzos et al.\ (\cite{Pra94}) and
Gustafsson et al.\ (\cite{Gus99}) found massive stars to be significant sites for the production of carbon. At
the present the situation is not clear.

This work presents a systematic, uniform and detailed study of the volatile elements C, S and Zn and of the refractory Cu
in two large samples, a set of planet-harbouring stars and a volume-limited comparison sample of stars with no known
planetary mass companions. We studied C, S, Zn and Cu abundances in several optical lines, by measuring {EW} in the case of
carbon and by the synthesis technique for the other elements. We compare our results with those of other recent studies 
(Santos et al.\ \cite{San00}; Gonzalez et al.\ \cite{Gonz01}; Takeda et al.\ \cite{Tak01}; Sadakane et al.\ \cite{Sad02}). 

\begin{table}
\caption[]{Observing log for the new set of data. The S/N ratio is provided at 6050 \AA\ and 5375 \AA}
\begin{center}
\begin{tabular}{lcccr}
\hline
\noalign{\smallskip}
Star & $V$ & Obser. run & S/N & Date \\
\hline
\hline
\noalign{\smallskip}
\object{HD\,7570}   & 5.0 & UVES &  850 & July 2001 \\
\object{HD\,9826}   & 4.1 & UES  &  500 & Feb.\ 2001 \\
\object{HD\,16141}  & 6.8 & UES  &  800 & Oct.\ 2001 \\
\object{HD\,17051}  & 5.4 & UVES &  950 & July 2001 \\
\object{HD\,19994}  & 5.1 & UVES & 1100 & Aug.\ 2001 \\
\object{HD\,22049}  & 3.7 & UES  &  500 & Oct.\ 2001 \\
\object{HD\,52265}  & 6.3 & UVES & 1200 & Mar.\ 2001 \\
\object{HD\,75289}  & 6.4 & UVES & 1000 & Mar.\ 2001 \\
\object{HD\,82943}  & 6.5 & UVES &  950 & Mar.\ 2001 \\
\object{HD\,89744}  & 5.7 & UES  &  680 & Feb.\ 2001 \\
\object{HD\,108147} & 7.0 & UVES &  950 & Mar.\ 2001 \\
\object{HD\,114762} & 7.3 & UVES & 1350 & Mar.\ 2001 \\
\object{HD\,120136} & 4.5 & UVES & 1100 & Mar.\ 2001 \\
\object{HD\,121504} & 7.6 & UVES &  900 & Mar.\ 2001 \\
\object{HD\,195019} & 6.9 & UES  &  800 & Oct.\ 2001 \\
\object{HD\,209458} & 7.7 & UVES & 1100 & June 2001 \\
\object{HD\,217107} & 6.2 & UES  &  500 & Oct.\ 2001 \\
\noalign{\smallskip}
\hline
\end{tabular}
\end{center}
\label{tab1}
\end{table}
\begin{table}[!]
\caption[]{Atomic parameters adopted for \ion{C}{i}, \ion{S}{i}, \ion{Zn}{i} and \ion{Cu}{i}  lines}
\begin{center}
\begin{tabular}{lccc}
\hline
\noalign{\smallskip}
Species & $\lambda$ (\AA) & $\chi_l$ (eV) & $\log{gf}$ \\
\hline 
\hline
\noalign{\smallskip}
\ion{C}{i} & 5380.340 & 7.68 & $-1.71$ \\
\ion{C}{i} & 5052.160 & 7.68 & $-1.42$ \\
\ion{S}{i} & 6743.440 & 7.87 & $-1.27$ \\
\ion{S}{i} & 6743.531 & 7.87 & $-0.92$ \\
\ion{S}{i} & 6743.640 & 7.87 & $-0.93$ \\
\ion{S}{i} & 6757.007 & 7.87 & $-0.81$ \\
\ion{S}{i} & 6757.177 & 7.87 & $-0.33$ \\
\ion{Zn}{i} & 4810.537 & 4.08 & $-0.13$ \\
\ion{Zn}{i} & 4722.160 & 4.03 & $-0.37$ \\
\ion{Cu}{i} & 5218.204 & 3.82 & $-1.33$ \\
\ion{Cu}{i} & 5218.206 & 3.82 & $-0.85$ \\
\ion{Cu}{i} & 5218.208 & 3.82 & $-0.98$ \\
\ion{Cu}{i} & 5218.210 & 3.82 & $-0.26$ \\
\ion{Cu}{i} & 5218.214 & 3.82 & $-0.48$ \\
\ion{Cu}{i} & 5218.216 & 3.82 & $-0.48$ \\
\ion{Cu}{i} & 5218.219 & 3.82 & $ 0.14$ \\
\ion{Cu}{i} & 5782.032 & 1.64 & $-3.53$ \\
\ion{Cu}{i} & 5782.042 & 1.64 & $-3.84$ \\
\ion{Cu}{i} & 5782.054 & 1.64 & $-3.14$ \\
\ion{Cu}{i} & 5782.064 & 1.64 & $-3.19$ \\
\ion{Cu}{i} & 5782.073 & 1.64 & $-3.49$ \\
\ion{Cu}{i} & 5782.084 & 1.64 & $-2.79$ \\
\ion{Cu}{i} & 5782.086 & 1.64 & $-3.14$ \\
\ion{Cu}{i} & 5782.098 & 1.64 & $-3.14$ \\
\ion{Cu}{i} & 5782.113 & 1.64 & $-2.79$ \\
\ion{Cu}{i} & 5782.124 & 1.64 & $-2.79$ \\
\ion{Cu}{i} & 5782.153 & 1.64 & $-2.69$ \\
\ion{Cu}{i} & 5782.173 & 1.64 & $-2.34$ \\
\noalign{\smallskip}
\hline	      
\end{tabular} 
\end{center}
\label{tab2}  
\end{table}

\begin{figure*}
\centering
\includegraphics[height=6cm]{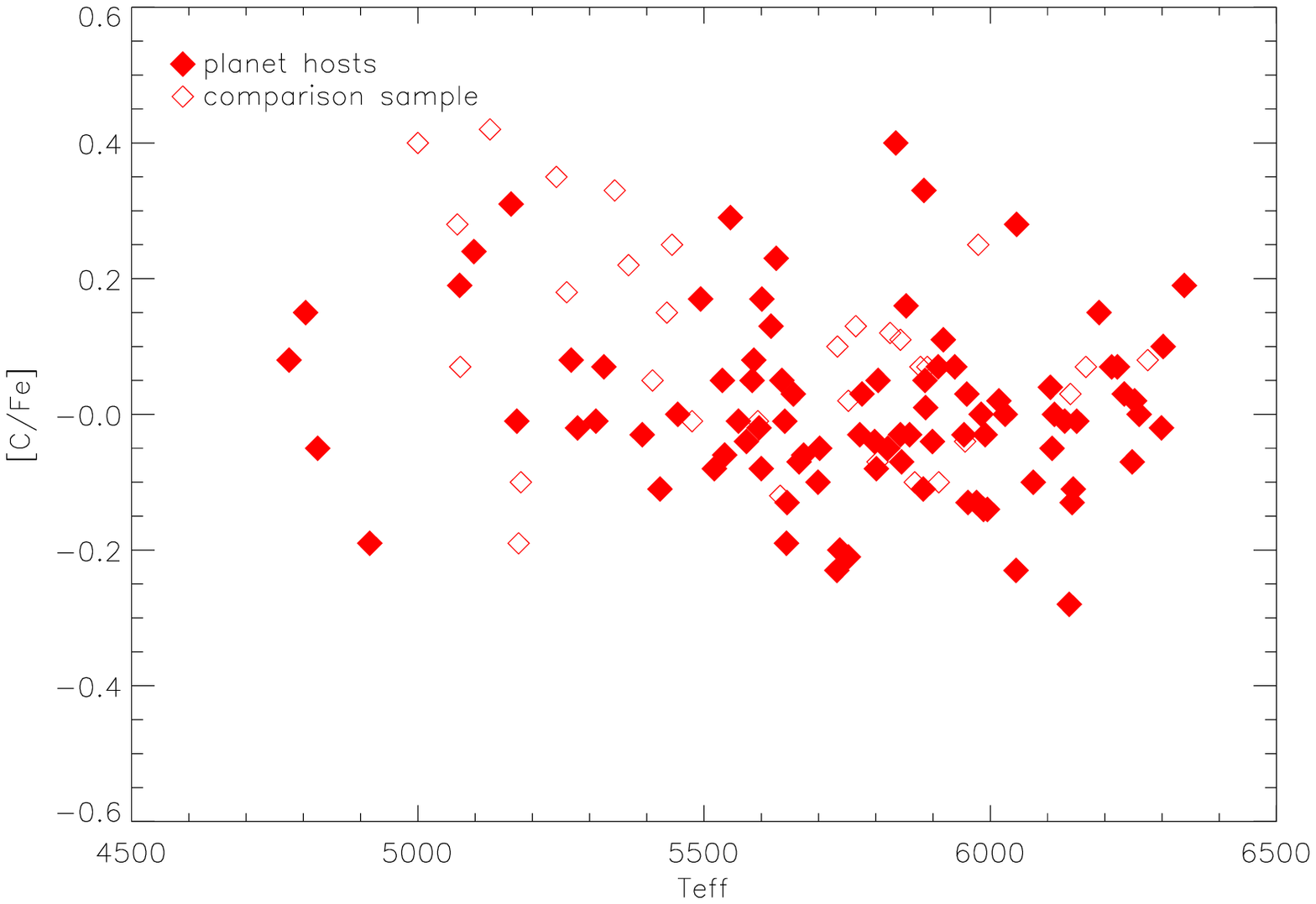}
\includegraphics[height=6cm]{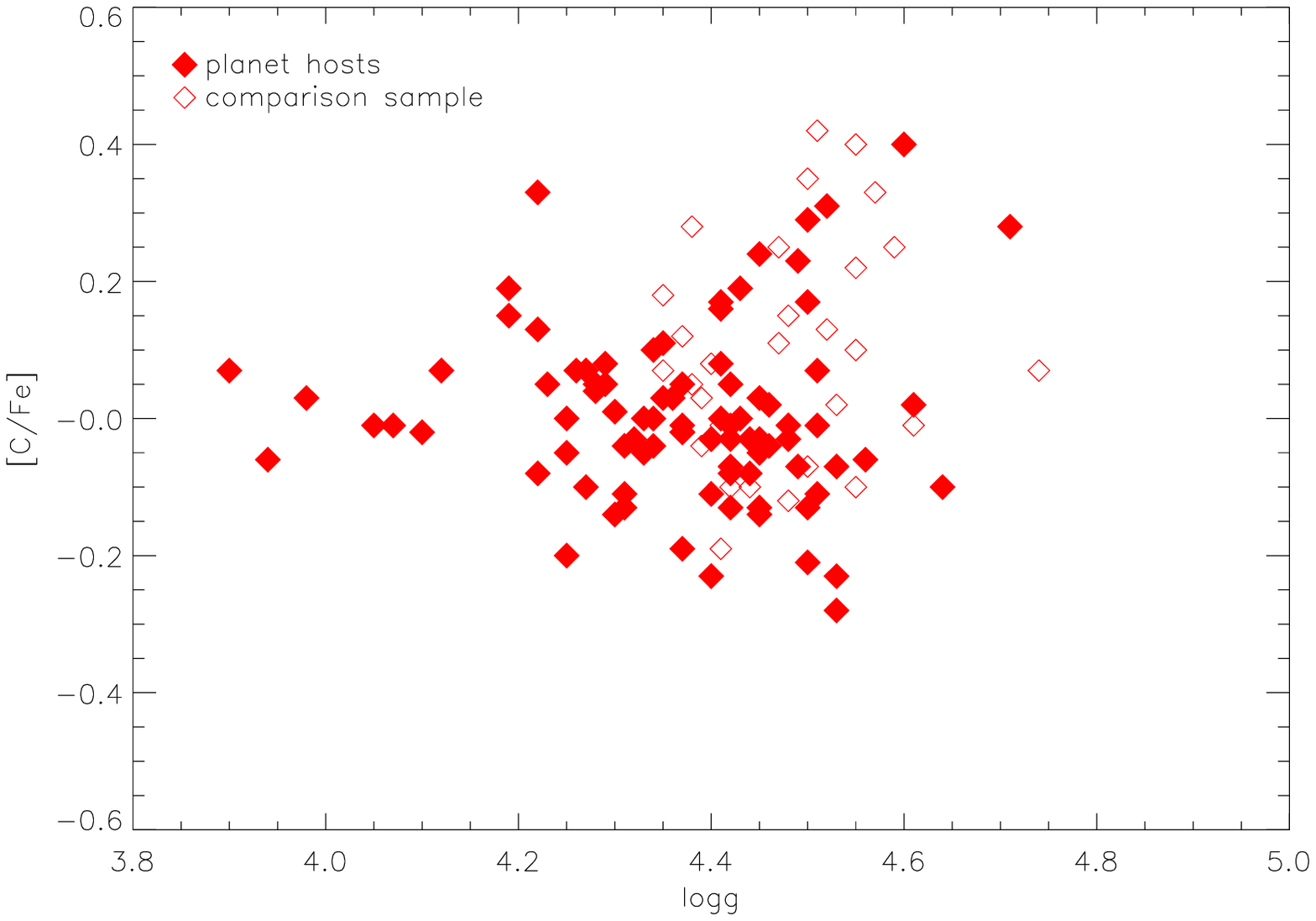}
\caption{[C/Fe] vs.\ $T_\mathrm{eff}$ and $\log {g}$. Filled and open symbols represent planet host and comparison sample 
stars, respectively.}
\label{fig4}
\end{figure*}

\begin{figure*}
\centering
\includegraphics[height=6cm]{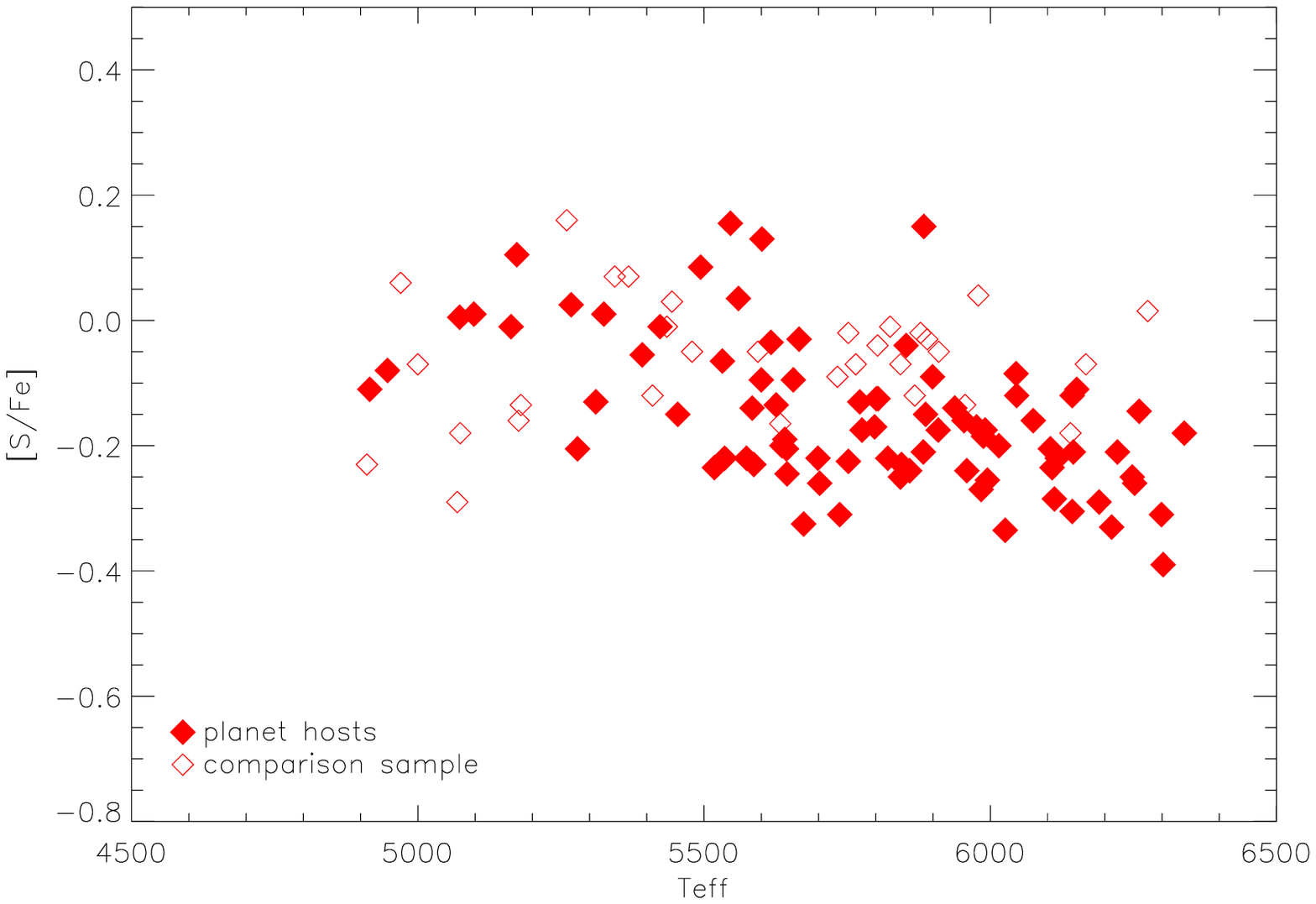}
\includegraphics[height=6cm]{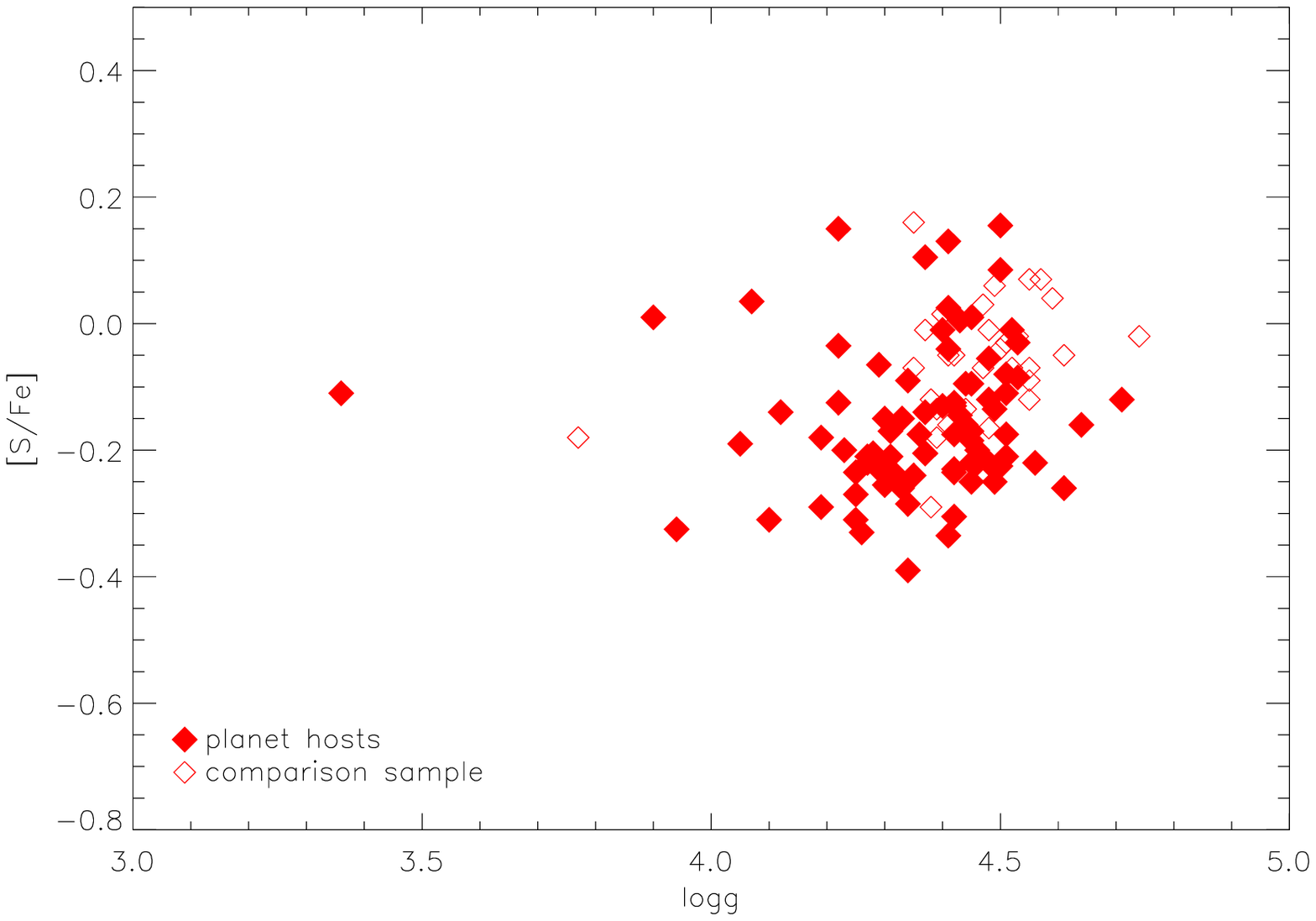}
\caption{[S/Fe] vs.\ $T_\mathrm{eff}$ and $\log {g}$. Filled and open symbols represent planet host and comparison sample 
stars, respectively.}
\label{fig5}
\end{figure*}

\begin{figure*}
\centering
\includegraphics[height=6cm]{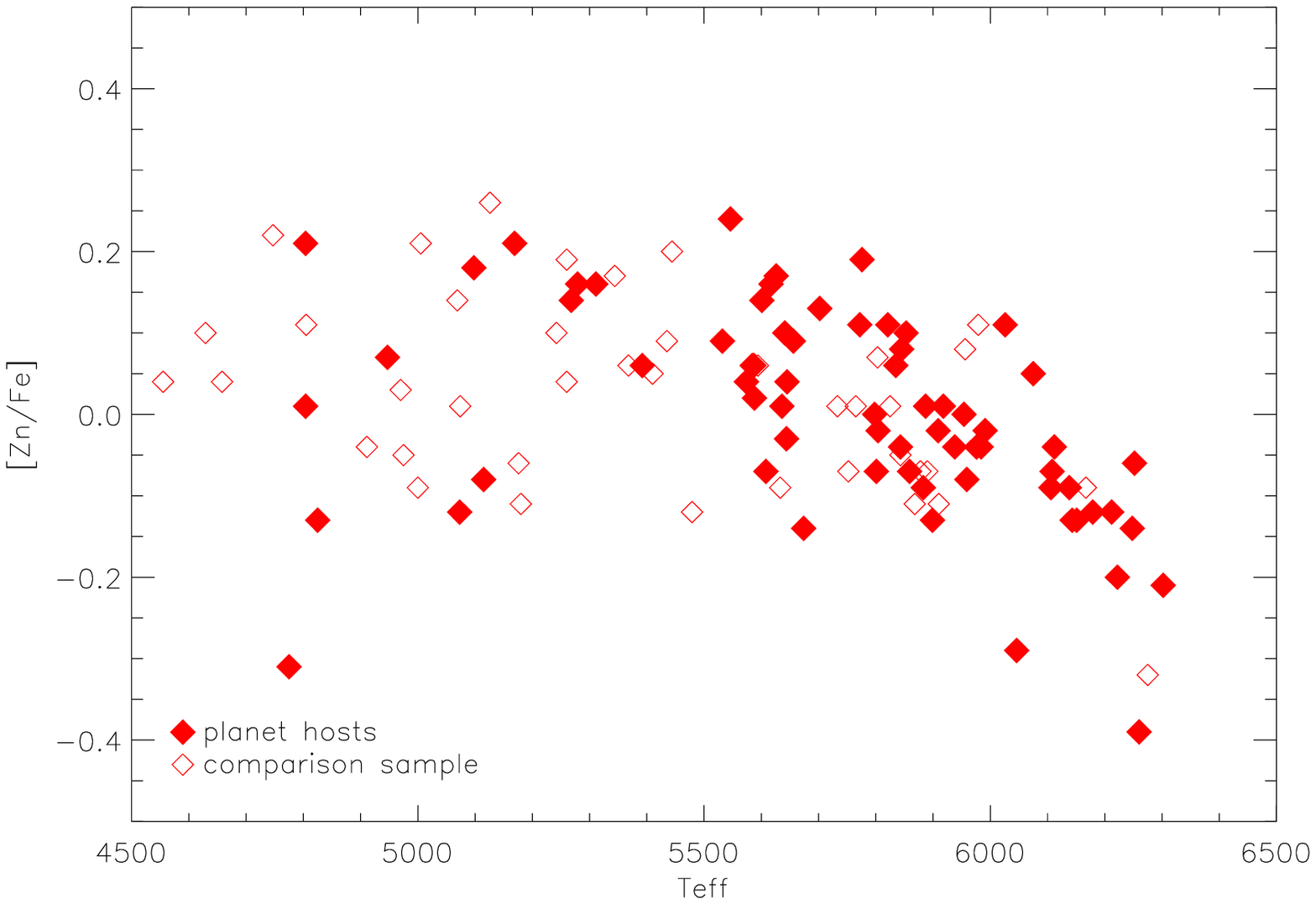}
\includegraphics[height=6cm]{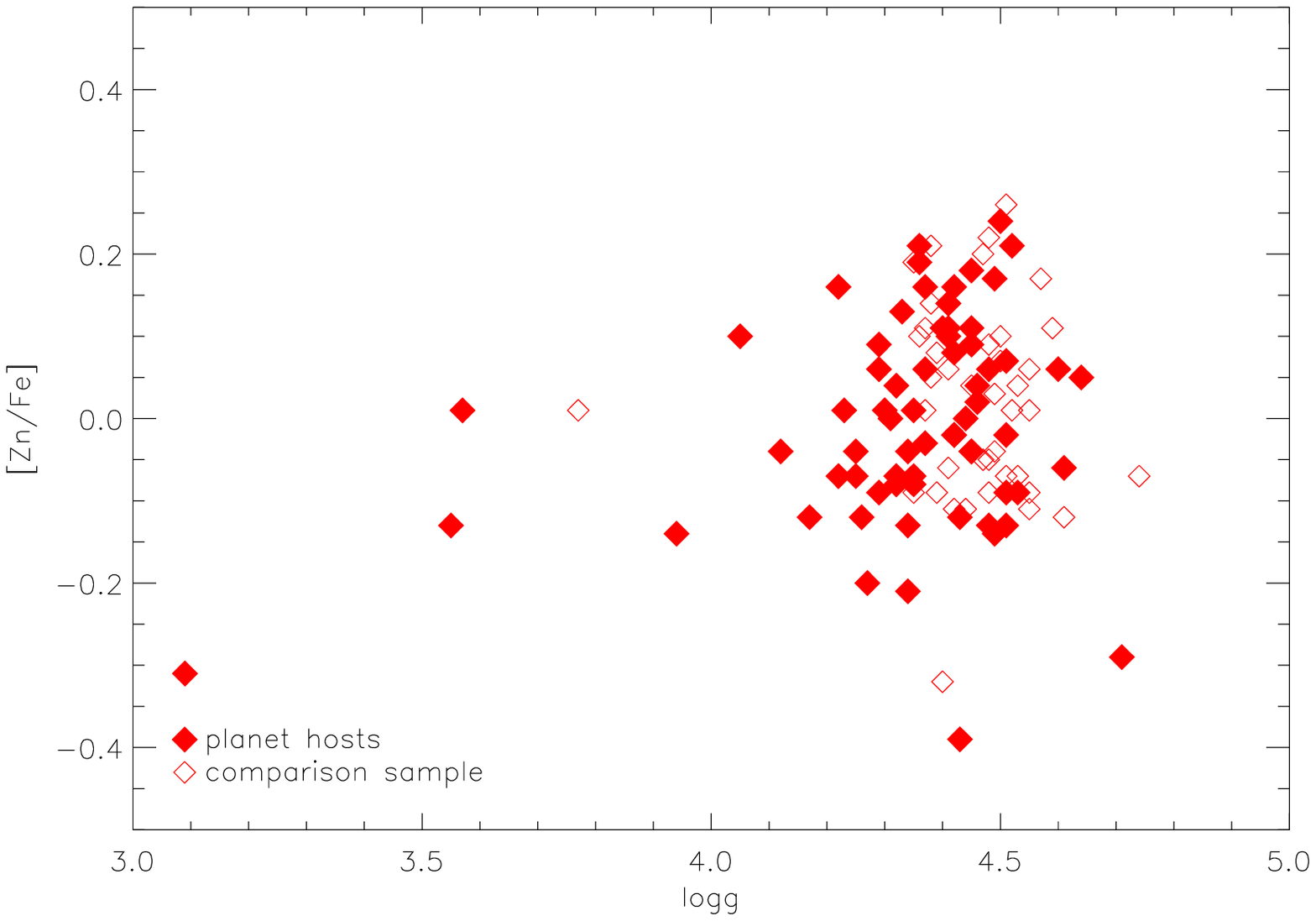}
\caption{[Zn/Fe] vs.\ $T_\mathrm{eff}$ and $\log {g}$. Filled and open symbols represent planet host and comparison sample 
stars, respectively.}
\label{fig6}
\end{figure*}

\begin{figure*}
\centering
\includegraphics[height=6cm]{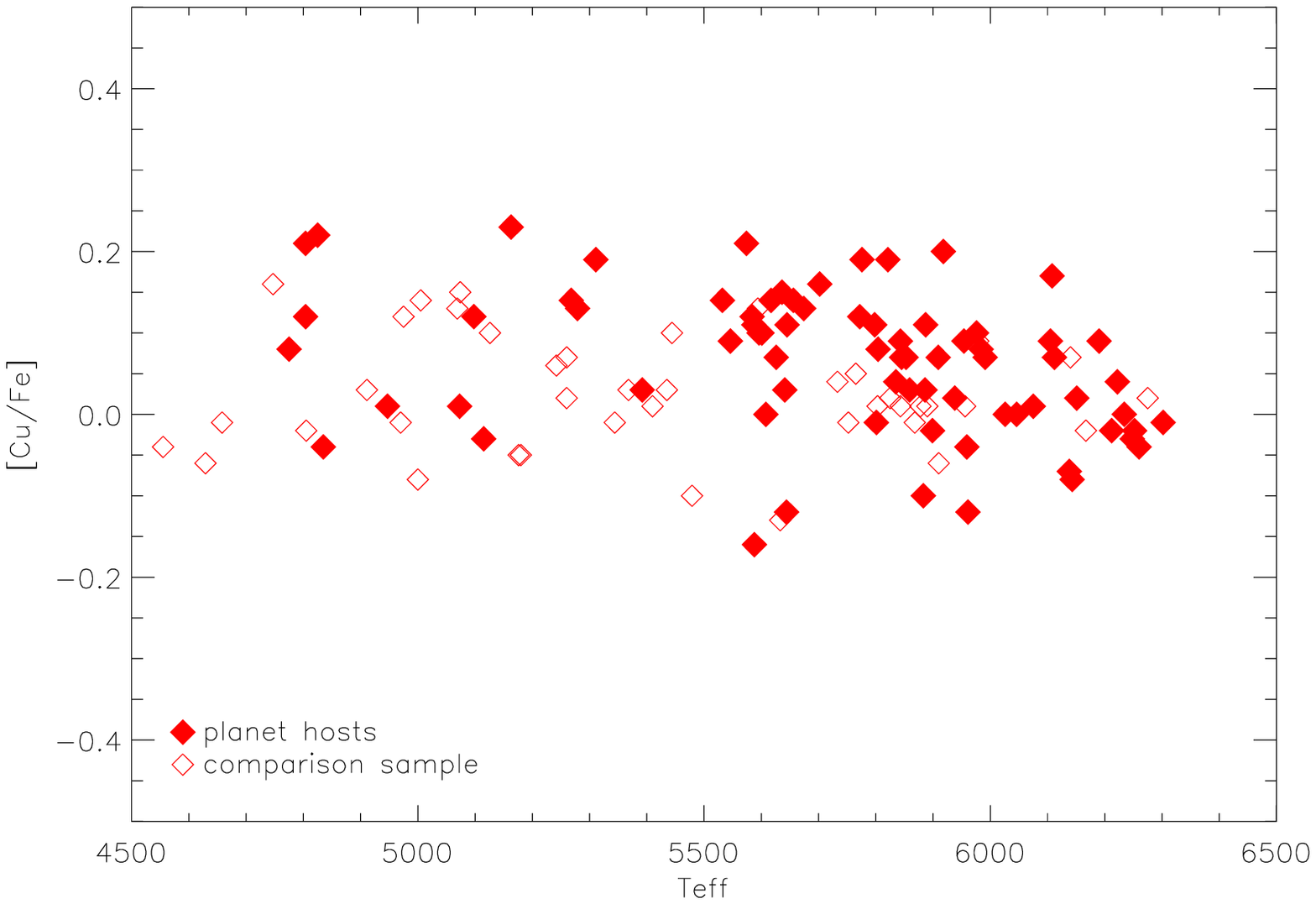}
\includegraphics[height=6cm]{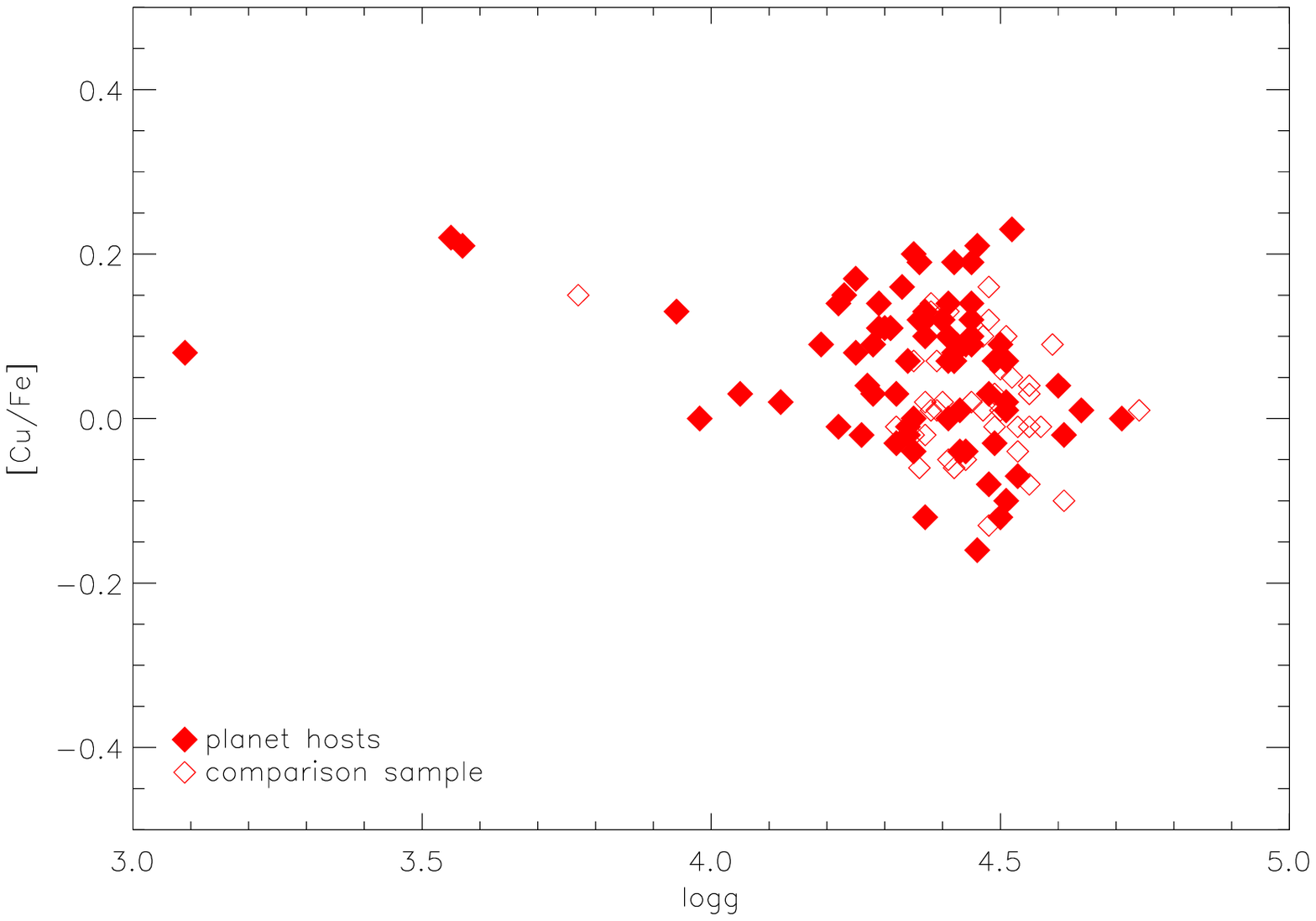}
\caption{[Cu/Fe] vs.\ $T_\mathrm{eff}$ and $\log {g}$. Filled and open symbols represent planet host and comparison sample 
stars, respectively.}
\label{fig7}
\end{figure*}

\section{Observations}
Most of the analysed spectra were collected during several observational campaigns with the CORALIE spectrograph on the 1.2 m 
Euler Swiss telescope, the FEROS spectrograph on the 2.2 m ESO/MPI telescope (both at La Silla, Chile), the UVES 
spectrograph on the VLT/UT2 Kueyen telescope (Paranal Observatory, ESO, Chile), the SARG spectrograph on the 3.5 m TNG and 
the UES spectrograph on the 4.2 m WHT  (both at the Roque de los Muchachos Observatory, La Palma, Spain). Part of these data 
were used in previous articles to derive precise stellar parameters (Santos et al.\ \cite{San01}, \cite{San03b}, \cite{San04a}) 
and abundances of C, O, S, Si, Ca and Ti (Santos et al.\ \cite{San00}), as well as to study trends of nine refractory 
elements (Bodaghee et al.\ \cite{Bod03}) and of the volatile nitrogen (Ecuvillon et al.\ \cite{Ecu04}). 
We refer the reader to these papers for a description of the data.

New spectra with high S/N ratios from the UVES and UES spectrographs, on the VLT/UT2 Kueyen telescope (Paranal 
Observatory, ESO, Chile) and on the 4.2 m WHT (Roque de los Muchachos Observatory, La Palma, Spain), respectively, were
used. These are listed in Table~\ref{tab1}. The S/N ratios for these spectra are in most cases higher than 800 at a
resolution of 55000 and 110000 for the UES and the UVES spectrograph, respectively.

\section{Analysis}
The abundance analysis was carried out in standard local thermodinamic equilibrium (LTE) using a revised version of the
spectral synthesis code MOOG (Sneden\ \cite{Sne73}) and a grid of Kurucz (\cite{Kur93}) ATLAS9 atmospheres.
All the atmospheric parameters, $T_\mathrm{eff}$, $\log {g}$, $\xi_t$ and [Fe/H], and the corresponding
uncertainties, were taken from Santos et al.\ (\cite{San04a}). 
The adopted solar abundances for  carbon, sulphur, zinc, copper and iron 
were respectively $\log{\epsilon}\,({\rm C})_{\odot}$ = 8.56\,dex, $\log{\epsilon}\,({\rm S})_{\odot}$ = 7.21\,dex , 
$\log{\epsilon}\,({\rm Zn})_{\odot}$ = 4.60\,dex, $\log{\epsilon}\,({\rm Cu})_{\odot}$ = 4.21\,dex (Anders \& Grevesse \cite{And89}) 
and $\log{\epsilon}\,({\rm Fe})_{\odot}$ = 7.47\,dex (as used in Santos et al.\ \cite{San04a}).

Carbon abundance analysis was carried out by {EW} measurement of two \ion{C}{i} lines  5380.3 \AA\ and at 5052.2 \AA.
Wavelength and excitation energies of the lower level were taken from VALD
(Kupka et al. \cite{Kup99}) and solar gf values were 
computed using {EW}s (21.0 m\AA\ and 34.2 m\AA, respectively) obtained from the Kurucz Solar Atlas (Kurucz et al.\ \cite{Kur84}), and a solar model with 
$T_\mathrm{eff}$ = 5777\,K, $\log{g}$ = 4.44 and $\xi_t$ = 1.0\,km\,s$^{-1}$. All the adopted parameters for each spectral 
line are presented in Table~\ref{tab2}. {EW}s were determined by Gaussian fitting using the SPLOT task of IRAF and 
abundances were computed with the ABFIND driver of MOOG.   
 
Sulphur, zinc and copper abundances were derived by fitting synthetic spectra to the data. For sulphur, we analysed two
\ion{S}{i} features: the first at 6743.5 \AA\ consisting of three \ion{S}{i} lines, at 6743.44 \AA, 6743.53 \AA\ and at
6743.64 \AA, and the second at 6757.1 \AA\ consisting of two \ion{S}{i} lines, at 6757.01 \AA\ and at 6757.18 \AA.
Fits were carried out in the spectral regions 6742.5--6744.5 \AA\ and  6756.0--6758.0 \AA. The line lists for each 
spectral region were taken from VALD (Kupka et al. \cite{Kup99}) and $\log{gf}$ values were slightly modified in order to achieve 
a good fit to the Kurucz Solar Atlas (Kurucz et al.\ \cite{Kur84}). The adopted atomic data are listed
in Table~\ref{tab2}. 
For the instrumental broadening we used a Gaussian function with {FWHM} of 0.07 \AA\ and 0.11 \AA, depending on the
instrumental resolution, 
and a rotational broadening function with $v\sin{i}$ values from the CORALIE database. For targets without 
a CORALIE $v\sin{i}$ determination, we chose the $v\sin{i}$ values which fitted the best the spectral 
lines in the synthesized regions. These values were between 1 and 5\,km\,s$^{-1}$, since all our targets are
slow rotators. Two examples of the fitting of the two features are shown in Figure~\ref{fig1}. The red 
side of the S\,I feature at 6757 \AA\ could not be perfectly reproduced because of an unknown feature
between 6757.25 and 6757.45 \AA\ (there are no spectral lines in the VALD database at these wavelengths). 
Our differential analysis relative to the Sun cancels out this source of uncertainty since all our targets 
are affected in the same way as the Sun. Moreover, we have verified that an $EW$ analysis yields the same 
final abundances, and therefore we consider this effect to be negligible.

Zinc abundances were derived from the two \ion{Zn}{i} lines at 4810.5 \AA\ and at 4722.2 \AA. We computed 5\AA\ wide 
synthetic spectra. The atomic lists for elements other than zinc were provided by VALD (Kupka et al. \cite{Kup99}), while atomic 
data for the \ion{Zn}{i} lines were taken from Gurtovenko \& Kostyk (\cite{Gurt89}). Copper abundances were derived from 
the \ion{Cu}{i} lines at 5218.2 \AA\ and at 5782.1 \AA, by the synthesis of the spectral regions 5213--5220 \AA\ and 
5780--5785\AA. The atomic lists for elements other than copper were provided by VALD (Kupka et al. \cite{Kup99}), while atomic 
data (hyperfine structure components) for the \ion{Cu}{i} lines were taken from Steffen (\cite{Ste85}). A differential 
analysis relative to the Sun was carried out for the two elements. Some wavelengths were slightly modified in
order to fit  the solar spectrum better. Table~\ref{tab2} presents the adopted values for both elements. 
Figures~\ref{fig2} and \ref{fig3} show 
two examples of fits for the \ion{Zn}{i} and \ion{Cu}{i} lines, respectively.

Uncertainties in the atmospheric parameters are of the order of 50\,K in $T_\mathrm{eff}$, 0.12\,dex in $\log {g}$, 
0.08\,km\,s$^{-1}$ in $\xi_t$ and 0.05\,dex in [Fe/H] (see Santos et al.\ \cite{San04a}). We estimated 
the effect of changes in atmospheric parameters on the resulting abundances in the following way.
For each atmospheric parameter, we selected a set of three stars having
similar values of all the parameters, excepting the considered one, which
must vary within the sample. 
We then tested abundance sensitivity to changes in the parameter ($\pm$100\,K for $T_\mathrm{eff}$, $\pm$0.3\,dex for $\log {g}$ and
[Fe/H], $\pm$0.5\,km\,s$^{-1}$ for $\xi_t$). The results are shown in Table~\ref{tab3}. 
Uncertainties in the abundances of all the elements were determined adding in quadrature the dispersion 
of each abundance from the mean value, 0.05 dex of 
the continuum determination and the errors due to the abundance sensitivities to changes in the 
atmospheric parameters.

Dependences on $T_\mathrm{eff}$ and on $\log {g}$ of [C/Fe], [S/Fe], [Zn/Fe] and [Cu/Fe] are represented in Fig.~\ref{fig4},
Fig.~\ref{fig5}, Fig.~\ref{fig6} and Fig.~\ref{fig7}, respectively. We note that no characteristic trends appear in 
any of the cases. This means that our results are almost free from systematic errors. Only in the case of Zn
does [Zn/Fe] decrease 
for $T_\mathrm{eff}$ greater than 6000\,K. This could be due to an NLTE effect, usually not very strong in metal-rich stars 
for complex atoms such as iron (see Edvardsson et al.\ \cite{Edv93}; Th\'evenin \& Idiart \cite{The99}).
We verified that using only stars with $T_\mathrm{eff}$ lower than 6000\,K would not change the resulting
trends.								    

\begin{table*}[t]
\caption[]{Sensitivity of the abundances of C, S, Zn and Cu to changes of 100\,K in effective temperature, 0.3\,dex in
gravity and metallicity, and 0.5\,km\,s$^{-1}$ in microturbulence}
\begin{center}
\begin{tabular}{ccccc}
\hline
\noalign{\smallskip}
 & Star & \object{HD\,22049} & \object{HD\,168746} & \object{HD\, 136118} \\ 
 & ($T_\mathrm{eff}$; $\log {g}$ ; [Fe/H] ; $\xi_t$) & (5073; 4.43; -0.13; 1.05) & (5601; 4.41; -0.08; 0.99) & (6222; 4.27; -0.04; 1.79)\\
\noalign{\smallskip}
\hline 
\noalign{\smallskip}
C: & $\Delta T_\mathrm{eff}=\pm100$\,K & $\mp0.09$ & $\mp0.07$ & $\mp0.05$ \\
\noalign{\smallskip}
\hline 
\noalign{\smallskip}
S: & $\Delta T_\mathrm{eff}=\pm100$\,K & $\mp0.10$ & $\mp0.10$ & $\mp0.05$ \\
\noalign{\smallskip}
\hline 
\noalign{\smallskip}
Zn: & $\Delta T_\mathrm{eff}=\pm100$\,K & $\pm0.03$ & $\pm0.04$ & $\pm0.05$ \\
\noalign{\smallskip}
\hline 
\noalign{\smallskip}
Cu: & $\Delta T_\mathrm{eff}=\pm100$\,K & $\pm0.03$ & $\pm0.05$ & $\pm0.05$ \\
\noalign{\smallskip}
\hline 
\noalign{\bigskip}
\hline
\noalign{\smallskip}
 & Star & \object{HD\,10697} & \object{HD\,16141} & \object{HD\,28185} \\ 
 & ($T_\mathrm{eff}$; $\log {g}$ ; [Fe/H]; $\xi_t$) & (5641; 4.05; 0.14; 1.13) & (5801; 4.22; 0.15; 1.34) & (5656; 4.45; 0.22; 1.01)\\
\noalign{\smallskip}
\hline 
\noalign{\smallskip}
C: & $\Delta \log{g}=\pm0.3$\,dex & $\mp0.10$ & $\mp0.10$ & $\mp0.10$ \\
\noalign{\smallskip}
\hline 
\noalign{\smallskip}
S: & $\Delta \log{g}=\pm0.3$\,dex & $\mp0.10$ & $\mp0.10$ & $\mp0.10$ \\
\noalign{\smallskip}
\hline 
\noalign{\smallskip}
Zn: & $\Delta \log{g}=\pm0.3$\,dex & $\pm0.07$ & $\pm0.07$ & $\pm0.06$ \\
\noalign{\smallskip}
\hline 
\noalign{\smallskip}
Cu: & $\Delta \log{g}=\pm0.3$\,dex & $\pm0.02$ & $\pm0.03$ & $\pm0.03$ \\
\noalign{\smallskip}
\hline 
\noalign{\bigskip}
\hline
\noalign{\smallskip}
 & Star & \object{HD\,6434} & \object{HD\,95128} & \object{HD\,4203} \\ 
 & ($T_\mathrm{eff}$; $\log {g}$ ; [Fe/H]; $\xi_t$) & (5835; 4.60; -0.52; 1.53) & (5954; 4.44; 0.06; 1.30) & (5636; 4.23; 0.40; 1.12)\\
\noalign{\smallskip}
\hline 
\noalign{\smallskip}
C: & $\Delta$[Fe/H]$ =\pm0.3$\,dex & $\mp0.03$ & $\mp0.02$ & $\mp0.02$ \\
\noalign{\smallskip}
\hline 
\noalign{\smallskip}
S: & $\Delta$[Fe/H] $=\pm0.3$\,dex & $\mp0.00^1$ & $\mp0.00^1$ & $\mp0.00^1$ \\
\noalign{\smallskip}
\hline 
\noalign{\smallskip}
Zn: & $\Delta$[Fe/H] $=\pm0.3$\,dex & $\mp0.20$ & $\mp0.20$ & $\mp0.20$ \\
\noalign{\smallskip}
\hline 
\noalign{\smallskip}
Cu: & $\Delta$[Fe/H] $=\pm0.3$\,dex & $\mp0.30$ & $\mp0.20$ & $\mp0.20$ \\
\noalign{\smallskip}
\hline 
\noalign{\bigskip}
\hline
\noalign{\smallskip}
 & Star & \object{HD\,49674} & \object{HD\,73256} & \object{HD\,19994} \\ 
 & ($T_\mathrm{eff}$; $\log {g}$ ; [Fe/H]; $\xi_t$) & (5644; 4.37; 0.33; 0.89) & (5518; 4.42; 0.26; 1.22) & (6190; 4.19; 0.24; 1.54)\\
\noalign{\smallskip}
\hline 
\noalign{\smallskip}
C: & $\Delta \xi_t=\pm0.5$km\,s$^{-1}$ & $\mp0.01$ & $\mp0.01$ & $\mp0.01$ \\
\noalign{\smallskip}
\hline 
\noalign{\smallskip}
S: & $\Delta \xi_t=\pm0.5$km\,s$^{-1}$ & $\mp0.00^1$ & $\mp0.00^1$ & $\mp0.00^1$ \\
\noalign{\smallskip}
\hline 
\noalign{\smallskip}
Zn: & $\Delta \xi_t=\pm0.5$km\,s$^{-1}$ & $\mp0.10$ & $\mp0.10$ & $\mp0.10$ \\
\noalign{\smallskip}
\hline 
\noalign{\smallskip}
Cu: & $\Delta \xi_t=\pm0.5$km\,s$^{-1}$ & $\mp0.10$ & $\mp0.10$ & $\mp0.10$ \\
\noalign{\smallskip}
\hline
\end{tabular}
\end{center}
\footnotesize{$^1$ These sensitivities are based on the method described in Section\,3. The values can be
slightly larger if more explicit calculations are carried out.}
\label{tab3}
\end{table*}

\section{Results}
\label{Res}
Several studies have been recently published about abundances of elements other than iron in planet host stars
(e.g. Santos et al.\ \cite{San00}, \cite{San02}; Gonzalez et al.\ \cite{Gonz01}; Takeda et al.\ \cite{Tak01}; 
Sadakane et al.\ \cite{Sad02}; Bodaghee et al.\ \cite{Bod03}; Israelian et al.\ \cite{Isr04}; Ecuvillon et al.\ \cite{Ecu04}). 
Studies concerning such volatile elements as CNO, S and Zn were 
carried out, but most of them did not count on a large set of planet-harbouring targets and a homogeneous comparison sample
of stars without known planetary companions. 

Santos et al.\ (\cite{San00}) compared the abundances of C, O and S for eight
planet host stars with  results for field dwarfs from the literature without finding significant differences. They 
pointed out the need to use a homogeneous comparison sample of stars without planetary mass companions. 
Also Gonzalez et al.\ (\cite{Gonz01}) found no deviation of the C and O abundances of twenty planet hosts from the 
trends traced by field stars taken from the literature. Takeda et al.\ (\cite{Tak01}) and Sadakane et al.\ (\cite{Sad02}) 
presented abundances of twelve and nineteen elements, respectively, among which the volatiles C, N, O, S and Zn and the refractory Cu, 
for fourteen and twelve planet-harbouring stars, respectively. Their results  revealed no anomalies between planet host and 
comparison samples.    
Ecuvillon et al.\ (\cite{Ecu04}) presented a homogeneous analysis of N abundances in 91 solar-type stars, 66 with and 25 
without known planets, and confirmed that the two samples show the same behaviour.

In this paper we analysed two optical lines of \ion{C}{i} in 91 targets with planets and in 31 comparison sample stars. All atmospheric
parameters, {EW} values with uncertainties and abundance results for both samples are listed in 
Tables~\ref{tab4},~\ref{tab5} and~\ref{tab6}. For sulphur, zinc and copper, we synthesized two optical lines in 84, 68 and 73 planet
host stars, and in 31, 41 and 41 comparison sample stars, respectively. Atmospheric parameters, abundance results and 
 the corresponding uncertainties, are listed in Tables~\ref{tab7} to~\ref{tab15}.  

\begin{figure*}
\centering
\includegraphics[height=6cm]{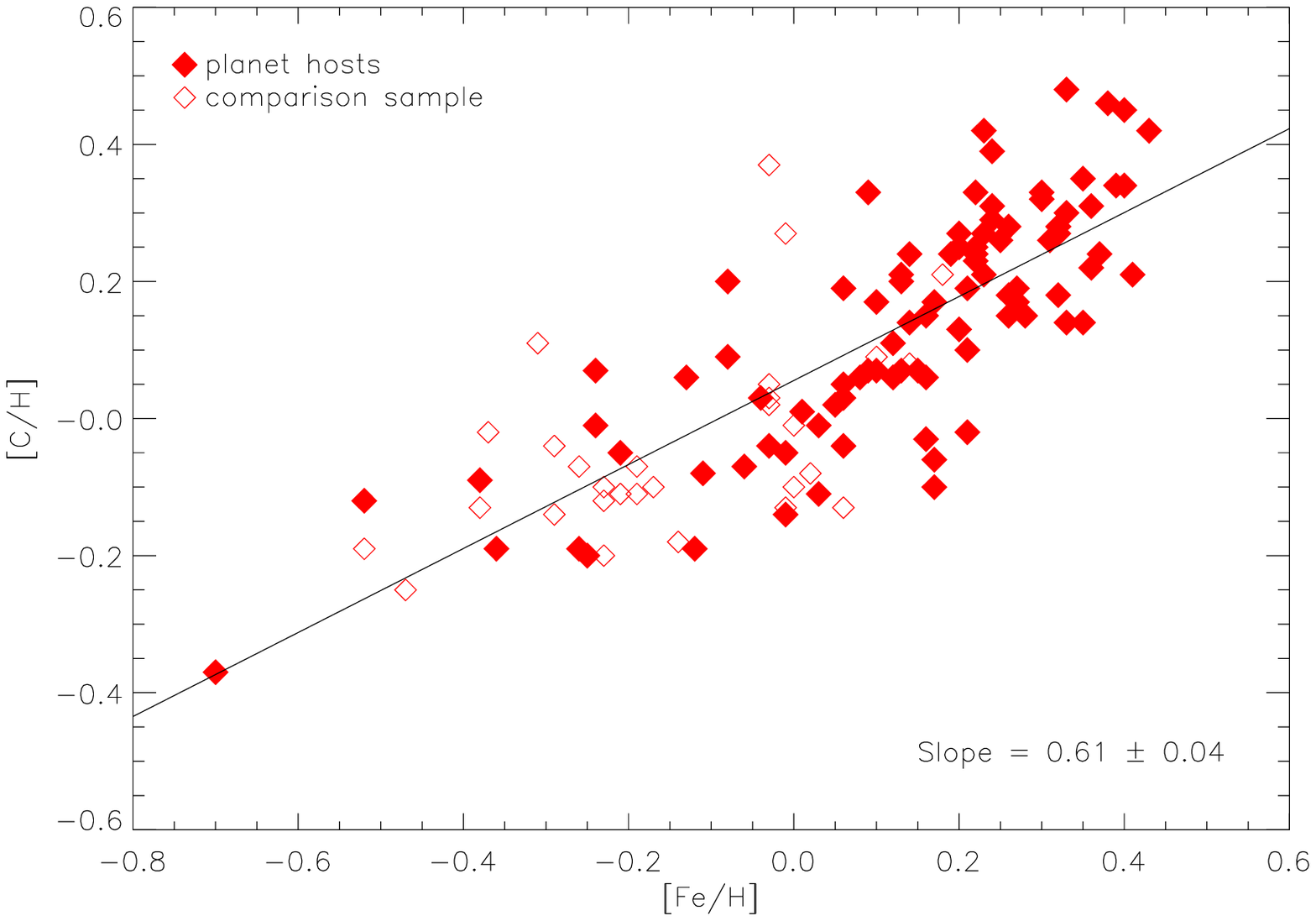}
\includegraphics[height=6cm]{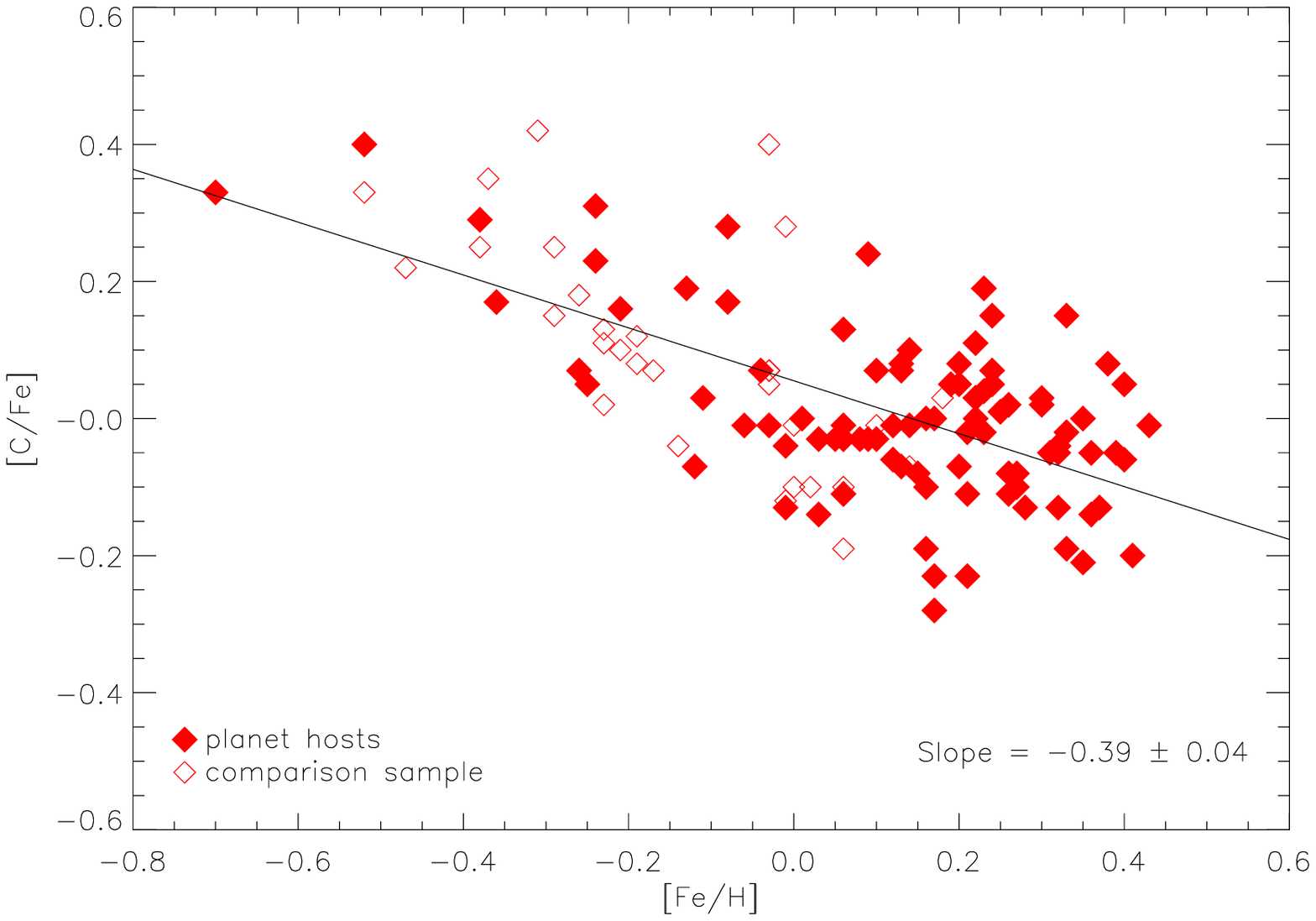}
\caption{[C/H] and [C/Fe] vs.\ [Fe/H] plots. Filled and open symbols represent planet
 host and comparison sample stars, respectively. Linear least-squares fits to both samples together are represented and 
slope values are indicated at the bottom of each plot.}
\label{fig8}
\end{figure*}

\begin{figure*}
\centering
\includegraphics[height=6cm]{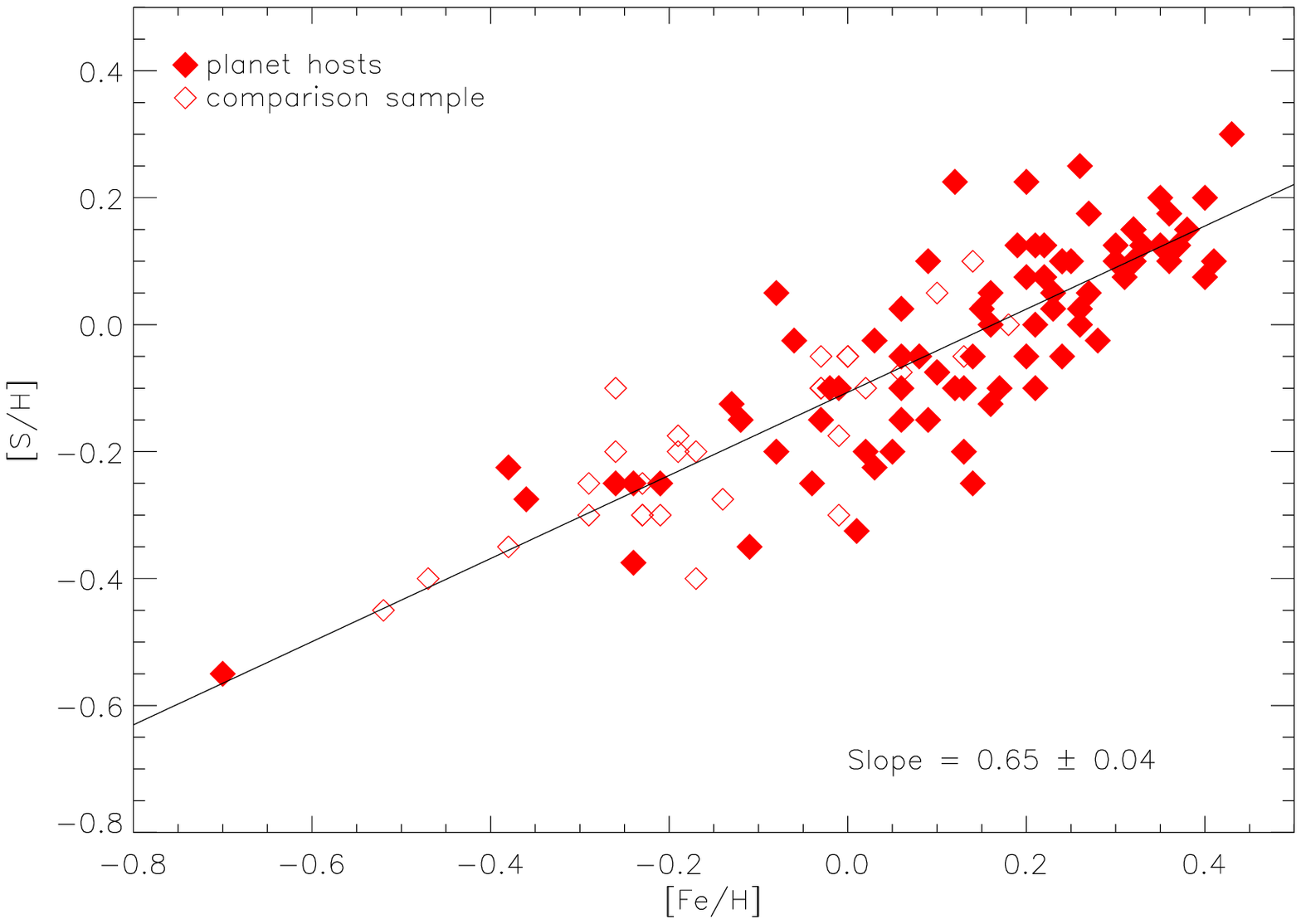}
\includegraphics[height=6cm]{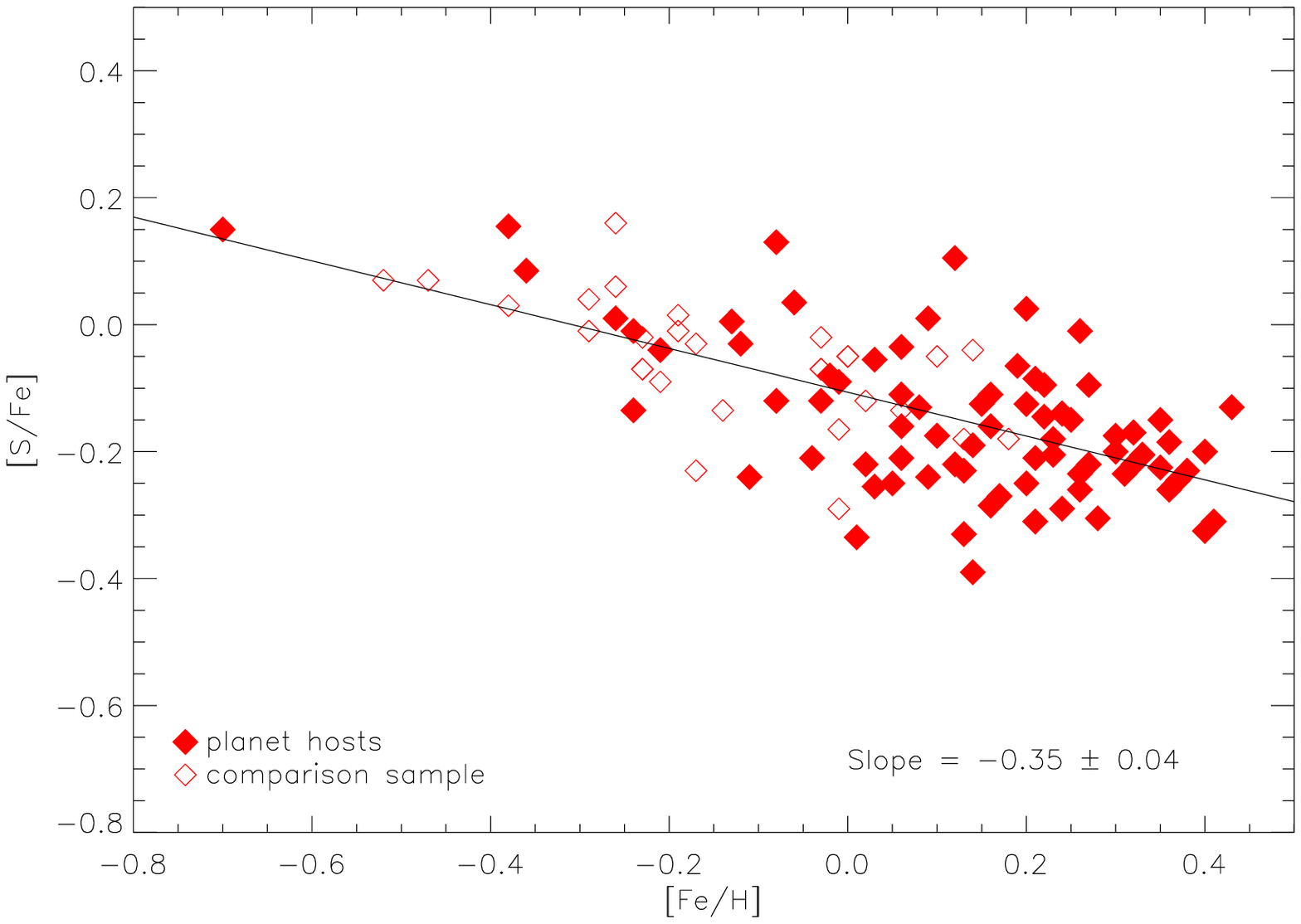}
\caption{[S/H] and [S/Fe] vs.\ [Fe/H] plots. Filled and open symbols represent planet
 host and comparison sample stars, respectively. Linear least-squares fits to both samples together are represented and 
slope values are indicated at the bottom of each plot.}
\label{fig9}
\end{figure*}

\begin{figure*}
\centering
\includegraphics[height=6cm]{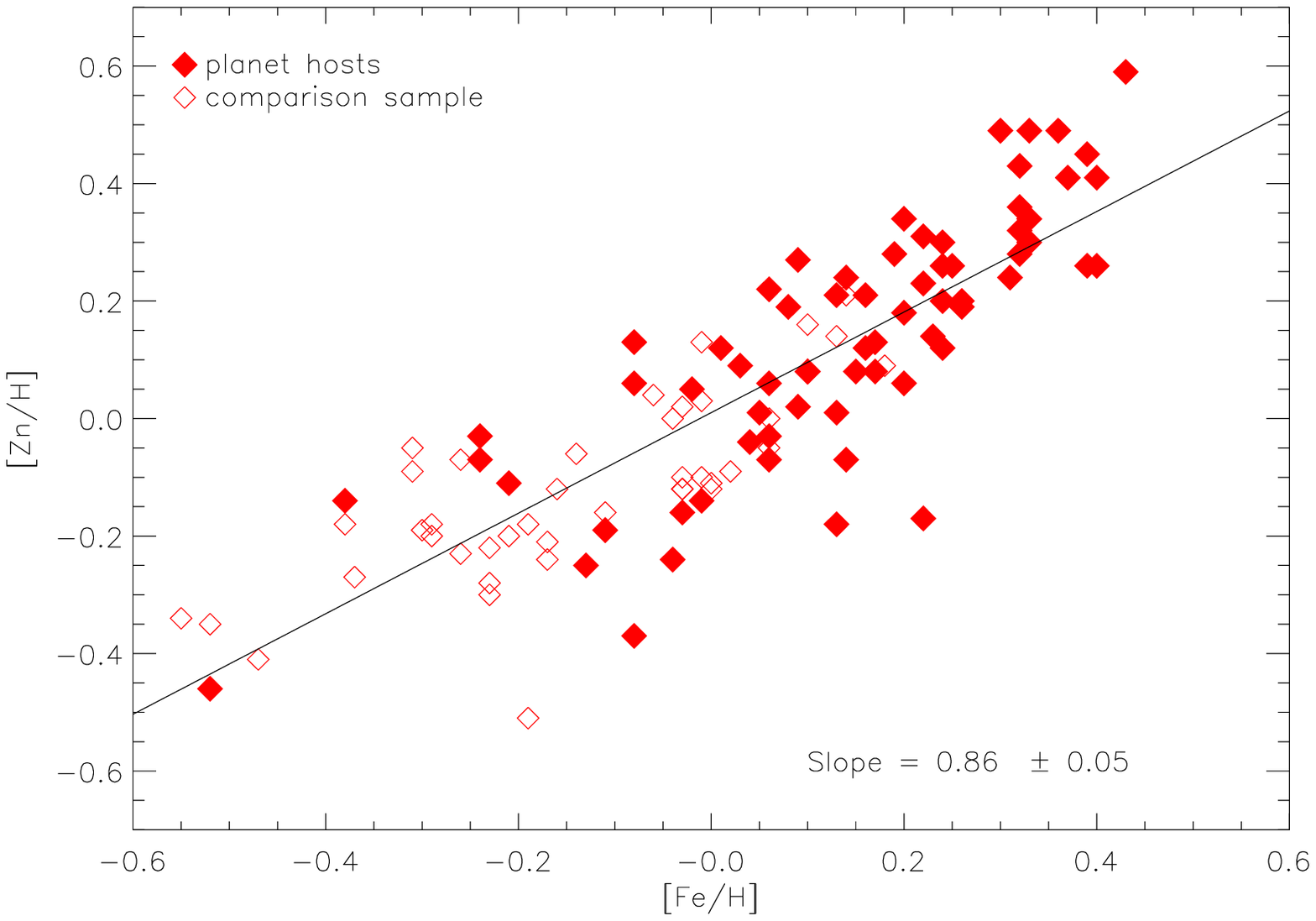}
\includegraphics[height=6cm]{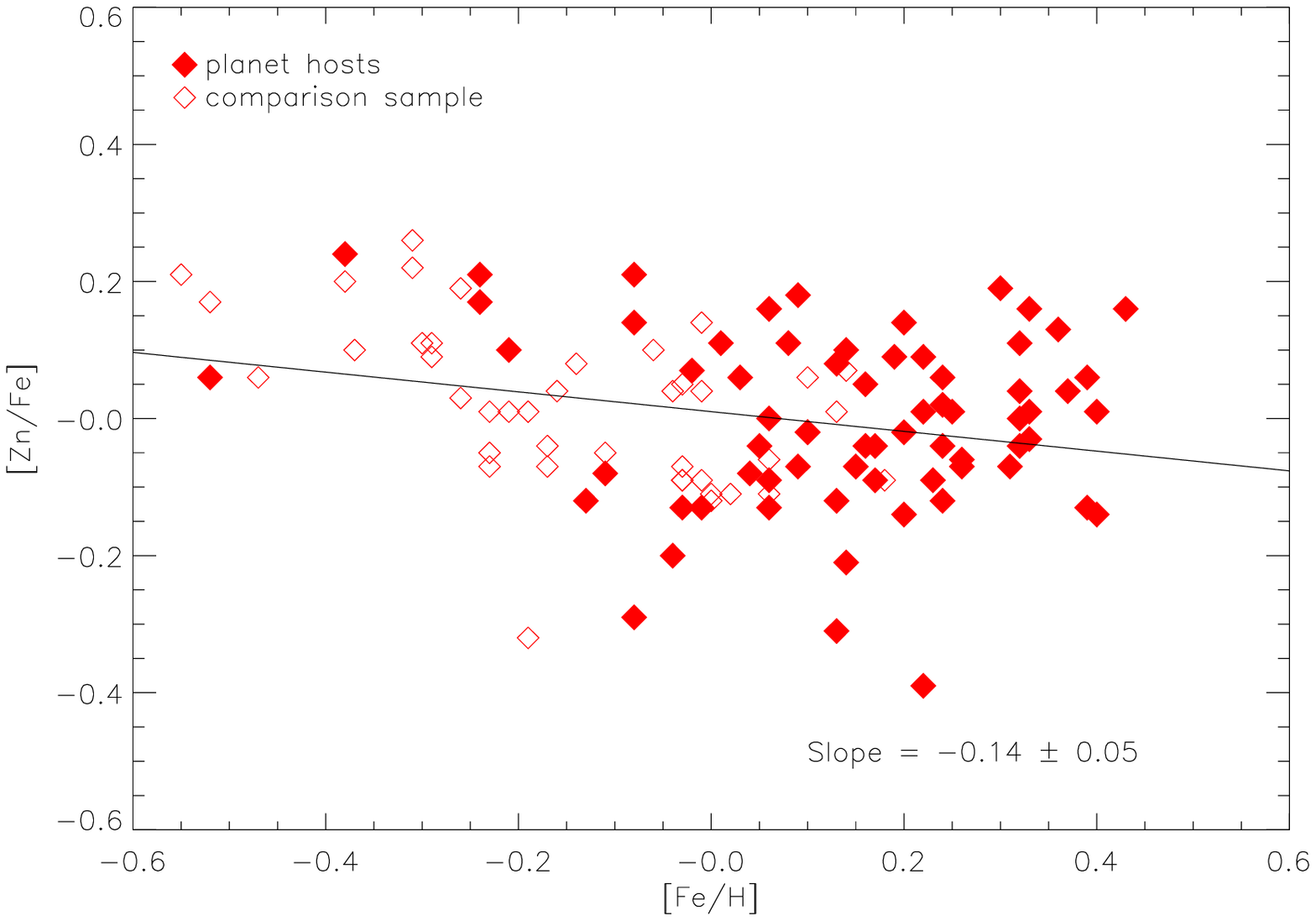}
\caption{[Zn/H] and [Zn/Fe] vs.\ [Fe/H] plots. Filled and open symbols represent planet
 host and comparison sample stars, respectively. Linear least-squares fits to both samples together are represented and 
slope values are indicated at the bottom of each plot.}
\label{fig10}
\end{figure*}

Figure~\ref{fig8} presents [C/H] and [C/Fe] as functions of [Fe/H]. Both samples, stars with and without planets, behave
quite similarly. Since targets with planets are on average more metal rich than comparison sample stars, their
abundance distributions correspond to the extensions of the comparison sample trends at high [Fe/H]. Planet host stars do 
not present anomalies in carbon abundances with respect to comparison sample dwarfs.

Similar results are obtained for sulphur and zinc. Figures~\ref{fig9} and~\ref{fig10} show the [$X$/H] and [$X$/Fe] vs.\ 
[Fe/H] plots for the two elements. The abundance trends of sulphur and zinc in planet host stars are similar to those of the 
comparison sample. In fact, no discontinuity is seen in the overlap region of both samples. 

[$X$/Fe] vs.\ [Fe/H] plots for the three volatiles show decreasing trends with increasing [Fe/H]. Although this effect is more 
evident for [C/Fe] and [S/Fe] than for [Zn/Fe], they all show a negative slope . This means that, unlike nitrogen (see
Ecuvillon et al.\ \cite{Ecu04}), none of these elements keeps pace with iron; an excess of carbon,  sulphur and zinc exists 
in the more metal-poor tail with respect to stars with high [Fe/H]. Since no differences appear 
to be related to the presence of 
planets, the 
observed slopes may be a by-product of Galactic chemical evolution. However, no detailed models are available
to explain trends such as these. 

\begin{figure*} 
\centering
\includegraphics[height=6cm]{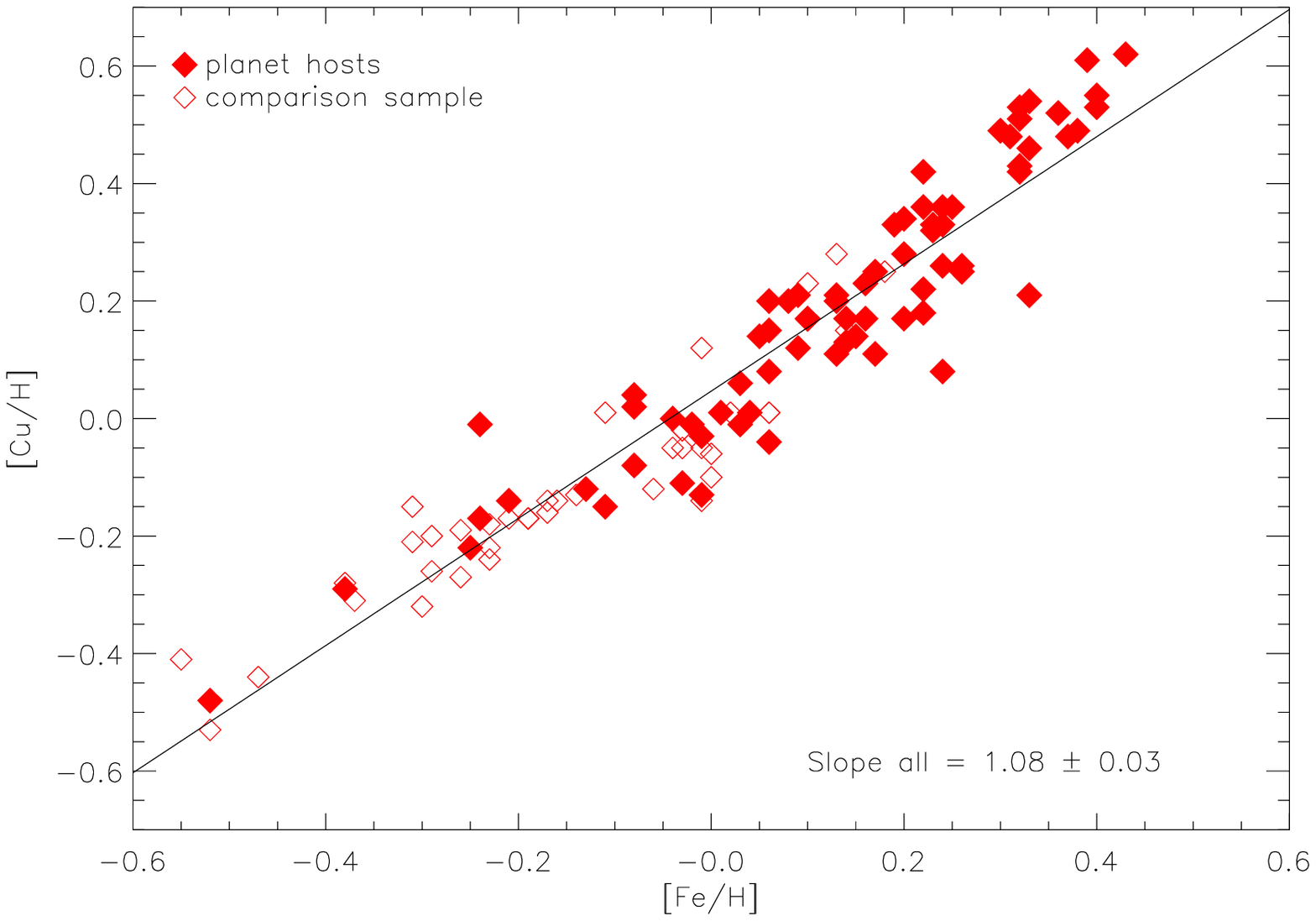}
\includegraphics[height=6cm]{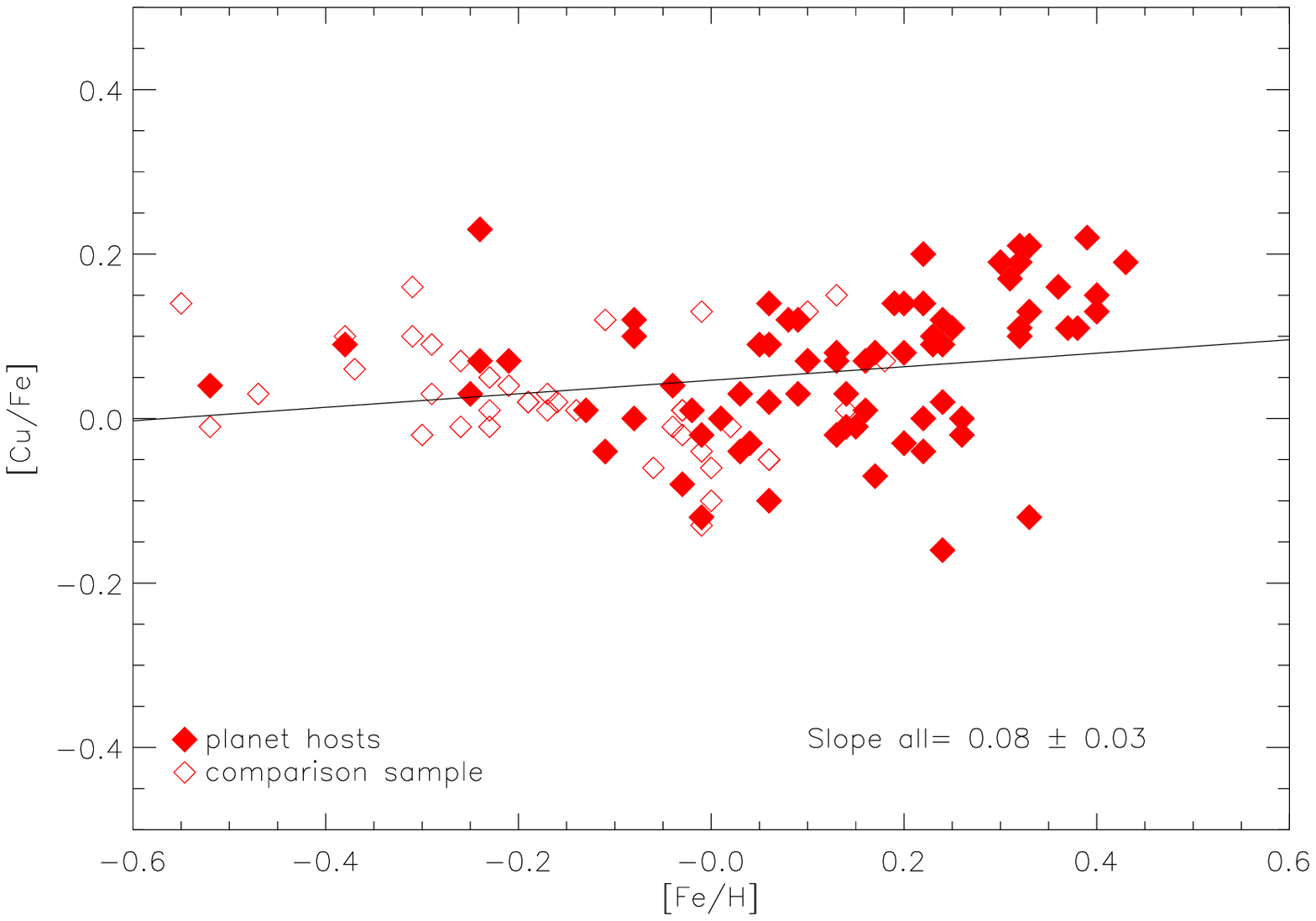}
\caption{[Cu/H] and [Cu/Fe] vs.\ [Fe/H] plots. Filled and open symbols represent planet
 host and comparison sample stars, respectively. Linear least-squares fits to both samples together are 
 represented and slope values are indicated at the bottom of each plot.}
\label{fig11}
\end{figure*}

[Cu/H] and [Cu/Fe] vs.\ [Fe/H] trends are provided in Figure~\ref{fig11}. No discontinuities appear between 
the two samples. 
In the [Cu/Fe] vs.\ [Fe/H] plot, the planet host star set seems to produce a steeper fit than that 
corresponding to the comparison sample. Although this may be seen as a signature of the presence of planets, it seems more likely that the 
[Cu/Fe] slope increases at [Fe/H] $>$ 0.1, and that its relation to the presence of planets is solely due to the high 
metallicity of the parent stars. The [Cu/Fe] trend corresponding to both samples is on average slightly 
overabundant with respect to [Fe/H].   

\begin{figure*}
\centering 
\includegraphics[height=6cm]{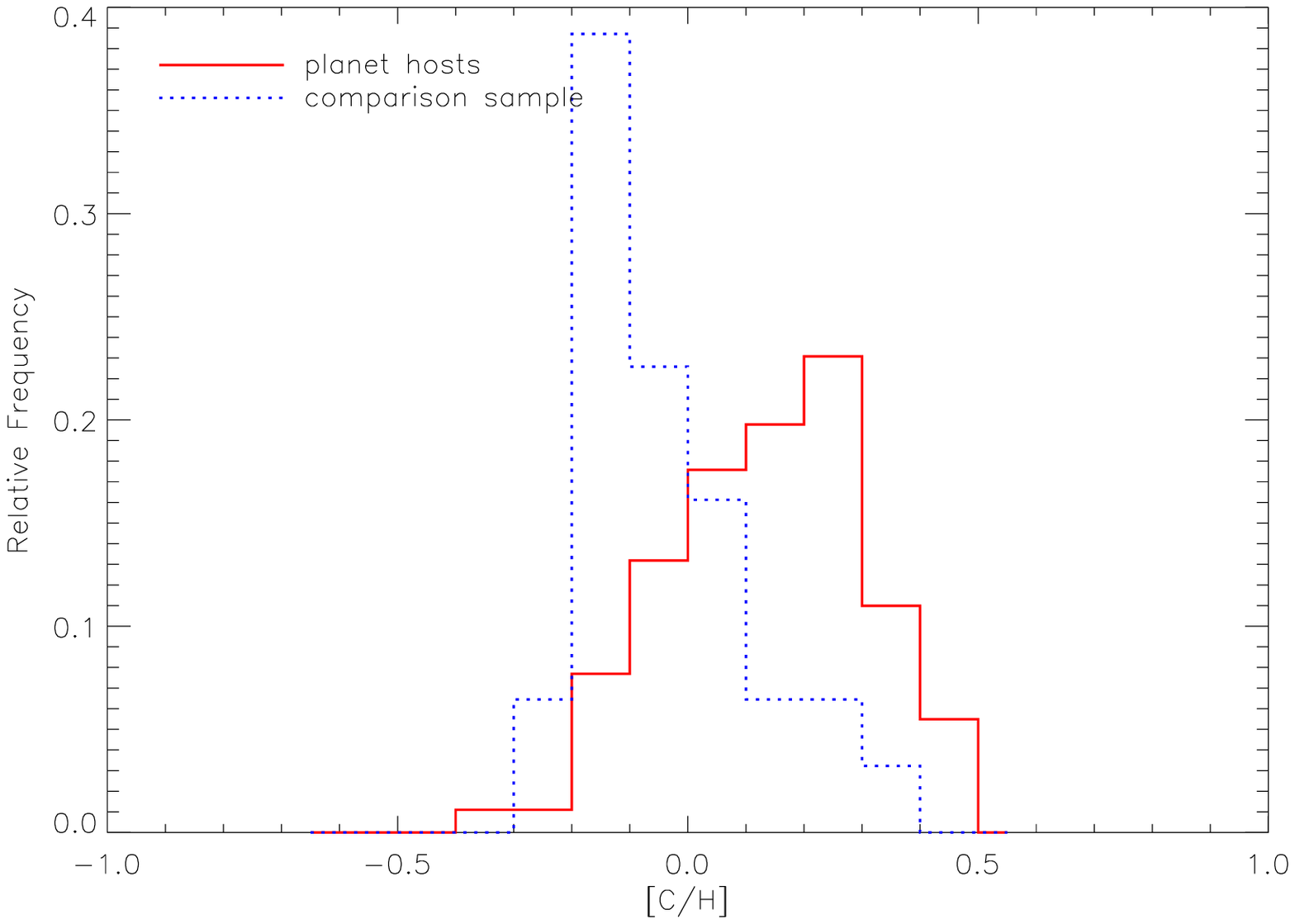}
\includegraphics[height=6cm]{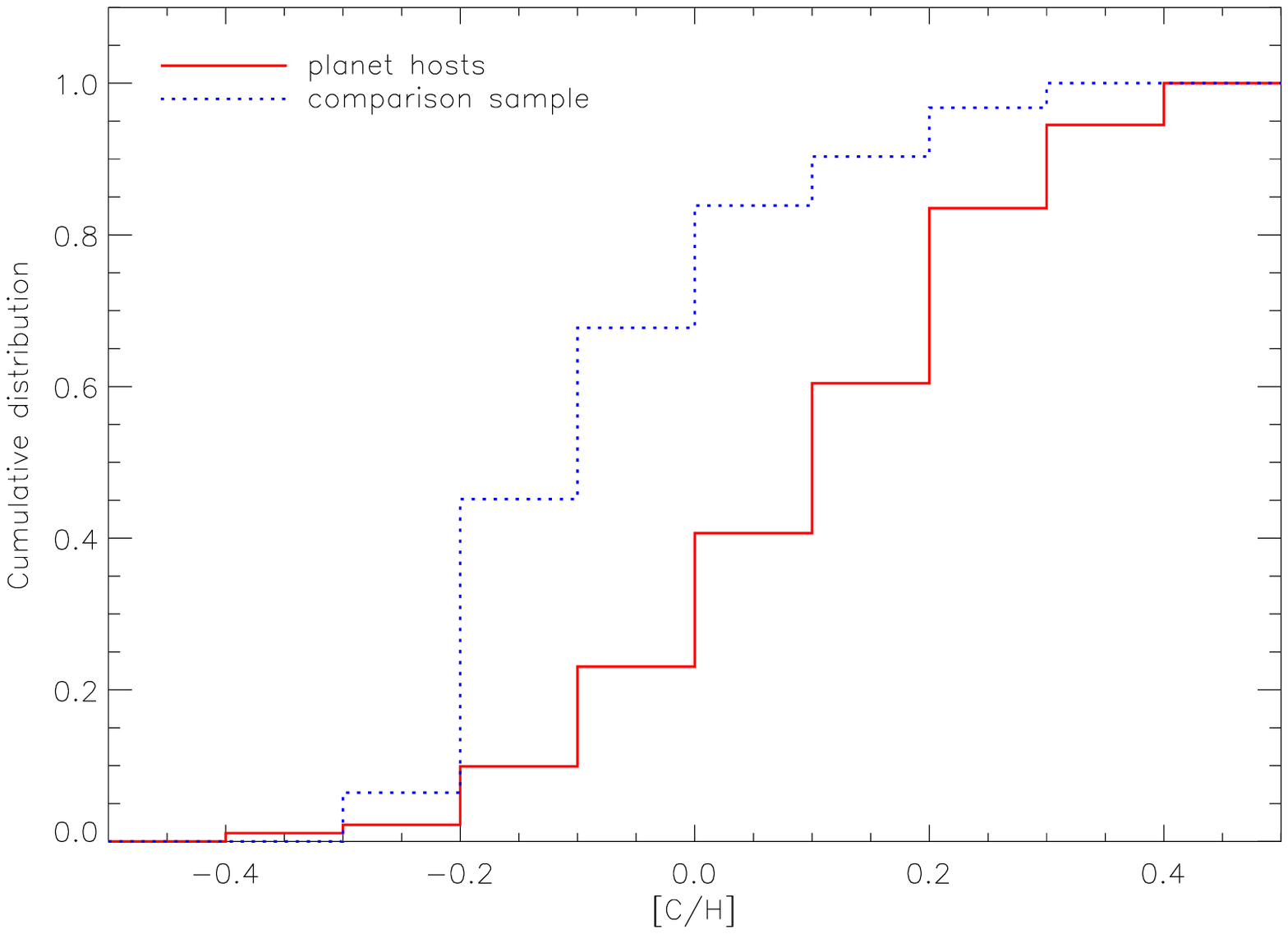}
\centering
\includegraphics[height=6cm]{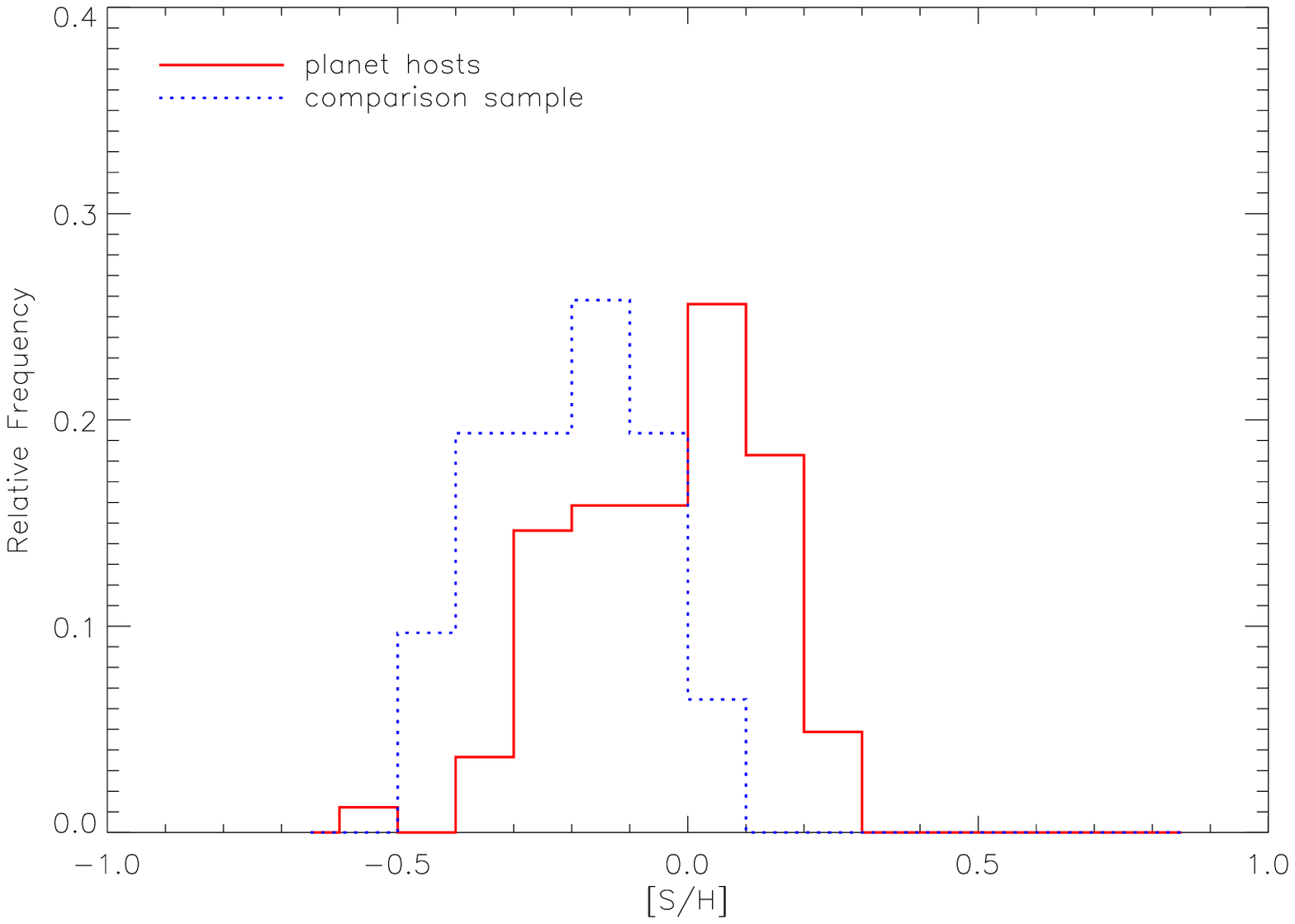}
\includegraphics[height=6cm]{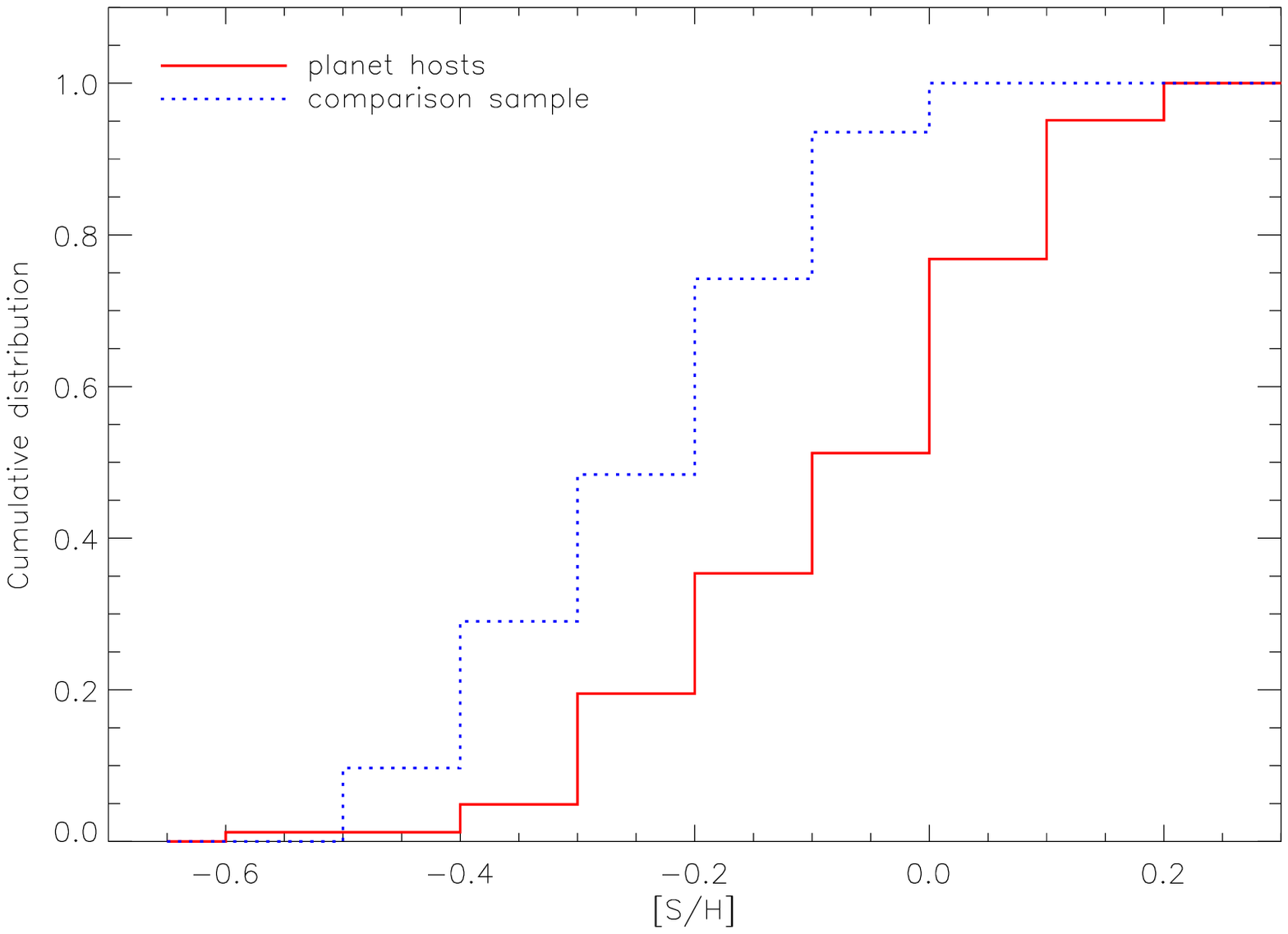}
\centering
\includegraphics[height=6cm]{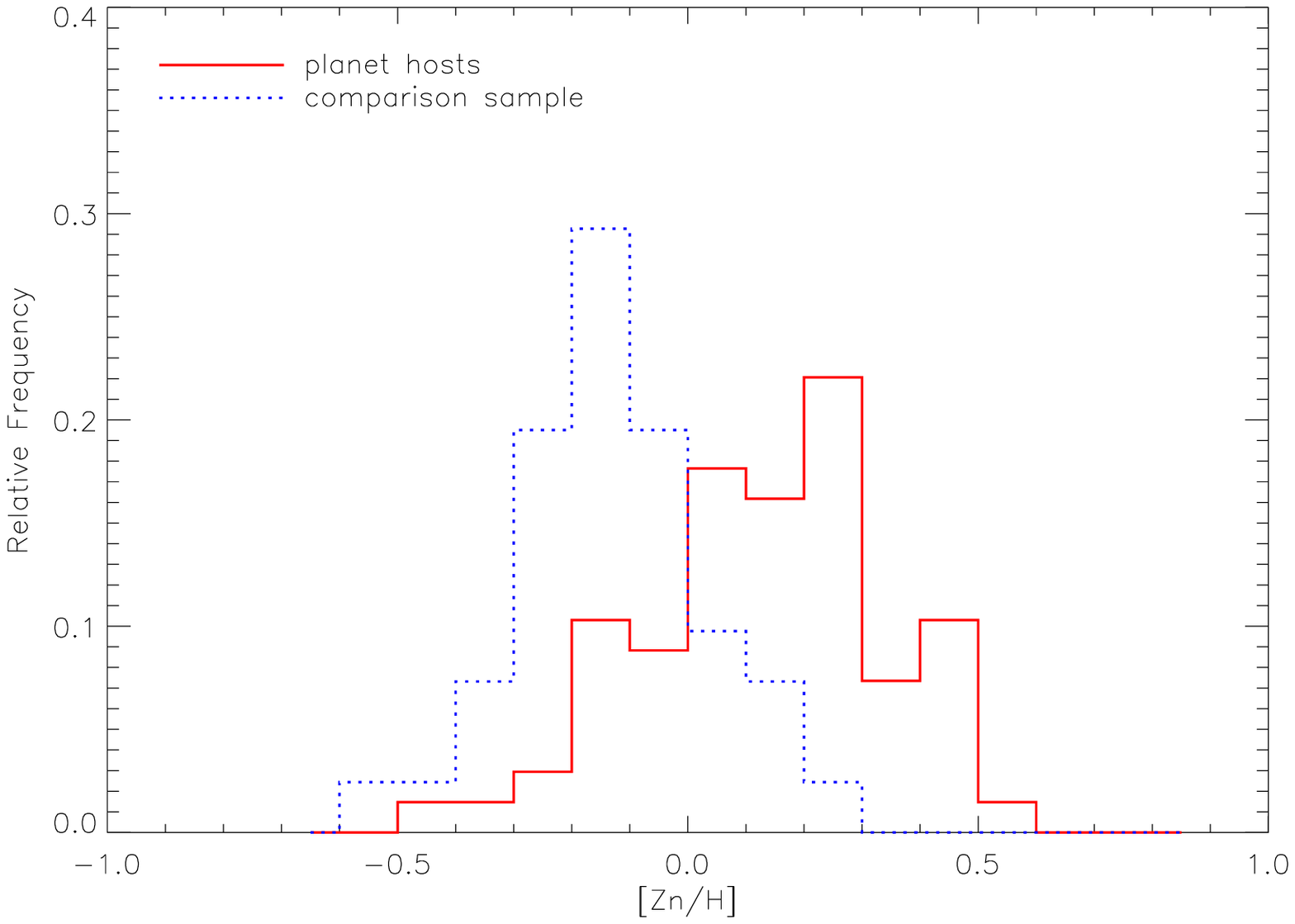}
\includegraphics[height=6cm]{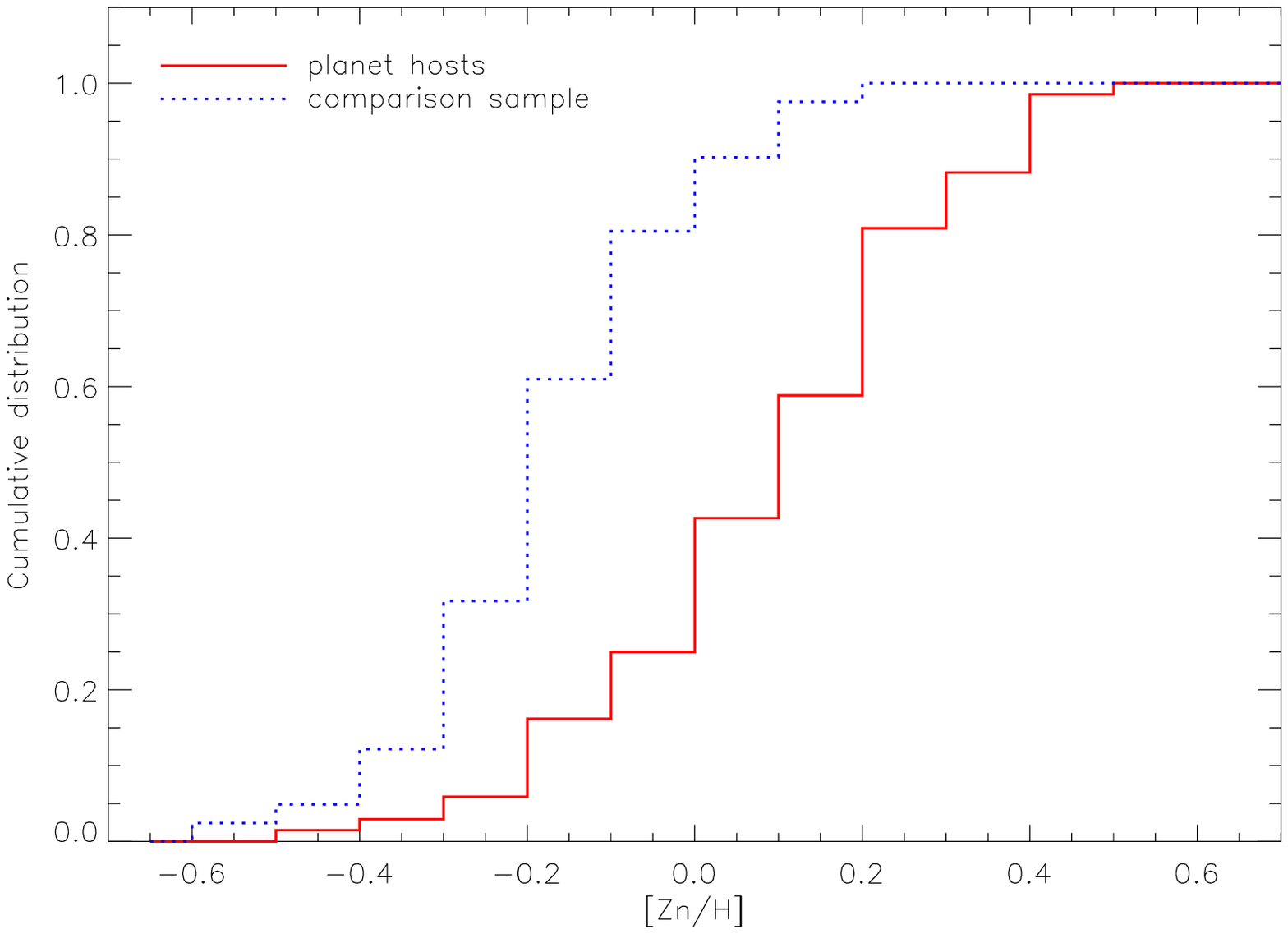}
\centering
\includegraphics[height=6cm]{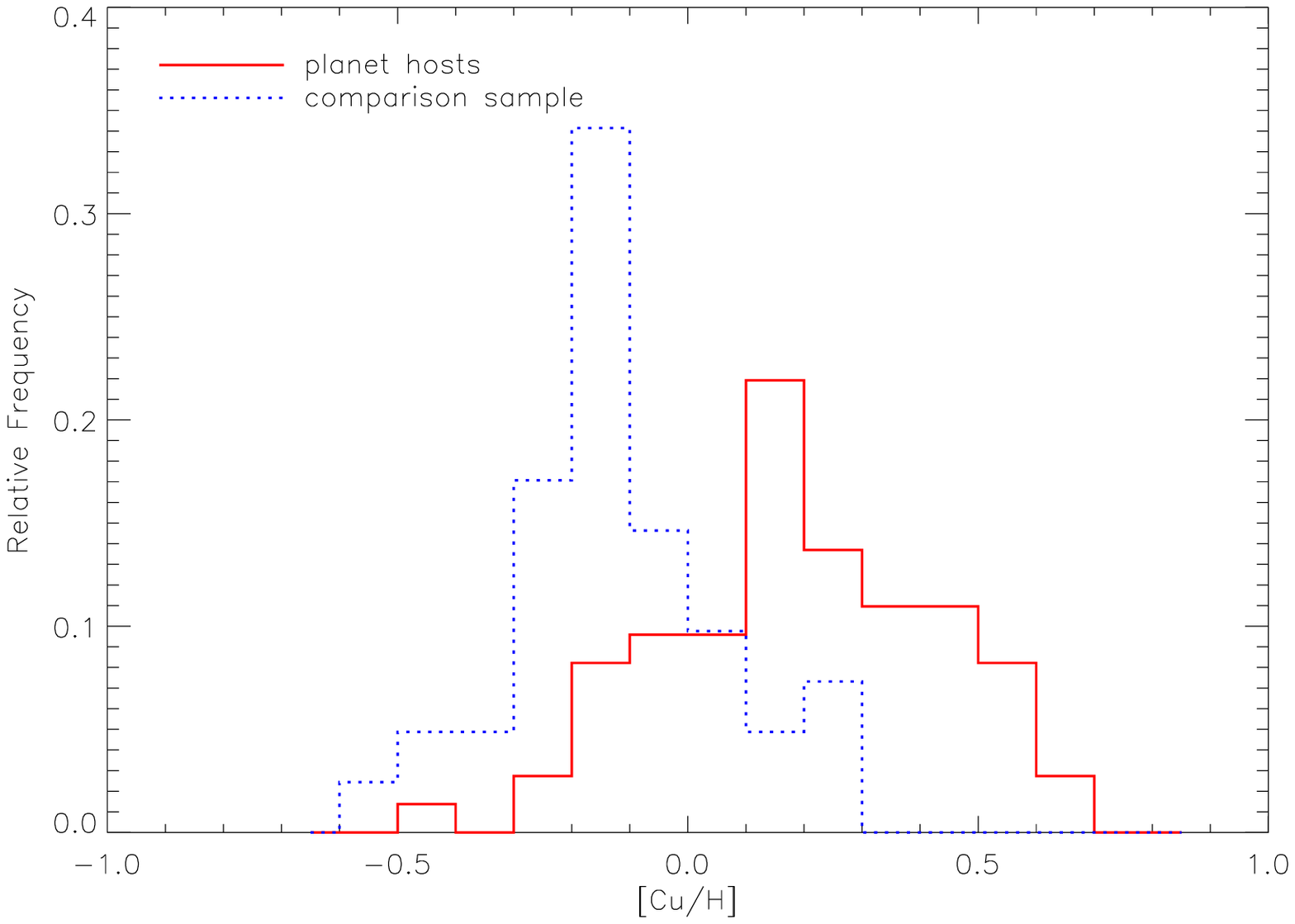}
\includegraphics[height=6cm]{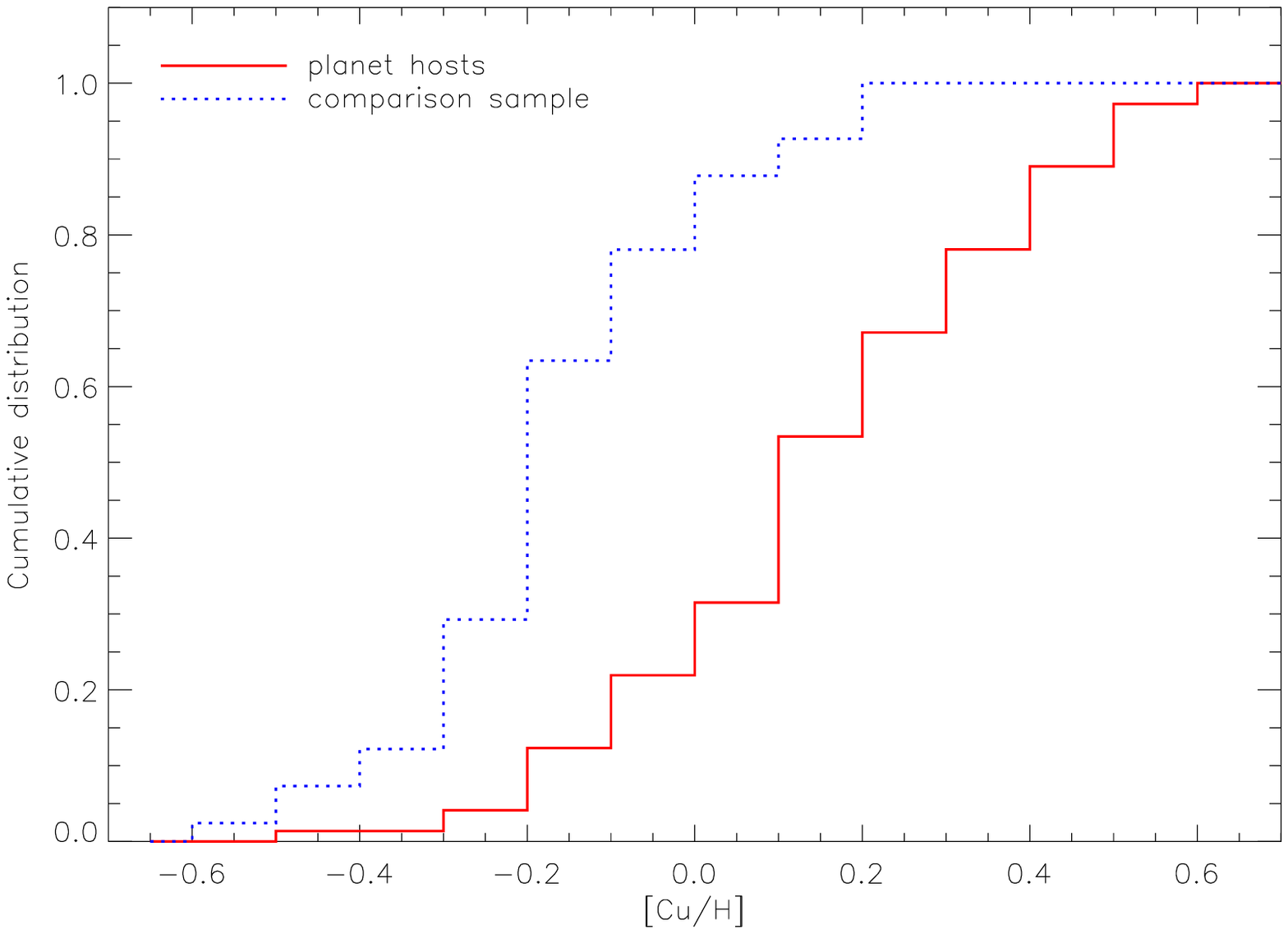}
\caption{[$X$/H] distributions for C, S, Zn and Cu. The
solid and dotted lines represent planet host and comparison sample stars,
respectively.}
\label{fig12}
\end{figure*}

The [$X$/H] distributions for the four elements are presented in Figure~\ref{fig12}. Volatile (C, S and Zn) distributions differ 
in their shapes. The planet host distributions are strongly asymmetrical, with peaks at high [$X$/H]. 
The comparison samples show different behaviours: the same asymmetric distribution for sulphur and a mirror shape for carbon.
 In the case of zinc, the comparison sample distribution looks more symmetrical. The [Cu/H] distributions for the two sets, stars 
with and without planets, present similar shapes, both quite symmetric. The cumulative distributions show that
differences between planet host and comparison sample stars are statistically significant.

The average values of [$X$/H] for the samples with and without planets for each element, as well as the rms 
dispersions and the differences between the mean [$X$/H] values, are listed in Table~\ref{TabAve}. The 
differences vary from 0.16 to 0.31\,dex. These discrepancies are not very significant because of the high 
dispersion around the mean value. 

We note that the differences between average values of [$X$/H] for the two samples are smaller for carbon and sulphur 
than for zinc and copper. For nitrogen, another volatile element, a difference of the order of 0.3\,dex was found between the average 
[N/H] values for planet host and comparison sample stars (Ecuvillon et al.\ \cite{Ecu04}). Even though these discrepancies are not 
statistically significant, we cannot completely exclude that this characteristic, if confirmed, may be related with the different 
condensation temperatures of the elements.

\begin{figure*}
\centering 
\includegraphics[height=6cm]{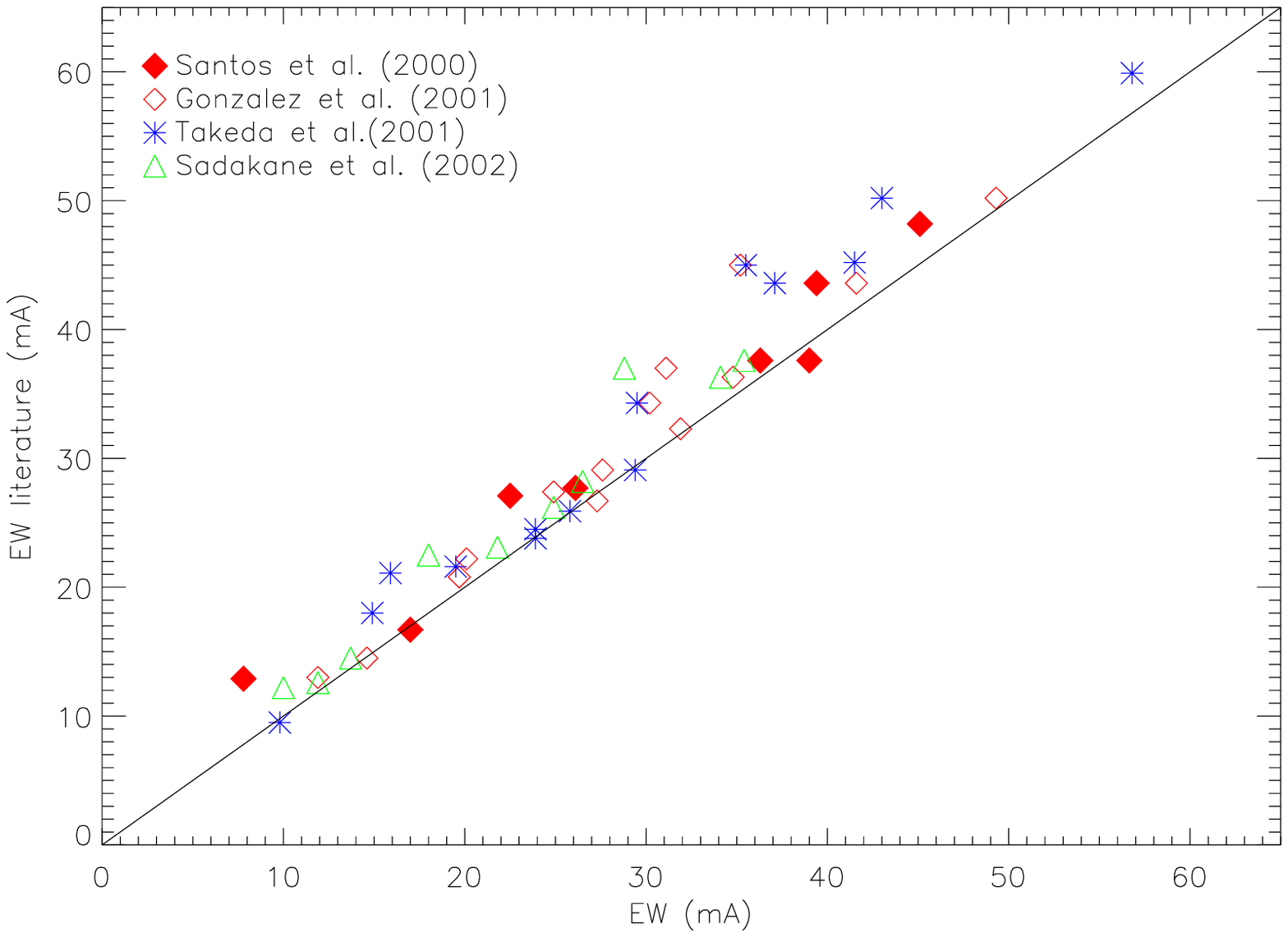}
\includegraphics[height=6cm]{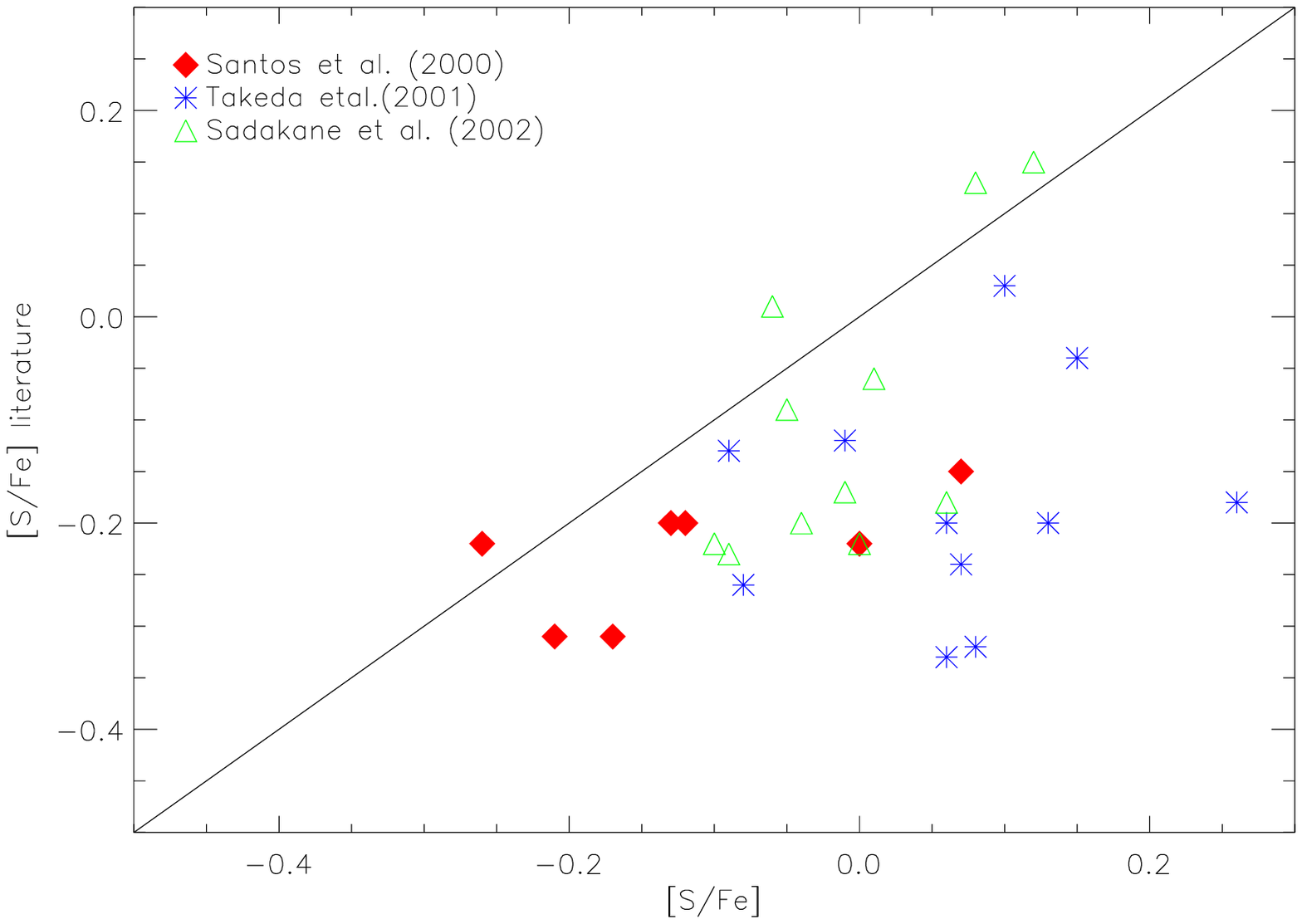}
\centering 
\includegraphics[height=6cm]{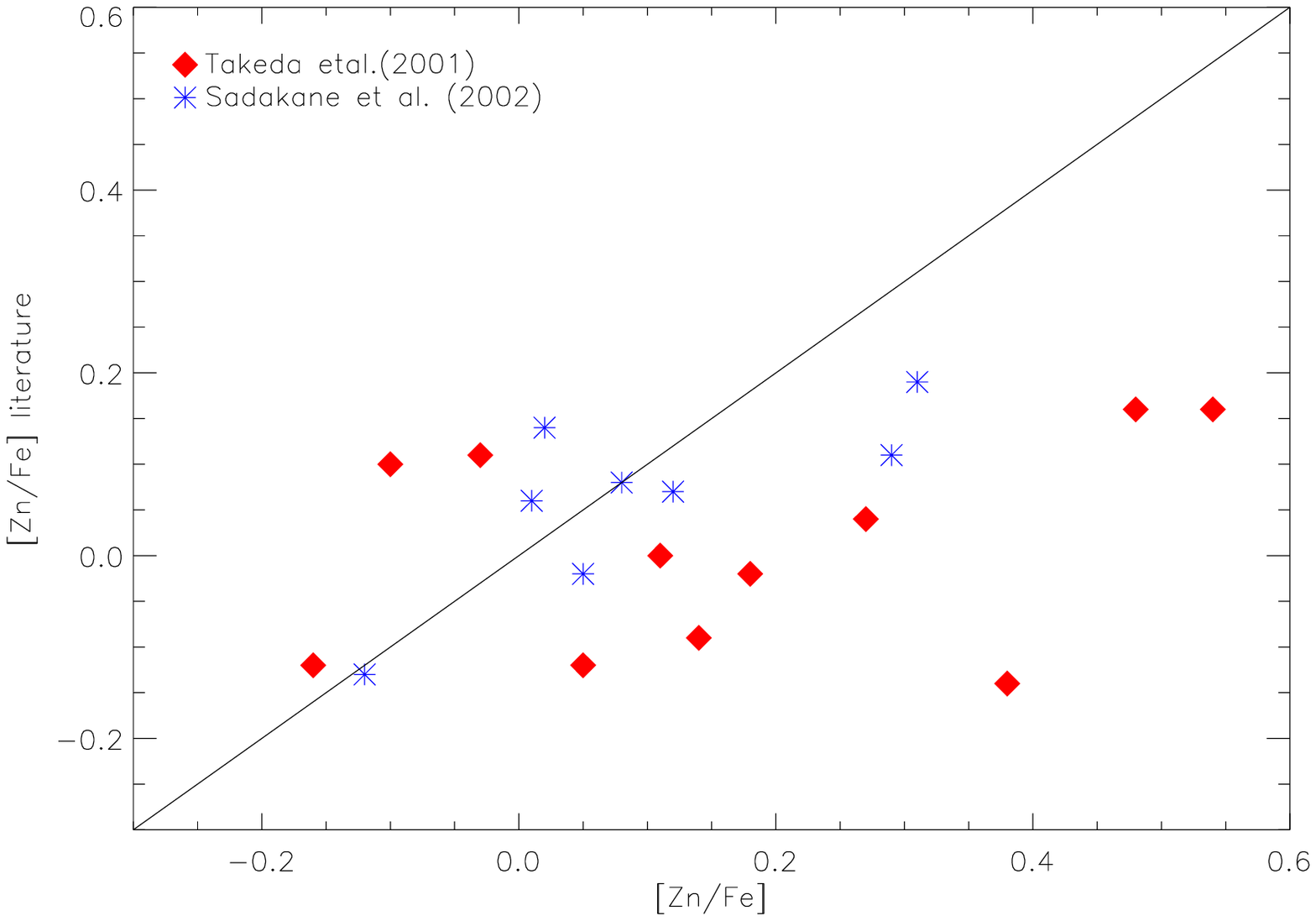}
\includegraphics[height=6cm]{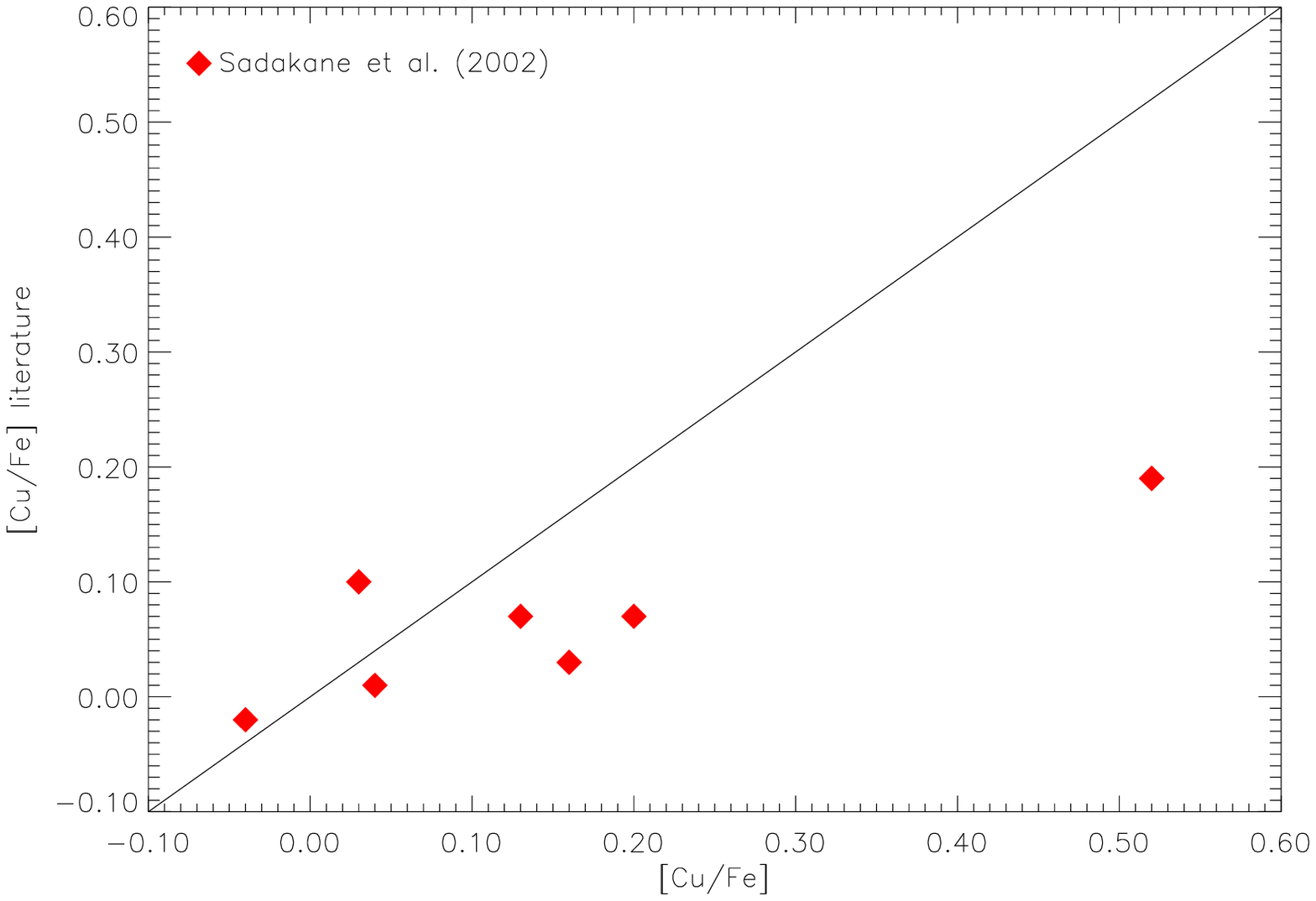}
\caption{{\it Top left}: comparison of the measured {\it EW}s of the \ion{C}{i} line at 5380 \AA\ with data 
given in Santos et al.\ (\cite{San00}, filled diamonds), in Gonzalez et al.\
(\cite{Gonz01}, open diamonds), 
in Takeda et al.\ (\cite{Tak02}, asterisks) and in Sadakane et al.\ (\cite{Sad02}, triangles). {\it Top
right}: comparison of [S/Fe] ratios from our work and from Santos et al.\ (\cite{San00}, filled symbols), 
Takeda et al.\ (\cite{Tak02}, asterisks) and Sadakane et al.\ (\cite{Sad02}, triangles). {\it Bottom left}:
comparison of [Zn/Fe] ratios from our work and from Takeda et al.\
(\cite{Tak02}, filled diamonds) and 
Sadakane et al.\ (\cite{Sad02},asterisks).{\it Bottom right}: comparison of [Cu/Fe] ratios from our work and 
from Sadakane et al.\ (\cite{Sad02},filled diamonds).}
\label{fig13}
\end{figure*}

\section{Comparison with the literature}
Previous studies concerning abundances of refractory and volatiles elements (see Sect.~\ref{Res}) have analysed planet host
stars common to our study. In order to carry out a homogeneous comparison, we gathered all these abundance results and 
scaled them to our atmospheric parameters using the published sensitivities for each atomic line.

Santos et al.\ (\cite{San00}), Gonzalez et al.\ (\cite{Gonz01}), Takeda et al.\ (\cite{Tak01}) and Sadakane et al.\ 
(\cite{Sad02}) analysed one of the two \ion{C}{i} lines we used to compute carbon abundances, the \ion{C}{i} line at
5380 \AA.  We could therefore directly compare the measured {EW}s. Table~\ref{tab16} shows all the atmospheric parameters and
the {EW} values for the 46 compared targets. All the measurements are in good agreement, especially for values of 
Gonzalez et al.\ (\cite{Gonz01}) from Keck spectra, with differences tipically
lower than 2 m\AA. In some cases, our data differ from 
the values taken from Takeda et al.\ (\cite{Tak01}), but the other sources agree better with our results (see 
\object{HD\,52265} and \object{HD\,89744}). Only for \object{HD\,92788}  is our measurement 
larger than the two  
values in the literature. 

For sulphur, comparisons for 31 targets were possible. All the atmospheric parameters and S
abundances are listed in Table~\ref{tab17}. [S/H] values from Santos et al.\ (\cite{San00}) and Sadakane et al.\ 
(\cite{Sad02}) are in good agreement with our results in most cases, with discrepancies lower than 0.2\,dex. However, 
significant disagreements, of the order of 0.4\,dex, appear when comparing with S abundances from Takeda et al.\ 
(\cite{Tak01}).
Using different lines and atomic parameters could be the cause of this discrepancy. 
Sadakane et al.\ (\cite{Sad02}) synthesized 
the same \ion{S}{i} lines as we did. Our results agree very well. Since  Gonzalez et al.\ 
(\cite{Gonz01}) did not include a sensitivity study in their work, we could not scale their results to our atmospheric 
parameters, thus preventing a homogeneous comparison.

Zinc abundances were compared with literature values in nineteen targets (see Table~\ref{tab18}). We obtained similar results in 
almost all cases, with differences lower than 0.2\,dex. Only \object{HD\,38529}, \object{HD\,75732} and \object{HD\,145675} 
show clear discrepancies with results from Takeda et al.\ (\cite{Tak01}). As for S, the cause of these differences could be
that different methods, with different sets of lines, were used.

Sadakane et al.\ (\cite{Sad02}) obtained copper abundances in eight planet host stars common to our work. The comparison is
presented in Table~\ref{tab19}. The results are in very good agreement, with differences lower than 0.13\,dex, except for 
\object{HD\,92788} and \object{HD\,134987}. In these two cases, the results may diverge because of the different methods used. 
All the comparisons are represented in Fig.~\ref{fig13}.

\section{Discussion and conclusions}
This article presents the abundances of one refractory (Cu) and three volatile (C, S and Zn)   elements in a large set of 
planet-harbouring stars and in an unbiased volume-limited comparison sample of solar-type dwarfs with no known planetary 
mass companions. An independent and uniform study of the two samples was performed for each element using atmospheric 
parameters derived from a detailed spectroscopic analysis by Santos et al. (\cite{San04a}). We have carried out a
careful comparison of our results with those already published by several authors and obtained a good agreement on the
whole. The result is a significant study for the completeness and homogeneity of the four  elements analysed. 

Comparing trends and searching for differences between the two samples, stars with and without known planets, can provide 
clues toward clarifying the formation and evolution of planetary systems. In particular, the behaviour of volatile elements can be very informative
for discriminating between the ``self-enrichment'' and the ``primordial'' hypotheses, and the relative importance of the
differential accretion (Gonzalez \cite{Gonz97}; Santos et al.\ \cite{San00}, \cite{San01}; Smith et al.\ \cite{Smi01}). If
the accretion of large amounts of metal-rich rocky material were the principal reason for the observed enhancement in
iron abundance in planet host stars, then volatiles would not show as much overabundance as refractories do in these
targets.

Our results show that abundances of volatile elements have the same behaviour in stars with and without planets. The
abundance trends for planet host and comparison sample stars are nearly indistinguishable. The planet host distributions are
the natural extensions of the comparison sample trends toward high metallicities. This may imply that the accretion of rocky
materials is not the principal cause of the metal-rich nature of stars with planets. Although the possibility of 
pollution is not excluded (see Laws \& Gonzalez \cite{Law01}; Israelian et al.\ \cite{Isr01}, \cite{Isr03a}), the hypothesis 
of a primordial metal-rich cloud out of which planetary systems would have formed seems  likelier than the 
``self-enrichment'' scenario, as an explanation for the observed metallicity enhancement. 

Most of the evidence suggests a ``primordial'' origin for the metallicity excess (Pinsonneault et al.\ \cite{Pin01};
Santos et al.\ \cite{San01}, \cite{San03b}, \cite{San04a}).
Previous studies of volatiles have already led to the same conclusion. Santos et al.\ (\cite{San00}) found no 
significant differences in [C/Fe] and [O/Fe] trends between field stars from literature and eight planet host stars. Takeda et 
al.\ (\cite{Tak01}) and Sadakane et al.\ (\cite{Sad02}) arrived at the same 
conclusion by comparing refractory and volatile elements 
and searching for a $T_{\rm C}$ dependence. Ecuvillon et al.\ (\cite{Ecu04}) found that nitrogen abundances show the same
behaviour in a large set of stars with and without known planets.  

Furthermore, we have obtained negative slopes in the [$X$/Fe] trends for the three volatiles, while the [N/Fe]
trend was previously found flat (Ecuvillon et al.\ \cite{Ecu04}). Contrary to results
obtained in previous 
studies  (see Takeda et al.\ \cite{Tak01}; Sadakane et al.\ \cite{Sad02}), C, S and Zn abundances do not scale
with that of iron. Since no differences are seen between the two samples, this behaviour must be  evidence of the 
chemical evolution of the Galactic disc, with no link to the presence of planets. In fact, several studies of 
abundances in the Galactic disc have revealed just such a clear linear behaviour for carbon (Friel \& Boesgaard \cite{Fri92}; 
Andersson \& Edvardsson \cite{And94}; Gustafsson et al.\ \cite{Gus99}; Shi et al.\ \cite{Shi02}) and for sulphur 
(Takada-Hidai et al.\ \cite{Tak02}; Chen et al.\ \cite{Che02}) in the metallicity range $-0.8<$ [Fe/H] $<$ 0.3. 
The [Zn/Fe] vs.\ [Fe/H] plot shows a decreasing trend for $-0.6<$ [Fe/H] $<$ 0 and a slight rise at 
metallicities above solar. A 
similar result has been recently obtained by Bensby et al.\ (\cite{Ben03}) for 66 disc stars.

Concerning the refractory Cu, [Cu/Fe] vs.\ [Fe/H] plot reveals a behaviour similar to Zn, with a slight decrease at 
$-$0.6 $<$ [Fe/H] $<$ 0 and a rise above the solar metallicity, steeper than for Zn. [Cu/Fe] is on average overabundant with 
respect to the solar value. The larger value of the slope resulting from fitting only stars with planets 
might be due to the 
observed [Cu/Fe] rise at [Fe/H] $>$ 0, probably related to  Galactic chemical evolutionary effects rather than to the 
presence of planets. In fact, planet host stars do not show different behaviour with respect to the comparison sample. 
Since all the available studies of Cu Galactic trends do not include the solar metallicity range, we cannot compare our 
results with those of the literature for this element.  
  
The [$X$/Fe] trends must be compared with detailed models of Galactic chemical evolution to distinguish possible effects due 
to the presence of planets.
In this framework, our work provides new additional data of C, S, Zn and Cu abundances to check 
Galactic chemical evolution models in the high [Fe/H] region. Because of the lack of detailed abundance studies reaching 
solar metallicities, our results can be very informative to improve our present understanding of stellar nucleosynthesis and
chemical evolution.

In the future, it will be very important to manage uniform studies of other volatile and refractory elements, with a wide
range of condensation temperatures $T_{\rm C}$. Possible systematic trends of [$X$/H] in terms of $T_{\rm C}$ will 
give us conclusive
information about the relative importance of differential accretion on the metallicity excess as a whole.

\begin{acknowledgements}
IRAF is distributed by the National Optical Astronomy Observatories, operated by the Association of Universities for 
Research in Astronomy, Inc., under contract with the National Science Foundation, USA.
\end{acknowledgements}


\begin{table*}[p]
\caption[]{Carbon abundances from CI lines at 5380 \AA\, and 5052 \AA\, for a
set of stars with planets and brown dwarf companions}
\begin{center}
\begin{tabular}{lccccccr}
\hline
\noalign{\smallskip}
Star & $T_\mathrm{eff}$ & $\log {g}$ & $\xi_t$ & [Fe/H] & EW$_{5380}$ & EW$_{5052}$ & [C/Fe] \\
 & (K) & (cm\,s$^{-2}$) & (km\,s$^{-1}$) & & (m\AA) & (m\AA)  & \\
\hline 
\hline
\noalign{\smallskip}
\object{HD\,142}     & $6302\pm56$ & $4.34\pm0.13$ & $ 1.86\pm0.17$ & $ 0.14\pm0.07$ & 52.1 & 69.1 & $ 0.10\pm0.08$ \\
\object{HD\,1237}    & $5536\pm50$ & $4.56\pm0.12$ & $ 1.33\pm0.06$ & $ 0.12\pm0.06$ & 16.7 & 27.6 & $-0.06\pm0.05$ \\
\object{HD\,2039}    & $5976\pm51$ & $4.45\pm0.10$ & $ 1.26\pm0.07$ & $ 0.32\pm0.06$ & 33.3 & 52.8 & $-0.13\pm0.05$ \\
\object{HD\,3651}    & $5173\pm35$ & $4.37\pm0.12$ & $ 0.74\pm0.05$ & $ 0.12\pm0.04$ & 12.9 & 20.1 & $-0.01\pm0.07$ \\
\object{HD\,4203}    & $5636\pm40$ & $4.23\pm0.14$ & $ 1.12\pm0.05$ & $ 0.40\pm0.05$ & 41.0 & 55.9 & $ 0.05\pm0.07$ \\
\object{HD\,4208}    & $5626\pm32$ & $4.49\pm0.10$ & $ 0.95\pm0.06$ & $-0.24\pm0.04$ & 20.5 & 22.2 & $ 0.23\pm0.19$ \\
\hline
\end{tabular}
\end{center}
\label{tab4}
\end{table*}

\begin{table*}[p]
\caption[]{Continued Table 4.}
\begin{center}
\begin{tabular}{lccccccr}
\hline
\noalign{\smallskip}
Star & $T_\mathrm{eff}$ & $\log {g}$ & $\xi_t$ & [Fe/H] & EW$_{5380}$ & EW$_{5052}$  & [C/Fe] \\
 & (K) & (cm\,s$^{-2}$) & (km\,s$^{-1}$) & & (m\AA) & (m\AA)& \\
\hline 
\hline
\noalign{\smallskip}
\object{HD\,121504}  & $6075\pm40$ & $4.64\pm0.12$ & $ 1.31\pm0.07$ & $ 0.16\pm0.05$ & 29.7 & 41.8 & $-0.10\pm0.08$  \\
\object{HD\,130322}  & $5392\pm36$ & $4.48\pm0.06$ & $ 0.85\pm0.05$ & $ 0.03\pm0.04$ & 14.5 & 19.5 & $-0.03\pm0.11$  \\
\object{HD\,134987}  & $5776\pm29$ & $4.36\pm0.07$ & $ 1.09\pm0.04$ & $ 0.30\pm0.04$ & 36.3 & 55.2 & $ 0.03\pm0.03$  \\
\object{HD\,136118}  & $6222\pm39$ & $4.27\pm0.15$ & $ 1.79\pm0.12$ & $-0.04\pm0.05$ & 36.9 & 53.4 & $ 0.07\pm0.06$  \\
\object{HD\,137759}  & $4775\pm113$ & $3.09\pm0.40$ & $ 1.78\pm0.11$ & $0.13\pm0.14$ & 17.3 & 25.5 & $ 0.08\pm0.16$  \\
\object{HD\,141937}  & $5909\pm39$ & $4.51\pm0.08$ & $ 1.13\pm0.06$ & $ 0.10\pm0.05$ & 28.2 & 49.4 & $ 0.07\pm0.06$  \\
\hline
\end{tabular}
\end{center}
\label{tab5}
\end{table*}

\begin{table*}[p]
\caption[]{Carbon abundances from CI lines at 5380 \AA\, and 5052 \AA\, for a
set of comparison stars.}
\begin{center}
\begin{tabular}{lccccccr}
\hline
\noalign{\smallskip}
Star & $T_\mathrm{eff}$ & $\log {g}$ & $\xi_t$ & [Fe/H] & EW$_{5380}$ & EW$_{5052}$  & [C/Fe] \\
 & (K) & (cm\,s$^{-2}$) & (km\,s$^{-1}$) & & (m\AA) & (m\AA) &  \\
\hline 
\hline
\noalign{\smallskip}
\object{HD\,1581 }   & $ 5956\pm44$ & $4.39\pm0.13$ & $1.07\pm0.09$ & $-0.14\pm0.05$ & 18.8 & 30.7 & $-0.04\pm0.05$  \\
\object{HD\,4391 }   & $ 5878\pm53$ & $4.74\pm0.15$ & $1.13\pm0.10$ & $-0.03\pm0.06$ & 18.9 & 35.8 & $ 0.07\pm0.08$  \\
\object{HD\,7570 }   & $ 6140\pm41$ & $4.39\pm0.16$ & $1.50\pm0.08$ & $ 0.18\pm0.05$ & 39.9 & 62.2 & $ 0.03\pm0.07$  \\
\object{HD\,10700}   & $ 5344\pm29$ & $4.57\pm0.09$ & $0.91\pm0.06$ & $-0.52\pm0.04$ &  7.0 & 12.6 & $ 0.33\pm0.04$  \\
\object{HD\,14412}   & $ 5368\pm24$ & $4.55\pm0.05$ & $0.88\pm0.05$ & $-0.47\pm0.03$ &  7.5 & 10.2 & $ 0.22\pm0.12$  \\
\object{HD\,17925}   & $ 5180\pm56$ & $4.44\pm0.13$ & $1.33\pm0.08$ & $ 0.06\pm0.07$ &  9.2 & 15.6 & $-0.10\pm0.07$  \\
\hline
\end{tabular}
\end{center}
\label{tab6}
\end{table*}

\begin{table*}[p]
\caption[]{Sulphur abundances from SI lines at 6743 \AA\ and 6757 \AA\ for a
set of stars with planets and brown dwarf companions.}
\begin{center}
\begin{tabular}{lcccccccr}
\hline
\noalign{\smallskip}
Star & $T_\mathrm{eff}$ & $\log {g}$ & $\xi_t$ & [Fe/H] & [S/Fe]$_1$ & [S/Fe]$_2$ & [S/Fe]\\ 
 & (K) & (cm\,s$^{-2}$) & (km\,s$^{-1}$) & &  &  & \\
\hline 
\hline
\noalign{\smallskip}
  \object{HD\,142}  & $6302\pm56$ & $4.34\pm0.13$ & $1.86\pm0.17$ & $ 0.14\pm0.07$ &  $-$    & $-0.39$ & $ -0.39 \pm  0.09 $\\
  \object{HD\,1237} & $5536\pm50$ & $4.56\pm0.12$ & $1.33\pm0.06$ & $ 0.12\pm0.06$ &  $-$    & $-0.22$ & $ -0.22 \pm  0.08 $\\
  \object{HD\,2039} & $5976\pm51$ & $4.45\pm0.10$ & $1.26\pm0.07$ & $ 0.32\pm0.06$ &  $-$    & $-0.17$ & $ -0.17 \pm  0.08 $\\
  \object{HD\,3651} & $5173\pm35$ & $4.37\pm0.12$ & $0.74\pm0.05$ & $ 0.12\pm0.04$ & $ 0.13$ & $ 0.08$ & $  0.11 \pm  0.08 $\\
  \object{HD\,4203} & $5636\pm40$ & $4.23\pm0.14$ & $1.12\pm0.05$ & $ 0.40\pm0.05$ & $-0.20$ & $-0.20$ & $ -0.20 \pm  0.08 $\\
  \object{HD\,4208} & $5626\pm32$ & $4.49\pm0.10$ & $0.95\pm0.06$ & $-0.24\pm0.04$ & $-0.16$ & $-0.11$ & $ -0.14 \pm  0.07 $\\
\hline
\end{tabular}
\end{center}
\label{tab7}
\end{table*}

\begin{table*}[p]
\caption[]{Continued Table 7.}
\begin{center}
\begin{tabular}{lccccccr}
\hline
\noalign{\smallskip}
Star & $T_\mathrm{eff}$ & $\log {g}$ & $\xi_t$ & [Fe/H] & [S/Fe]$_1$ & [S/Fe]$_2$ & [S/Fe] \\
 & (K) & (cm\,s$^{-2}$) & (km\,s$^{-1}$) & &  &  & \\
\hline 
\hline
\noalign{\smallskip}
\object{HD\,141937} & $5909\pm39$ & $4.51\pm0.08$ & $1.13\pm0.06$ & $ 0.10\pm0.05$ & $-0.15$ & $-0.20$ & $ -0.18 \pm  0.07 $\\
\object{HD\,142415} & $6045\pm44$ & $4.53\pm0.08$ & $1.12\pm0.07$ & $ 0.21\pm0.05$ & $-0.06$ & $-0.11$ & $ -0.09 \pm  0.08 $\\
\object{HD\,143761} & $5853\pm25$ & $4.41\pm0.15$ & $1.35\pm0.07$ & $-0.21\pm0.04$ & $-0.04$ & $-0.04$ & $ -0.04 \pm  0.07 $\\
\object{HD\,145675} & $5311\pm87$ & $4.42\pm0.18$ & $0.92\pm0.10$ & $ 0.43\pm0.08$ & $-0.13$ & $-0.13$ & $ -0.13 \pm  0.12 $\\
\object{HD\,147513} & $5883\pm25$ & $4.51\pm0.05$ & $1.18\pm0.04$ & $ 0.06\pm0.04$ & $-$     & $-0.21$ & $ -0.21 \pm  0.06 $\\
\object{HD\,160691} & $5798\pm33$ & $4.31\pm0.08$ & $1.19\pm0.04$ & $ 0.32\pm0.04$ & $-0.17$ & $-0.17$ & $ -0.17 \pm  0.07 $\\
\hline
\end{tabular}
\end{center}
\label{tab8}
\end{table*}

\begin{table*}[p]
\caption[]{Sulphur abundances from SI lines 6743 \AA\ and 6757 \AA\ for a
set of comparison stars.}
\begin{center}
\begin{tabular}{lccccccr}
\hline
\noalign{\smallskip}
Star & $T_\mathrm{eff}$ & $\log {g}$ & $\xi_t$ & [Fe/H] & [S/Fe]$_1$ & [S/Fe]$_2$ & [S/Fe] \\
 & (K) & (cm\,s$^{-2}$) & (km\,s$^{-1}$) & &  &  & \\
\hline 
\hline
\noalign{\smallskip}
\object{HD\,1581  } & $5956\pm44$ & $4.39\pm0.13$ & $1.07\pm0.09$ & $-0.14\pm0.05$ & $-0.16$ & $-0.11$ & $ -0.28 \pm  0.08 $\\
\object{HD\,4391  } & $5878\pm53$ & $4.74\pm0.15$ & $1.13\pm0.10$ & $-0.03\pm0.06$ & $-$     & $-0.02$ & $ -0.02 \pm  0.09 $\\
\object{HD\,5133  } & $4911\pm54$ & $4.49\pm0.18$ & $0.71\pm0.11$ & $-0.17\pm0.06$ & $-0.23$ & $-0.23$ & $ -0.23 \pm  0.09 $\\
\object{HD\,7570  } & $6140\pm41$ & $4.39\pm0.16$ & $1.50\pm0.08$ & $ 0.18\pm0.05$ & $-0.18$ & $-0.18$ & $ -0.18 \pm  0.08 $\\
\object{HD\,10360 } & $4970\pm40$ & $4.49\pm0.10$ & $0.76\pm0.07$ & $-0.26\pm0.04$ & $-$     & $ 0.06$ & $  0.06 \pm  0.07 $\\
\object{HD\,10700 } & $5344\pm29$ & $4.57\pm0.09$ & $0.91\pm0.06$ & $-0.52\pm0.04$ & $ 0.02$ & $ 0.12$ & $  0.07 \pm  0.08 $\\
\hline
\end{tabular}
\end{center}
\label{tab9}
\end{table*}

\begin{table*}[p]
\caption[]{Zn abundances from ZnI lines at 4810 \AA\ and 4722 \AA\ for a set of stars with planets and brown dwarf companions.}
\begin{center}
\begin{tabular}{lcccccccr}
\hline
\noalign{\smallskip}
Star & $T_\mathrm{eff}$ & $\log {g}$ & $\xi_t$ & [Fe/H] & [Zn/Fe]$_1$ & [Zn/Fe]$_2$ & [Zn/Fe] \\
 & (K) & (cm\,s$^{-2}$) & (km\,s$^{-1}$) & & & & & \\
\hline 
\hline
\noalign{\smallskip}
\object{HD\,142   }&$6302\pm56 $&$ 4.34\pm0.13 $&$1.86\pm0.17 $&$  0.14\pm0.07 $&$-0.25 $&$  -0.16 $&$ -0.21 \pm  0.10 $\\   
\object{HD\,2039  }&$5976\pm51 $&$ 4.45\pm0.10 $&$1.26\pm0.07 $&$  0.32\pm0.06 $&$-0.05 $&$  -0.03 $&$ -0.04 \pm  0.07 $\\   
\object{HD\,4203  }&$5636\pm40 $&$ 4.23\pm0.14 $&$1.12\pm0.05 $&$  0.40\pm0.05 $&$0.00	$&$  0.02  $&$  0.01 \pm  0.07 $\\   
\object{HD\,4208  }&$5626\pm32 $&$ 4.49\pm0.10 $&$0.95\pm0.06 $&$ -0.24\pm0.04 $&$0.12  $&$  0.22  $&$  0.17 \pm  0.08 $\\   
\object{HD\,6434  }&$5835\pm59 $&$ 4.60\pm0.12 $&$1.53\pm0.27 $&$ -0.52\pm0.08 $&$0.05  $&$  0.07  $&$  0.06 \pm  0.10 $\\   
\object{HD\,8574  }&$6151\pm57 $&$ 4.51\pm0.10 $&$1.45\pm0.15 $&$  0.06\pm0.07 $&$-0.17 $&$  -0.08 $&$ -0.13 \pm  0.10 $\\
\hline										  
\end{tabular}									  
\end{center}									  
\label{tab10} 	
\end{table*}									  
\clearpage
\begin{table*}[p]
\caption[]{Continued Table 10.}
\begin{center}
\begin{tabular}{lccccccr}
\hline
\noalign{\smallskip}
Star & $T_\mathrm{eff}$ & $\log {g}$ & $\xi_t$ & [Fe/H] & [Zn/Fe]$_1$ & [Zn/Fe]$_2$ & [Zn/Fe] \\
 & (K) & (cm\,s$^{-2}$) & (km\,s$^{-1}$) & &  &  & \\
\hline 
\hline
\noalign{\smallskip}
\object{HD\,187123}&$5845\pm22 $&$ 4.42\pm0.07 $&$1.10\pm0.03 $&$  0.13\pm0.03 $&$0.10	$&$  0.05  $&$  0.08 \pm  0.06 $\\   
\object{HD\,190360}&$5584\pm36 $&$ 4.37\pm0.06 $&$1.07\pm0.05 $&$  0.24\pm0.05 $&$0.05  $&$  0.07  $&$  0.06 \pm  0.07 $\\   
\object{HD\,192263}&$4947\pm58 $&$ 4.51\pm0.20 $&$0.86\pm0.09 $&$ -0.02\pm0.06 $&$0.06  $&$  0.07  $&$  0.07 \pm  0.09 $\\   
\object{HD\,195019}&$5859\pm31 $&$ 4.32\pm0.07 $&$1.27\pm0.05 $&$  0.09\pm0.04 $&$-0.05 $&$  -0.08 $&$ -0.07 \pm  0.06 $\\   
\object{HD\,196050}&$5918\pm44 $&$ 4.35\pm0.13 $&$1.39\pm0.06 $&$  0.22\pm0.05 $&$0.00	$&$  0.02  $&$  0.01 \pm  0.07 $\\   
\object{HD\,210277}&$5532\pm00 $&$ 4.29\pm0.00 $&$1.03\pm0.00 $&$  0.19\pm0.00 $&$0.05  $&$  0.12  $&$  0.09 \pm  0.07 $\\
\hline										  
\end{tabular}									  
\end{center}									  
\label{tab11} 	
\end{table*}	
\begin{table*}[p]
\caption[]{Zn abundances from ZnI lines at 4810 \AA\ and 4722 \AA\ for a set of comparison stars.}
\begin{center}
\begin{tabular}{lccccccr}
\hline
\noalign{\smallskip}
Star & $T_\mathrm{eff}$ & $\log {g}$ & $\xi_t$ & [Fe/H] & [Zn/Fe]$_1$ & [Zn/Fe]$_2$ & [Zn/Fe] \\
 & (K) & (cm\,s$^{-2}$) & (km\,s$^{-1}$) & &  &  & \\
\hline 
\hline
\noalign{\smallskip}
\object{HD\,1581  }&$5956\pm44 $&$ 4.39\pm0.13 $&$1.07\pm0.09 $&$ -0.14\pm0.05 $&$0.09  $&$  0.06  $&$  0.08 \pm  0.08 $\\ 
\object{HD\,4391  }&$5878\pm53 $&$ 4.74\pm0.15 $&$1.13\pm0.10 $&$ -0.03\pm0.06 $&$-0.05 $&$  -0.08 $&$ -0.07 \pm  0.08 $\\ 
\object{HD\,5133  }&$4911\pm54 $&$ 4.49\pm0.18 $&$0.71\pm0.11 $&$ -0.17\pm0.06 $&$-0.05 $&$  -0.03 $&$ -0.04 \pm  0.08 $\\ 
\object{HD\,7570  }&$6140\pm41 $&$ 4.39\pm0.16 $&$1.50\pm0.08 $&$  0.18\pm0.05 $&$-0.05 $&$  -0.13 $&$ -0.09 \pm  0.08 $\\ 
\object{HD\,10360 }&$4970\pm40 $&$ 4.49\pm0.10 $&$0.76\pm0.07 $&$ -0.26\pm0.04 $&$0.00	$&$  0.05  $&$  0.03 \pm  0.07 $\\ 
\object{HD\,10700 }&$5344\pm29 $&$ 4.57\pm0.09 $&$0.91\pm0.06 $&$ -0.52\pm0.04 $&$0.15  $&$  0.19  $&$  0.17 \pm  0.07 $\\
\hline
\end{tabular}
\end{center}
\label{tab12}
\end{table*}

\begin{table*}[p]
\caption[]{Cu abundances from CuI lines at 5218 \AA\ and 5782 \AA\ for a set of stars with planets and brown dwarf companions.}
\begin{center}
\begin{tabular}{lccccccr}
\hline
\noalign{\smallskip}
Star & $T_\mathrm{eff}$ & $\log {g}$ & $\xi_t$ & [Fe/H] & [Cu/Fe]$_1$ & [Cu/Fe]$_2$ & [Cu/Fe] \\
 & (K) & (cm\,s$^{-2}$) & (km\,s$^{-1}$) & &  &  &  \\
\hline 
\hline
\noalign{\smallskip}
\object{HD\,142   }&$6302\pm56 $&$ 4.34\pm0.13 $&$1.86\pm0.17 $&$  0.14\pm0.07 $&$-0.06 $&$  0.04 $&$-0.01 \pm0.10$\\ 
\object{HD\,2039  }&$5976\pm51 $&$ 4.45\pm0.10 $&$1.26\pm0.07 $&$  0.32\pm0.06 $&$ 0.08 $&$  0.12 $&$ 0.10 \pm0.08$\\ 
\object{HD\,4203  }&$5636\pm40 $&$ 4.23\pm0.14 $&$1.12\pm0.05 $&$  0.40\pm0.05 $&$ 0.20	$&$ 0.10  $&$ 0.15 \pm0.08$\\ 
\object{HD\,4208  }&$5626\pm32 $&$ 4.49\pm0.10 $&$0.95\pm0.06 $&$ -0.24\pm0.04 $&$ 0.06	$&$ 0.07  $&$ 0.07 \pm0.06$\\ 
\object{HD\,6434  }&$5835\pm59 $&$ 4.60\pm0.12 $&$1.53\pm0.27 $&$ -0.52\pm0.08 $&$ 0.00	$&$ 0.08  $&$ 0.04 \pm0.11$\\ 
\object{HD\,8574  }&$6151\pm57 $&$ 4.51\pm0.10 $&$1.45\pm0.15 $&$  0.06\pm0.07 $&$ 0.00 $&$  0.04 $&$ 0.02 \pm0.09$\\
\hline																    
\end{tabular}															    
\end{center}															    
\label{tab13}	
\end{table*}
\begin{table*}[p]
\caption[]{Continued Table 13.}
\begin{center}
\begin{tabular}{lccccccr}
\hline
\noalign{\smallskip}
Star & $T_\mathrm{eff}$ & $\log {g}$ & $\xi_t$ & [Fe/H] & [Cu/Fe]$_1$ & [Cu/Fe]$_2$ & [Cu/Fe] \\
 & (K) & (cm\,s$^{-2}$) & (km\,s$^{-1}$) & &  &  &  \\
\hline 
\hline
\noalign{\smallskip}
\object{HD\,168746}&$5601\pm33 $&$ 4.41\pm0.12 $&$0.99\pm0.05 $&$ -0.08\pm0.05 $&$0.10	$&$ 0.10  $&$ 0.10 \pm0.07$\\ 
\object{HD\,177830}&$4804\pm77 $&$ 3.57\pm0.17 $&$1.14\pm0.09 $&$  0.33\pm0.09 $&$ 0.17  $&$  0.24 $&$  0.21 \pm0.10 $\\
\object{HD\,178911}&$5588\pm115$&$ 4.46\pm0.20 $&$0.82\pm0.14 $&$  0.24\pm0.10 $&$ -0.14 $&$ -0.17 $&$ -0.16 \pm0.11 $\\
\object{HD\,179949}&$6260\pm43 $&$ 4.43\pm0.05 $&$1.41\pm0.09 $&$  0.22\pm0.05 $&$ -0.05 $&$ -0.03 $&$ -0.04 \pm0.07 $\\
\object{HD\,186427}&$5772\pm25 $&$ 4.40\pm0.07 $&$1.07\pm0.04 $&$  0.08\pm0.04 $&$ 0.12  $&$  0.12 $&$  0.12 \pm0.06 $\\
\object{HD\,187123}&$5845\pm22 $&$ 4.42\pm0.07 $&$1.10\pm0.03 $&$  0.13\pm0.03 $&$ 0.05  $&$  0.09 $&$  0.07 \pm0.06 $\\
\hline																    
\end{tabular}															    
\end{center}															    
\label{tab14}	
\end{table*}

\begin{table*}[p]
\caption[]{Cu abundances from CuI lines at 5218 \AA\ and 5782 \AA\ a for a set of comparison stars.}
\begin{center}
\begin{tabular}{lccccccr}
\hline
\noalign{\smallskip}
Star & $T_\mathrm{eff}$ & $\log {g}$ & $\xi_t$ & [Fe/H] & [Cu/Fe]$_1$ & [Cu/Fe]$_2$ & [Cu/Fe] \\
 & (K) & (cm\,s$^{-2}$) & (km\,s$^{-1}$) & &  &  &  \\
\hline 
\hline
\noalign{\smallskip}
\object{HD\,1581  }&$5956\pm44 $&$ 4.39\pm0.13 $&$1.07\pm0.09 $&$ -0.14\pm0.05 $&$-0.01$&$0.03  $&$  0.01 \pm0.07 $\\
\object{HD\,4391  }&$5878\pm53 $&$ 4.74\pm0.15 $&$1.13\pm0.10 $&$ -0.03\pm0.06 $&$ 0.00$&$0.01  $&$  0.01 \pm0.08 $\\
\object{HD\,5133  }&$4911\pm54 $&$ 4.49\pm0.18 $&$0.71\pm0.11 $&$ -0.17\pm0.06 $&$0.05 $&$0.00  $&$  0.03 \pm0.08 $\\
\object{HD\,7570  }&$6140\pm41 $&$ 4.39\pm0.16 $&$1.50\pm0.08 $&$  0.18\pm0.05 $&$0.05 $&$0.09  $&$  0.07 \pm0.07 $\\
\object{HD\,10360 }&$4970\pm40 $&$ 4.49\pm0.10 $&$0.76\pm0.07 $&$ -0.26\pm0.04 $&$0.01 $&$-0.03 $&$ -0.01 \pm0.07 $\\
\object{HD\,10700 }&$5344\pm29 $&$ 4.57\pm0.09 $&$0.91\pm0.06 $&$ -0.52\pm0.04 $&$0.02 $&$-0.03 $&$ -0.01 \pm0.07 $\\
\hline
\end{tabular}
\end{center}
\label{tab15}
\end{table*}

\begin{table*}[p]
\caption[]{Average [X/H] values, with the corresponding dispersions, for the set of planet host stars and the 
comparison sample.}
\begin{center}
\begin{tabular}{lccc}
\hline
\noalign{\smallskip}
Species & $<$[X/H]$>\pm$ rms & $<$[X/H]$>\pm$ rms & Difference \\
 &  (comp. sample) & (planet hosts) & \\
\noalign{\smallskip}
\hline
\hline
\noalign{\smallskip}
C  & $-0.03\pm0.14$ & $0.14\pm0.17$ & 0.17 \\
S  & $-0.18\pm0.14$ & $-0.02\pm0.17$ & 0.16 \\
Zn & $-0.12\pm0.16$ & $0.13\pm0.22$ & 0.25 \\
Cu & $-0.12\pm0.17$ & $0.19\pm0.23$ & 0.31 \\
\noalign{\smallskip}
\hline
\end{tabular}
\end{center}
\label{TabAve}
\end{table*}

\begin{table*}[p]
\caption[]{Comparison of measured {\it EW} of \ion{C}{i} line at 5380 \AA\ from our data and from other authors. }
\begin{center}
\begin{tabular}{lccccccr}
\hline
\noalign{\smallskip}
Star & $T_\mathrm{eff}$ & $\log {g}$       & $\xi_t$   & [Fe/H]  & EW & EW$_{liter}$ & Ref. \\
     & (K)              &  (cm\,s$^{-2}$) & (km\,s$^{-1}$) &  &  &  &  \\
\noalign{\smallskip}
\hline
\hline
\noalign{\smallskip}
\object{HD\,1237}   & $5536\pm50$ & $4.56\pm0.12$ & $1.33\pm0.06$ & $ 0.12\pm0.06$ & $16.7$ & $17.0$ & 1 \\
\object{HD\,9826}   & $6212\pm64$ & $4.26\pm0.13$ & $1.69\pm0.16$ & $ 0.13\pm0.08$ & $45.2$ & $41.5$ & 3 \\
\object{HD\,10697}  & $5641\pm28$ & $4.05\pm0.05$ & $1.13\pm0.03$ & $ 0.14\pm0.04$ & $27.4$ & $24.9$ & 2 \\
\object{HD\,12661}  & $5702\pm36$ & $4.33\pm0.08$ & $1.05\pm0.04$ & $ 0.36\pm0.05$ & $32.3$ & $31.9$ & 2K \\
\object{HD\,13445}  & $5163\pm00$ & $4.52\pm0.00$ & $0.72\pm0.00$ & $-0.24\pm0.00$ & $12.9$ & $ 7.8$ & 1 \\
\object{HD\,22049}  & $5073\pm42$ & $4.43\pm0.08$ & $1.05\pm0.06$ & $-0.13\pm0.05$ & $ 9.5$ & $ 9.8$ & 3 \\
\hline
\end{tabular}
\end{center}
\footnotetext{}{$^1$ from Santos et al. (\cite{San00})}\\
\footnotetext{}{$^2$ from Gonzalez et al. (\cite{Gonz01}); 2K refers to data from Keck spectra}\\
\footnotetext{}{$^3$ from Takeda et al.\ (\cite{Tak01})}\\
\footnotetext{}{$^4$ from Sadakane et al.\ (\cite{Sad02})}\\
\label{tab16}
\end{table*}

\begin{table*}[p]
\caption[]{Comparison of sulphur abundances from our results and from other authors. }
\begin{center}
\begin{tabular}{lccccccr}
\hline
\noalign{\smallskip}
Star & $T_\mathrm{eff}$ & $\log {g}$       & $\xi_t$   & [Fe/H]  & [S/Fe] & [S/Fe]$_{liter}$ & Ref. \\
     & (K)              &  (cm\,s$^{-2}$) & (km\,s$^{-1}$) &  &  &  &  \\
\noalign{\smallskip}
\hline
\hline
\noalign{\smallskip}
\object{HD\,1237}   & $5536\pm50$ & $4.56\pm0.12$ & $1.33\pm0.06$ & $ 0.12\pm0.06$ & $-0.22\pm0.08$ & $ 0.00$ & 1 \\
\object{HD\,9826}   & $6212\pm64$ & $4.26\pm0.13$ & $1.69\pm0.16$ & $ 0.13\pm0.08$ & $-0.33\pm0.09$ & $ 0.06$ & 2 \\
\object{HD\,22049}  & $5073\pm42$ & $4.43\pm0.08$ & $1.05\pm0.06$ & $-0.13\pm0.05$ & $ 0.00\pm0.08$ & $ 0.32$ & 2 \\
\object{HD\,38529}  & $5674\pm40$ & $3.94\pm0.12$ & $1.38\pm0.05$ & $ 0.40\pm0.06$ & $-0.32\pm0.08$ & $ 0.08$ & 2 \\ 
\object{HD\,52265}  & $6105\pm00$ & $4.28\pm0.00$ & $1.36\pm0.00$ & $ 0.23\pm0.00$ & $-0.20\pm0.06$ & $ 0.06$ & 2 \\
\object{HD\,52265}  & $6105\pm00$ & $4.28\pm0.00$ & $1.36\pm0.00$ & $ 0.23\pm0.06$ & $-0.20\pm0.06$ & $-0.12$ & 1 \\
\hline
\end{tabular}
\end{center}
\footnotetext{}{$^1$ from Santos et al. (\cite{San00})}\\
\footnotetext{}{$^2$ from Takeda et al.\ (\cite{Tak01})}\\
\footnotetext{}{$^3$ from Sadakane et al.\ (\cite{Sad02})}\\
\label{tab17}
\end{table*}

\begin{table*}[p]
\caption[]{Comparison of zinc abundances from our results and from other authors. }
\begin{center}
\begin{tabular}{lccccccr}
\hline
\noalign{\smallskip}
Star & $T_\mathrm{eff}$ & $\log {g}$       & $\xi_t$   & [Fe/H]  & [Zn/Fe] & [Zn/Fe]$_{liter}$ & Ref. \\
     & (K)              &  (cm\,s$^{-2}$) & (km\,s$^{-1}$) &  &  &  &  \\
\noalign{\smallskip}
\hline
\hline
\noalign{\smallskip}
\object{HD\,9826}  & $6212\pm64$ & $4.26\pm0.13$ & $1.69\pm0.16$ & $ 0.13\pm0.08$ & $-0.12\pm0.10$ & $ 0.05$ & 1 \\ 
\object{HD\,22049} & $5073\pm42$ & $4.43\pm0.08$ & $1.05\pm0.06$ & $-0.13\pm0.05$ & $-0.12\pm0.08$ & $-0.16$ & 1 \\
\object{HD\,38529} & $5674\pm40$ & $3.94\pm0.12$ & $1.38\pm0.05$ & $ 0.40\pm0.06$ & $-0.14\pm0.08$ & $ 0.38$ & 1 \\
\object{HD\,52265} & $6105\pm00$ & $4.28\pm0.00$ & $1.36\pm0.00$ & $ 0.23\pm0.00$ & $-0.09\pm0.07$ & $ 0.14$ & 1 \\
\object{HD\,75732} & $5279\pm62$ & $4.37\pm0.18$ & $0.98\pm0.07$ & $ 0.33\pm0.07$ & $ 0.16\pm0.09$ & $ 0.48$ & 1 \\
\object{HD\,92788} & $5821\pm41$ & $4.45\pm0.06$ & $1.16\pm0.05$ & $ 0.32\pm0.05$ & $ 0.11\pm0.07$ & $ 0.29$ & 2 \\
\hline
\end{tabular}
\end{center}
\footnotetext{}{$^1$ from Takeda et al.\ (\cite{Tak01})}\\
\footnotetext{}{$^2$ from Sadakane et al.\ (\cite{Sad02})}\\
\label{tab18}
\end{table*}

\begin{table*}[p]
\caption[]{Comparison of copper abundances from our results and from other authors. }
\begin{center}
\begin{tabular}{lccccccr}
\hline
\noalign{\smallskip}
Star & $T_\mathrm{eff}$ & $\log {g}$       & $\xi_t$   & [Fe/H]  & [Cu/Fe] & [Cu/Fe]$_{liter}$ & Ref. \\
     & (K)              &  (cm\,s$^{-2}$) & (km\,s$^{-1}$) &  &  &  &  \\
\noalign{\smallskip}
\hline
\hline
\noalign{\smallskip}
\object{HD\,92788}  & $5821\pm41$ & $4.45\pm0.06$ & $1.16\pm0.05$ & $ 0.32\pm0.05$ & $ 0.19\pm0.07$ & $ 0.52$ & 1 \\
\object{HD\,106252} & $5899\pm35$ & $4.34\pm0.07$ & $1.08\pm0.06$ & $-0.01\pm0.05$ & $-0.02\pm0.07$ & $-0.04$ & 1 \\
\object{HD\,130322} & $5392\pm36$ & $4.48\pm0.06$ & $0.85\pm0.05$ & $ 0.03\pm0.04$ & $ 0.03\pm0.07$ & $ 0.16$ & 1 \\
\object{HD\,134987} & $5776\pm29$ & $4.36\pm0.07$ & $1.09\pm0.04$ & $ 0.30\pm0.04$ & $ 0.19\pm0.06$ & $ 0.52$ & 1 \\
\object{HD\,141937} & $5909\pm39$ & $4.51\pm0.08$ & $1.13\pm0.06$ & $ 0.10\pm0.05$ & $ 0.07\pm0.07$ & $ 0.13$ & 1 \\
\object{HD\,168746} & $5601\pm33$ & $4.41\pm0.12$ & $0.99\pm0.05$ & $-0.08\pm0.05$ & $ 0.10\pm0.07$ & $ 0.03$ & 1 \\
\object{HD\,187123} & $5845\pm22$ & $4.42\pm0.07$ & $1.10\pm0.03$ & $ 0.13\pm0.03$ & $ 0.07\pm0.06$ & $ 0.20$ & 1 \\
\object{HD\,192263} & $4947\pm58$ & $4.51\pm0.20$ & $0.86\pm0.09$ & $-0.02\pm0.06$ & $ 0.01\pm0.08$ & $ 0.04$ & 1 \\
\noalign{\smallskip}
\hline
\end{tabular}
\end{center}
\footnotetext{}{$^1$ from Sadakane et al.\ (\cite{Sad02})}\\
\label{tab19}
 \end{table*}

\end{document}